\documentclass[fleqn,usenatbib]{mnras}

\usepackage{newtxtext,newtxmath}

\usepackage[T1]{fontenc}

\DeclareRobustCommand{\VAN}[3]{#2}
\let\VANthebibliography\thebibliography
\def\thebibliography{\DeclareRobustCommand{\VAN}[3]{##3}\VANthebibliography}

\usepackage{graphicx}	
\usepackage{amsmath}	

\usepackage[dvipsnames]{xcolor}

\usepackage[dvipsnames]{xcolor}

\title[Unequal-mass mergers with SIDM]{Unequal-mass mergers of dark matter haloes with rare and frequent self-interactions}

\author[M. S. Fischer et al.]{Moritz S. Fischer,$^{1}$\thanks{E-mail: moritz.fischer@uni-hamburg.de (UHH)}
Marcus Br\"{u}ggen,$^{1}$
Kai Schmidt-Hoberg,$^{2}$
Klaus Dolag,$^{3,4}$
\newauthor{Antonio Ragagnin,$^{5,6}$
Andrew Robertson$^{7}$}
\\
$^{1}$Hamburger Sternwarte, Universit\"at Hamburg, Gojenbergsweg 112, D-21029 Hamburg, Germany\\
$^{2}$Deutsches  Elektronen-Synchrotron  DESY,  Notkestra{\ss}e  85,  D-22607  Hamburg,  Germany\\
$^{3}$Faculty of Physics, Universit\"ats-Sternwarte M\"unchen, Ludwig-Maximilians-Universität, Scheinerstr. 1, D-81679 M\"unchen, Germany\\
$^{4}$Max-Planck-Institut f\"ur Astrophysik, Karl-Schwarzschild-Str. 1, D-85748 Garching, Germany\\
$^{5}$INAF- Osservatorio Astronomico di Trieste, via G. B. Tiepolo 11, I-34143 Trieste, Italy\\
$^{6}$IFPU - Institute for Fundamental Physics of the Universe, Via Beirut 2, I-34014 Trieste, Italy\\
$^{7}$Institute for Computational Cosmology, Department of Physics, Durham University, South Road, Durham DH1 3LE, UK
}

\date{Accepted XXX. Received YYY; in original form ZZZ}

\pubyear{2021}

\begin{document}
\label{firstpage}
\pagerange{\pageref{firstpage}--\pageref{lastpage}}
\maketitle

\begin{abstract}
Dark matter (DM) self-interactions have been proposed to solve problems on small length scales within the standard cold DM cosmology.
Here, we investigate the effects of DM self-interactions in merging systems of galaxies and galaxy clusters with equal and unequal mass ratios.
We perform $N$-body DM-only simulations of idealized setups to study the effects of DM self-interactions that are elastic and velocity-independent.
We go beyond the commonly adopted assumption of large-angle (rare) DM scatterings, paying attention to the impact of small-angle (frequent) scatterings on astrophysical observables and related quantities.
Specifically, we focus on DM-galaxy offsets, galaxy--galaxy distances, halo shapes, morphology, and the phase--space distribution. 
Moreover, we compare two methods to identify peaks: one based on the gravitational potential and one based on isodensity contours. We find that the results are sensitive to the peak finding method, which poses a challenge for the analysis of merging systems in simulations and observations, especially for minor mergers.
Large DM-galaxy offsets can occur in minor mergers, especially with frequent self-interactions.
The subhalo tends to dissolve quickly for these cases. While clusters in late merger phases lead to potentially large differences between rare and frequent scatterings, we believe that these differences are non-trivial to extract from observations.
We therefore study the galaxy/star populations which remain distinct even after the DM haloes have coalesced. We find that these collisionless tracers behave differently for rare and frequent scatterings, potentially giving a handle to learn about the micro-physics of DM.
\end{abstract}

\begin{keywords}
astroparticle physics -- methods: numerical -- galaxies: haloes -- dark matter
\end{keywords}



\section{Introduction} \label{sec:introduction}

In the standard cosmological model, structures in the Universe such as galaxies and galaxy clusters are thought to form hierarchically, meaning that small objects merge to form larger ones \citep[e.g.][]{Lacey_1993}.
In particular, cosmological $N$-body simulations have been used to investigate structure formation within the standard $\Lambda\mathrm{CDM}$ model
\citep[e.g.][]{Springel_2005n, Boylan-Kolchin_2009, Klypin_2011, Pillepich_2017, Hopkins_2018}.
These simulations agree remarkably well with observations of the large-scale structure \citep[e.g.][]{Springel_2006} and allow us to explain the formation of a wide range of cosmic objects.
In the cosmological standard model, today's Universe mainly consists of dark energy and dark matter (DM), which is thought to be collisionless and cold (cold dark matter -- CDM).
The underlying nature of those two components is largely unknown.

Although the large-scale structure of our Universe is reproduced well by $N$-body simulations assuming only CDM, there seem to be discrepancies between predictions and observations on scales of galaxies.
This is often referred to as the small-scale crisis of $\Lambda\mathrm{CDM}$, which manifests itself in the core-cusp problem, the diversity problem, the too-big-to-fail-problem and the plane of satellites problem \citep[for a review see][]{Bullock_2017}.

Many solutions have been proposed to solve these problems, including alternative models for DM \citep[e.g.][]{Dodelson_1994, Hu_2000}.
One class of these models assumes that DM is not fully collisionless but that DM particles scatter off each other with a non-zero cross-section \citep{Spergel_2000}.
Such models with self-interacting dark matter (SIDM) provide a promising solution to the small-scale crisis \citep[for a review see][]{Tulin_2018}.
The dark sector might be strongly coupled to itself, while interactions with standard model particles are weak enough to fulfil current constraints from laboratory experiments. In several particle physics models, this would be natural, \cite[e.g.][]{Carlson_1992, Kusenko_2001, Mohapatra_2002, Frandsen_2011}.

A variety of SIDM models exist for which the self-interactions can be velocity-independent or velocity-dependent \citep[e.g.][]{Ackermann_2009, Buckley_2010, Loeb_2011, van_den_Aarssen_2012, Tulin_2013a} as several studies have assumed \citep[e.g.][]{Colin_2002, Vogelsberger_2012, Vogelsberger_2013, Vogelsberger_2014, Robertson_2017b, Banerjee_2020, Nadler_2020, Turner_2021}.
The self-interactions could be elastic or inelastic \citep[e.g.][]{Essig_2019, Huo_2019} and the differential cross-section may have different shapes.
The latter implies that the typical scattering angles can be of different sizes.
In the regime of rare self-interacting dark matter (rSIDM), the typical scattering angle is large, the scattering can be isotropic as most studies assumed, but there have also been a few studies with anisotropic scattering \citep[e.g.][]{Robertson_2017b, Banerjee_2020, Nadler_2020}.
In contrast, the regime of frequent self-interacting dark matter (fSIDM) has typical scattering angles that are tiny and thus transfer less momentum per scattering event compared to rSIDM \citep{Kahlhoefer_2014}.
There exists a variety of particle physics models with long-range interactions arising from a mass-less mediator. These models have a strongly anisotropic cross-section, which typically is velocity-dependent. This is the case for mirror DM \citep{Blinnikov_1983, Kolb_1985, Berezhiani_1996, Foot_2004}, atomic DM \citep{Kaplan_2010, Cline_2012, Cyr-Racine_2013} and some other hidden sector DM models \citep{Feng_2009, Foot_2015, Boddy_2016}.

In addition to cosmological simulations, a number of studies have focused on individual mergers to constrain the nature of DM.
Several studies simulate galaxy clusters in a $\Lambda$CDM cosmology including the intracluster medium \citep[ICM;][]{Pool_2006, ZuHone_2011, Machado_2015a, Zhang_2016, Doubrawa_2020, Moura_2020}.
Mergers between galaxy clusters provide important test cases for theories of DM. Dissociative mergers, where the intracluster gas becomes separated from the DM haloes, are especially interesting, with known examples including the Bullet Cluster \citep[e.g.][]{Springel_2007, Mastropietro_2008, Lage_2014}, the `El Gordo' cluster \citep[e.g.]{Donnert_2014, Molnar_2015, Zhang_2015}, the `Sausage' cluster \citep[e.g.]{Donnert_2017, Molnar_2017}, A1758N \citep{Machado_2015b, Monteiro-Oliveira_2017}, and ZwCl008.8+52 \citep{Molnar_2018}.

In the context of SIDM, merging galaxy clusters have attracted attention since offsets between the DM component and the galaxies may provide evidence for DM self-interactions.
There have been several observational studies on galaxy cluster offsets that sparked a debate on the reliability of the largest offsets measured \citep[e.g.][]{Bradac_2008, Dawson_2012, Dawson_2013, Jee_2014, Jee_2015, Harvey_2017, Peel_2017, Taylor_2017, Wittman_2018}. Here, we focus on offsets that could discriminate between rSIDM and fSIDM.

From the theoretical side, several numerical studies of galaxy cluster mergers with SIDM have been carried out.
Although SIDM encompasses a wide range of models, most studies assumed elastic velocity-independent isotropic scattering \citep{Randall_2008, Kim_2017b, Robertson_2017a}.
\cite{Robertson_2017b} performed the first study of anisotropic scattering.
Nevertheless, these studies have been limited to large-angle scattering. Meanwhile, frequent self-interactions have only very recently been implemented in $N$-body simulations based on an effective drag force~\citep{Fischer_2021}, which is more generally applicable than the description in terms of a heat conduction approach~\citep{Kummer:2019yrb}.

\cite{Kim_2017b} have performed a parameter study of equal-mass mergers that demonstrated that offsets between DM and galaxies can arise from rare self-interactions.
Building on this, \cite{Fischer_2021} showed that even larger offsets arise when fSIDM is considered, as expected by \cite{Kahlhoefer_2014}.
Unequal-mass mergers are interesting since they occur much more frequently than equal-mass mergers.
In rSIDM, they have been studied in simulations which reproduce the Bullet Cluster (1E 0657--56) \citep{Randall_2008, Robertson_2017a,Robertson_2017b}.
In addition, there are many studies on structure formation in rSIDM using cosmological simulations \citep[e.g.][]{Vogelsberger_2012, Vogelsberger_2014,  Vogelsberger_2013, Peter_2013, Rocha_2013, Despali_2019, Banerjee_2020, Nadler_2020, Robertson_2020, Vega-Ferrero_2020, Sameie_2021, Shen_2021}.

In this paper, we focus on idealized equal and unequal-mass mergers of galaxies and galaxy clusters using $N$-body simulations.
Although the ICM contributes significantly to the total mass of galaxy clusters, we neglect the contribution from gas and consider idealized systems that only consist of DM and galaxies.
Likewise, for the galaxy simulations, we also consider only DM  and stars.
In our model, the DM is subject to self-interactions that are elastic and velocity-independent, and we investigate, both, rare and frequent self-interactions.

In Section~\ref{sec:numerical_setup}, we briefly describe the simulation code with its implementation for DM self-interactions and explain our initial conditions.
In Section~\ref{sec:methods}, we present the methods of our analysis, especially for the peak finding.
Subsequently, we present the results of our simulations in Section~\ref{sec:results}.
In particular, we measure offsets and shapes of the merging systems, investigate the morphology and phase--space distribution of the mergers and compare peak finding methods.
In Section~\ref{sec:discussion}, we discuss our results and their physical implications in the light of the assumptions we have made.
Finally, we summarize and conclude in Section~\ref{sec:conclusion}.
Additional details and plots are provided in the appendices.

\section{Numerical setup} \label{sec:numerical_setup}

In this section, we describe our numerical setup including the simulation code and the description of the initial conditions.

\subsection{Simulation code and implementation of self-interactions}

For our simulations, we use the cosmological $N$-body code \textsc{gadget-3}, which is a successor of \textsc{gadget-2} \citep{gadget2}.
For rare and frequent self-interactions, we are using the implementation described in \cite{Fischer_2021}.
This means that for the rare self-interactions, a similar scheme to the one introduced by \cite{Rocha_2013} is used and the scheme for frequent self-interactions is based on an effective description employing a drag force \citep{Kahlhoefer_2014} and was introduced in \cite{Fischer_2021}.
In addition to the existing implementation, we added a time-step criterion and slightly modified the implementation of rSIDM as described below.

\subsubsection{Time-step criterion for self-interactions}

For both rare and frequent self-interactions, we implemented a time-step criterion that limits the maximum allowed time-step for each particle.
In this context, our explanation about the time-step scaling for fSIDM in section~2.4 \cite{Fischer_2021} might be of interest.

In \textsc{gadget-3}, particles are assigned an individual time-step and our new criterion does not allow it to be larger than $\Delta t_\mathrm{si}$, which is defined as
\begin{equation}
    \Delta t_\mathrm{si} = \tau \frac{h^3}{\omega_\mathrm{max} \, m} \, .
\end{equation}
Here, $h$ denotes the kernel size (which is used in the implementation of the self-interactions) and $m$ denotes the numerical particle mass. $\omega_\mathrm{max}$ is computed as follows
\begin{equation}
    \omega_\mathrm{max} = \max(\omega) \qquad \textnormal{with} \quad \omega = \frac{\sigma_\mathrm{\tilde{T}}}{m_\chi} \, \Delta v \, ,
\end{equation}
where $\Delta v$ denotes the relative velocity of two particles.
We compute $\omega$ for each particle interaction, and then determine $\omega_\mathrm{max}$ for a given particle as the maximum value of $\omega$ over all interactions involving that particle. The size of the time-step can be adjusted by the numerical factor $\tau$.
For frequent self-interactions, this enables the control of the relative velocity change per particle interaction.
The simulations in this paper were conducted with a value of $\tau = 0.1$.

\subsubsection{rSIDM -- relabelling of particles} \label{sec:numerics_relabelling}

In this paper, we study mergers where we know which halo each particle initially belongs to.
This information is used in the peak finding as described in Section~\ref{sec:method_peak_find_pot}.
However, when considering rSIDM, the question of which DM halo a particle belongs to is not as clear as it might seem.
Consider two indistinguishable particles that belong to different haloes.
If they scatter by an angle $\theta < \pi/2$, this cannot be distinguished from an event with a scattering angle $\pi-\theta > \pi/2$ where the two particles are exchanged.
Hence, we modify our rSIDM implementation such that particles are not allowed to scatter by angles larger than 90°, but use the smaller angle instead.
This has the same effect as relabelling (exchange of host halo labels) the particles for scattering angles larger than 90°.
In appendix~\ref{sec:rSIDMa_comp}, we study the effects of this modification.

\subsection{Initial conditions and simulation parameters}

In this paper, we perform a parameter study of head-on collisions of DM haloes. Initially, the individual haloes are assumed to follow a Navarro--Frenk--White (NFW) profile \citep{Navarro_1996}. 
As the total mass of an NFW halo is infinite, it needs to be truncated at some radius, which we set as 20 times the scale radius, $r_s$.
We ran simulations for galaxy and cluster-scale mergers, i.e.\ the main haloes have virial masses of $M_\mathrm{vir, main} = 10^{12}$ or $10^{15} \, \mathrm{M_\odot}$, respectively.
For all haloes, the concentration parameter in the NFW profile, $c$, was chosen according to \cite{Dutton_2014}. In Tab.~\ref{tab:halos}, we give the corresponding scale radius and scale density for our haloes.
All haloes have an equal number of DM and collisionless galaxy/star particles. Each of these components follow an NFW profile with the same value for $r_s$.
The cluster-scale simulations have a mass resolution of $m_\mathrm{DM} = 2 \cdot 10^{8} \, \mathrm{M_\odot}$ for the DM particles and $m_\mathrm{Gal} = 4 \cdot 10^{6} \, \mathrm{M_\odot}$ for the galaxy particles.
For the galaxy-scale simulations, the mass resolution is: $m_\mathrm{DM} = 2 \cdot 10^5 \, \mathrm{M_\odot} $ and $m_\mathrm{Star} = 4 \cdot 10^3 \, \mathrm{M_\odot}$.
In addition, the haloes contain one more massive collisionless particle at their centre.
For the cluster-scale simulations, it may be interpreted as the brightest cluster galaxy (BCG) although it has only a mass of $m_\mathrm{BCG} = 7 \cdot 10^{10} \, \mathrm{M_\odot}$.

\begin{table}
    \centering
    \begin{tabular}{c|c|c}
        $M_\mathrm{vir}$ & $r_s$ & $\rho_0$ \\
        $(\mathrm{M_\odot})$ & (kpc) & ($\mathrm{M_\odot}$ kpc$^{-3}$) \\ \hline
        $10^{15}$ & 389.31 & $1.33 \cdot 10^6$\\
        $2 \times 10^{14}$ & 194.76 & $1.91 \cdot 10^6$\\
        $10^{14}$ & 144.53 & $2.24 \cdot 10^6$\\
        $10^{12}$ & 19.92 & $6.56 \cdot 10^6$\\
        $2 \times 10^{11}$ & 9.97 & $9.64 \cdot 10^6$\\
        $10^{11}$ & 7.40 & $1.14 \cdot 10^7$\\
    \end{tabular}
    \caption{The scale radius $r_s$ and the density $\rho_0 \equiv 4 \, \rho(r_s)$ are given for our initial NFW haloes, which have the virial mass $M_\mathrm{vir}$.}
    \label{tab:halos}
\end{table}

All our mergers are head-on mergers, i.e.\ their impact parameters, $b$, equal zero.
Initially, they are separated by a distance $d_\mathrm{ini}$ and they have a relative velocity of $v_\mathrm{ini}$.
We simulate collisionless DM as well as rare and frequent interacting DM with several cross-sections.
The initial velocity is chosen such that the sub halo is still gravitationally bound to the main halo.
An overview of all runs is given in Tab.~\ref{tab:runs}.
In addition, we simulated the cluster-scale setup of the 1:10 merger employing a cross-section of $\sigma_\mathrm{\tilde{T}}/m = 0.5 \mathrm{cm}^2 \, \mathrm{g}^{-1}$ with half the resolution to check that our results are converged.

For the self-interactions, we use the momentum transfer cross-section defined as\footnote{Note that for the case of identical particles, as implicitly assumed here, this definition is equivalent to the one advocated by \cite{Robertson_2017b} and \cite{Kahlhoefer:2017umn}.
}
\begin{equation} \label{eq:momentum_transfer_cross_section}
    \sigma_\mathrm{\tilde{T}} = 4\pi \int_0^1 \frac{\mathrm{d} \sigma}{\mathrm{d} \Omega_\text{cms}} (1 - \cos \theta_\text{cms}) \mathrm{d} \cos \theta_\text{cms} \, .
\end{equation}

\begin{table}
    \centering
    \begin{tabular}{c|c|c|c|c|c|c}
        $M_\mathrm{vir, main}$ & MMR & $d_\mathrm{ini}$& $\Delta v_\mathrm{ini}$ & $\sigma_\mathrm{\tilde{T}}/m$\\
        $(\mathrm{M_\odot})$ & & (kpc) & (km s$^{-1}$) & $(\mathrm{cm}^2 \, \mathrm{g}^{-1})$ \\ \hline
        $10^{15}$ & 1:1 & 4000 & 1000 & 0.0, 0.1, 0.3, 0.5 \\
        $10^{15}$ & 1:5 & 4000 & 1000 & 0.0, 0.1, 0.3, 0.5 \\
        $10^{15}$ & 1:10 & 4000 & 1000 & 0.0, 0.1, 0.3, 0.5 \\
        $10^{12}$ & 1:1 & 500 & 140 & 0.0, 1.0, 2.0 \\
        $10^{12}$ & 1:5 & 500 & 140 & 0.0, 1.0, 2.0 \\
        $10^{12}$ & 1:10 & 500 & 140 & 0.0, 1.0, 2.0 \\
    \end{tabular}
    \caption{Initial condition and simulation parameters for the runs presented in this paper.
    $M_\mathrm{vir, main}$ denotes the virial mass of the main halo, MMR gives the merger mass ratio in terms of the virial mass.
    The initial separation of the two haloes centres is given by $d_\mathrm{ini}$, their initial relative velocity is $\Delta v_\mathrm{ini}$ and they are all head-on collisions.
    The self-interaction cross-section is $\sigma_\mathrm{\tilde{T}}$ (see equation~\ref{eq:momentum_transfer_cross_section}) and the given values have been simulated with rare and frequent self-interactions, except of $\sigma_\mathrm{\tilde{T}}=0.0$ which corresponds to CDM.
    }
    \label{tab:runs}
\end{table}

In addition, we employ a fixed gravitational softening length of $\epsilon = 1.2 \, \mathrm{kpc}$ for the cluster-scale simulations and $\epsilon = 0.06 \, \mathrm{kpc}$ for the galaxy-scale simulations.
We use an adaptive kernel size for the DM self-interactions, which varies to keep the number of neighbours within each particles' kernel, $N_\mathrm{ngb}$, equal to 64.
For fSIDM, a larger number would lead to more interactions, enabling the use of a larger time-step and reducing numerical noise, but at the same time it would reduce spatial resolution and would require the computation of more particle interactions.
As a compromise, we choose to use the same value as in \cite{Fischer_2021}.

\section{Methods} \label{sec:methods}

In this section, we describe how we determine the peaks of our particle distributions and then how we define offsets. As we will describe below, the peak finding is a complex issue and crucial for the study of self-interactions in halo mergers. Finally, we will explain how we measure halo shapes and compare time-scales across simulations.

\subsection{Peak finding} \label{sec:method_peak_find}

In the literature, one can find various methods to find peaks of particle distributions.
For instance, the shrinking spheres/circles method \citep{Power_2003} or parametric fits \citep{Robertson_2017a} or the search for density maxima based on kernel density estimates \citep{Kim_2017b}.
Finding peaks for an unequal-mass merger is more difficult than for an equal-mass merger as the peak for the less massive halo vanishes faster and is harder to detect.
In this paper, we use two methods to find peaks, as we explain in the following subsections.

\subsubsection{Gravitational potential based peaks} \label{sec:method_peak_find_pot}

We employed a peak search strategy that is based on the gravitational potential energy of the particles.
We use the information of which halo a particle initially belongs to, and perform the search for the most gravitationally bound particles of each halo and each particle type separately.
For instance, the DM potential based peak of the main halo is the location of the DM particle that experience the lowest potential originally from the main halo, where the potential at each particle is calculated with respect to only the other DM particles that were originally part of the main halo. While this approach does not directly map to observationally available information, it does give an insight into the underlying merger dynamics.

In order to speed up the peak search, we employ an octree-like structure to cluster the particles, where every node is required to contain no more particles than a given maximum.
In the first step, we compute the potential using the nodes of the tree and search for local minima.
Thus, we estimate the potential at the centre of mass of each node and compare it to the neighbouring nodes.

In a second step, we investigate particles close to the minima, i.e.\ particles that belong to the corresponding node and its neighbours.
For the computation of the binding energy of individual particles, we also use the tree nodes, such that distant particles are not evaluated individually, but are clustered in nodes.
For the main halo, we search only for the global minimum, which is usually the only minimum, although there can be more.
The main peak is then given by the coordinates of the particle where the potential is minimal.

For the subhalo we investigate the deepest minimum, but also the second deepest local minimum provided it exists.
Typically, a second potential minimum forms at the centre of the main halo and after a while it becomes the dominant peak as the subhalo merges with the main one.
It is worth mentioning that the formation of the second peak is strongly affected by the relabelling procedure for rSIDM, which is implemented via a limited scattering angle as described in Section~\ref{sec:numerics_relabelling}.
In Appendix~\ref{sec:rSIDMa_comp}, we demonstrate the effects from this procedure.

If the subhalo contains a second peak, we check whether the first one coincides with the peak of the main halo. If this is the case, we take the second one as the peak of the subhalo.
From some point in time on, this second peak is no longer present as a minimum in the potential and, consequently, we are no longer able to determine the peak position of the dissolving subhalo.

For our computation of the gravitational potential, we also employ a gravitational softening length to avoid artefacts from very close particles.
In contrast to the actual simulations and for the sake of simplicity, we use Plummer softening \citep[e.g.][]{Dyer_1993}.
A large softening length would impact the peak positions. Consequently, we choose a value that is small enough to obtain reliable peaks but large enough to avoid misdetections due to close particles. 
For the analysis of the cluster-scale simulations, we use $\epsilon = 1.2 \, \mathrm{kpc}$ and for the galaxy-scale simulations, we employ $\epsilon = 0.06 \, \mathrm{kpc}$.

Finally, we estimate the error for the peaks by bootstrapping the particle distribution 24 times and determine the peaks again.
We obtain the error on the peak position by simply using the standard deviation of the peaks obtained from the bootstrapped haloes. Finally, we set the error to a value that has at least the value of the softening length.

\begin{figure}
    \centering
    \includegraphics[width=\columnwidth]{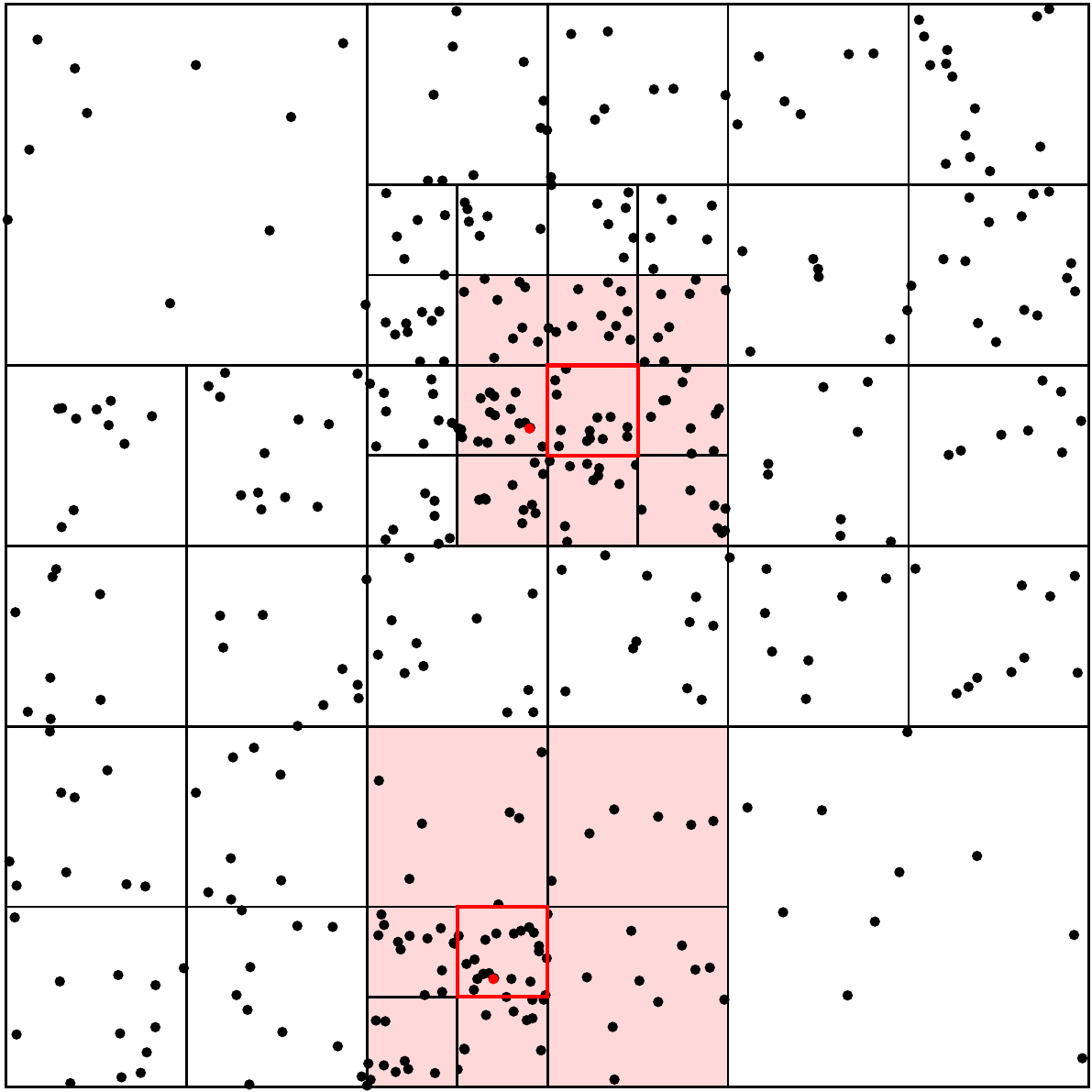}
    \caption{Illustration of the peak finding method.
    Particles (black dots) are clustered in cells (squares) of different sizes adapted to the particle number density.
    On the basis of the cells, a search for local potential minima (squares with red lines) is performed.
    The particles which have locally the lowest potential (red dots) are searched for in a neighbouring area around the minima (red shaded regions).}
    \label{fig:peakfind}
\end{figure}

The peak finding algorithm is illustrated in Fig.~\ref{fig:peakfind} and
can be broken down into five steps as follows:
\begin{enumerate}
    \item Generation of the mesh.
    \item Estimation of the potential at the positions of the cells.
    \item Search for local minima by comparison with neighbour cells.
    \item Determination of the particle with the lowest potential.
    \item Bootstrap distribution to obtain errors on peaks, redo (ii)--(iv).
\end{enumerate}

The peak finding method we are using is not affected by projection effects because we work in all three dimensions.
For peaks of collisionless particle components, this typically leads to a peak that coincides with the position of the single tracer particle placed initially at the halo centre.
Note that this is in contrast to methods that work in projection as in \cite{Kim_2017b} or \cite{Fischer_2021}.
However, the three-dimensional approach may not be ideal when comparing DM-galaxy offsets of simulations to observations, where the three-dimensional information is not available.
Moreover, our approach relies on knowledge to which halo each particle originally belongs, which can only be traced in simulations.

For a comparison with observations, fits of parametric models seem to be more interesting.
They are often used to analyse observational data.
Among other things, \cite{Robertson_2017b} employed this method to determine the positions of different components in simulations of a system that was designed to mimic the Bullet Cluster.
For isotropic scattering, their measured offsets arise solely from fitting the wake of scattered particles, whereas unscattered particles behave in the same way as the collisionless galaxy particles.
In consequence, measured offsets depend strongly on the chosen method and our potential based approach may lead to smaller offsets than parametric fits.
Moreover, spherically symmetric parametric models may not always provide a good description of the DM distribution and an asymmetric model is favourable \citep{Taylor_2017}.
This problem does not arise in other methods, such as the shrinking spheres method.

But the shrinking spheres or its two-dimensional analogue, the shrinking circles method suffers from more severe problems. It is highly sensitive to the starting position and radius \citep{Robertson_2017a}. 
Moreover, for a multiple peak search, extra guidance would be needed to detect the different peaks which is not necessary in the method based on the gravitational potential that can detect multiple peaks more easily.
Finally, the position of the subhalo can be strongly affected by the density gradient of the main halo \citep{Robertson_2017a}.

\subsubsection{Isodensity contour based peaks} \label{sec:method_peak_find_iso}

In addition to the peak finder based on the gravitational potential, we use a second method which may be more easily transferable to observations. In this second method, we determine the peaks of the particle distributions based on isodensity contours in projected two-dimensional density maps  \citep[similar to][]{Kahlhoefer_2014}. To this end, we project the density for each component onto a plane in which the merger axis lies and determine how many regions exceed a given threshold in surface density.
Initially, we choose a threshold close to the maximum surface density and lower it until we find two separate regions that exceed the threshold.
For each region, we compute the centroid of the particles that belong to this region, which gives us the peak position.
We perform this procedure for the DM and galactic component, separately.
In contrast to the peak finder described above, this method does not require any information about the origin of the particles.
In order to still be able to assign peaks to haloes, we linearly extrapolate the peak position of a halo from the past and compare it to the identified peaks.
The one which is closer to the linear prediction is associated with the halo.

If the centres of the haloes are too close, we are no longer able to identify them as separate peaks as we find only one region exceeding the density threshold.
In this case, we use the only peak found for the two haloes. 
As a result, we are not able to give accurate peak positions for very small separations and do not show quantities derived from the peaks if they are close.

Errors on the peak positions are computed via bootstrapping the particle distribution 24 times. In general, the obtained errors are tiny and thus usually not visible in our plots.

\subsection{Offsets} \label{sec:method_offset}

Here, we describe how we measure offsets between DM and galaxies, i.e.\ the distance between their respective peaks.
There are multiple ways that they could be defined, i.e.\ how their sign is chosen, but here we define the offsets between DM and the component $i$ by
\begin{equation} \label{eq:offset}
    \mathrm{offset} \equiv x_\mathrm{DM} - x_i \, ,
\end{equation}
where the coordinate along the merger axis is given by $x$.
Note that this definition is different from the one we used previously in \cite{Fischer_2021}.

\subsection{Halo shapes} \label{sec:method_shape}

In order to compute halo shapes we use the inertia tensor $\mathbf{I}$, with its moments of inertia, i.e. its eigenvalues $I_1$, $I_2$, and $I_3$. For $N$ point masses $m_n$ at position $\mathbf{r}_n$, the inertia tensor is
\begin{equation}
    \mathbf{I} \equiv \sum^N_n m_n\,[(\mathbfit{r}_n\cdot\mathbfit{r}_n)\,\mathbf{1}-\mathbfit{r}_n\otimes\mathbfit{r}_n] \,.
\end{equation}
Here, $I_1$ corresponds to the principal axis (or eigenvector) which is most closely aligned to the merger axis. 
The ratio of the moments of inertia gives us a shape variable

\begin{equation} \label{eq:shape}
   s \equiv \frac{2 \, I_1}{I_2+I_3} \, .
\end{equation}
For our head-on mergers, we expect $I_2 = I_3$ due to the symmetry of the system and initially our haloes are spherical, which implies $s = 1$. Values larger than one correspond to oblate haloes and values smaller than one to prolate haloes.
In Section~\ref{sec:results}, we compute the halo shape separately for the components of the haloes and with respect to the peaks determined according to the method based on the gravitational potential described in Section~\ref{sec:method_peak_find}. We consider only particles that are closer than twice the scale radius of the initial NFW profiles.

In contrast to our shape definition often a reduced inertia tensor is used \citep[e.g.][]{Allgood_2006, Bett_2012, Peter_2013, Vargya_2021}. This is in the context of measuring the shape as a function of distance. According to \cite{Zemp_2011}, the $1/r^2$ weighting of the reduced inertia tensor does not improve the shape measurement and they recommend using elliptical shells. However, for our work, we are only interested in an estimate of the shape that allows us to understand qualitative differences between DM models as a function of time. That is why we pursue a simplified approach.

\subsection{Merger times} \label{sec:method_time}

Self-interactions can change the merger time of a system, which can be problematic for a comparison between simulations using different cross-sections.
A system evolved with SIDM may have reached the second pericentre but when simulated with CDM after the same time, it could be in a phase before the second pericentre.
To allow comparison between the same stages in the evolution of a merger, we define an internal time $\tau$ of the system

\begin{equation} \label{eq:time}
    \tau \equiv \frac{t - t_\mathrm{first\,pericentre}}{t_m} , 
\end{equation}
where $t_m = t_\mathrm{second\,pericentre} - t_\mathrm{first\,pericentre}$ gives us the merger time.
By definition, $\tau=0$ corresponds to the first pericentre passage and $\tau=1$ to the second pericentre passage.
For the analysis of the simulation, we use the BCGs/BHs to compute the time $\tau$.
This has the advantage that the time is independent of the peak finding algorithm and thus always known.

\section{Results} \label{sec:results}

In this section, we present our results on equal and unequal-mass head-on mergers, both, for frequent and rare self-interactions.
In particular, we focus on the morphology (Section~\ref{sec:results_morphology}), DM-galaxy offsets (Section~\ref{sec:results_offsets}), shapes of the haloes (Section~\ref{sec:results_shapes}), compare fSIDM and rSIDM (Section~\ref{sec:results_comp}) as well as the peak finding methods (Section~\ref{sec:results_peakfind}) and examine the phase--space distribution (Section~\ref{sec:results_phase_space}).
The numerical setup used to produce the results is described in Section~\ref{sec:numerical_setup} and the methods employed to analyse the data are explained in Section~\ref{sec:methods}.

In the following, we call the more massive halo the `main halo' and the less massive one the `subhalo'.
We will use this terminology even in the case of an equal-mass merger in which case the assignment of the `main halo' and `subhalo' is arbitrary.

\begin{figure}
    \centering
    \includegraphics[width=\columnwidth]{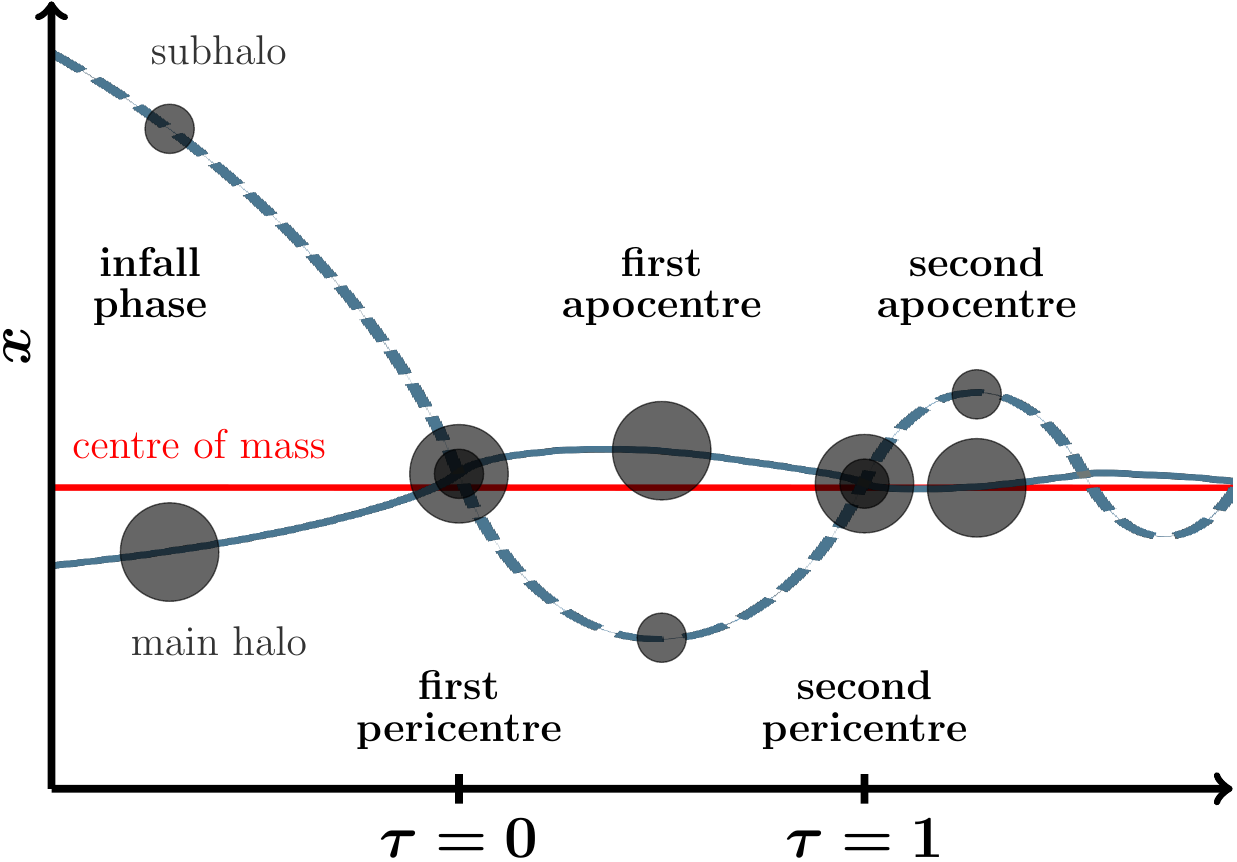}
    \caption{Illustration of the evolution of an unequal-mass merger without self-interactions. The halo position along the merger axis is shown as a function of the internal merger time $\tau$ (see Eq.~\ref{eq:time}).}
    \label{fig:merger_cartoon}
\end{figure}

In Fig.~\ref{fig:merger_cartoon}, we illustrate the evolution of an unequal-mass merger and indicate the different evolution stages of the system.
During the infall phase, the self-interactions do not affect the merger, apart from core formation in the two haloes.
At the first pericentre passage, self-interactions can be strong and decelerate the DM component.
This can lead to a smaller separation of the haloes at the first apocentre passage and a shorter merger time-scale as well as other phenomena such as offsets between the DM and galactic/stellar components.
Depending on the DM physics, the haloes coalesce at different rates, such that there can be further apocentre passages or not.
In the most extreme case, self-interactions are so strong that the haloes coalesce on contact.
Since stars or galaxies are not subject to self-interactions, they behave differently from SIDM but are affected by the overall gravitational potential.
This can lead to differences in the distribution of stars and galaxies between different DM models.
These differences tend to grow with time as we will see in the following analysis of our merger simulations.

To simplify the discussion, we will largely concentrate on a 1:10 cluster-scale merger for CDM, rSIDM, and fSIDM in the following. We will fix the cross-section to $\sigma_\mathrm{\Tilde{T}}/m = 0.5 \, \mathrm{cm}^2 \, \mathrm{g}^{-1}$ for the self-interacting cases, before we come to a comparison of the different mass ratios and cross-sections. Further details on additional runs with other parameters can be found in the Appendices.

\subsection{Morphology} \label{sec:results_morphology}

\begin{figure*}
    \centering
    \includegraphics[width=\columnwidth]{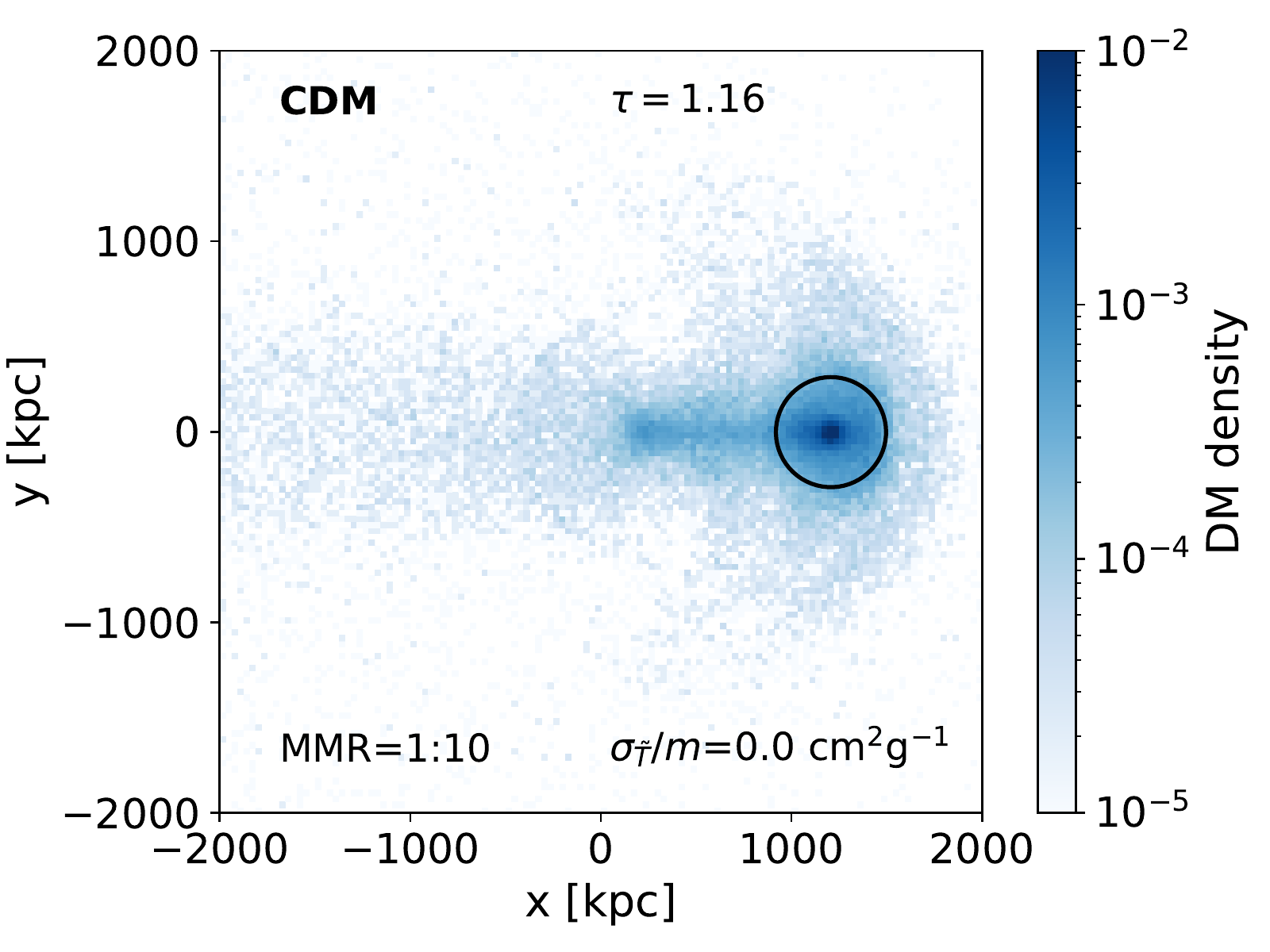}
    \includegraphics[width=\columnwidth]{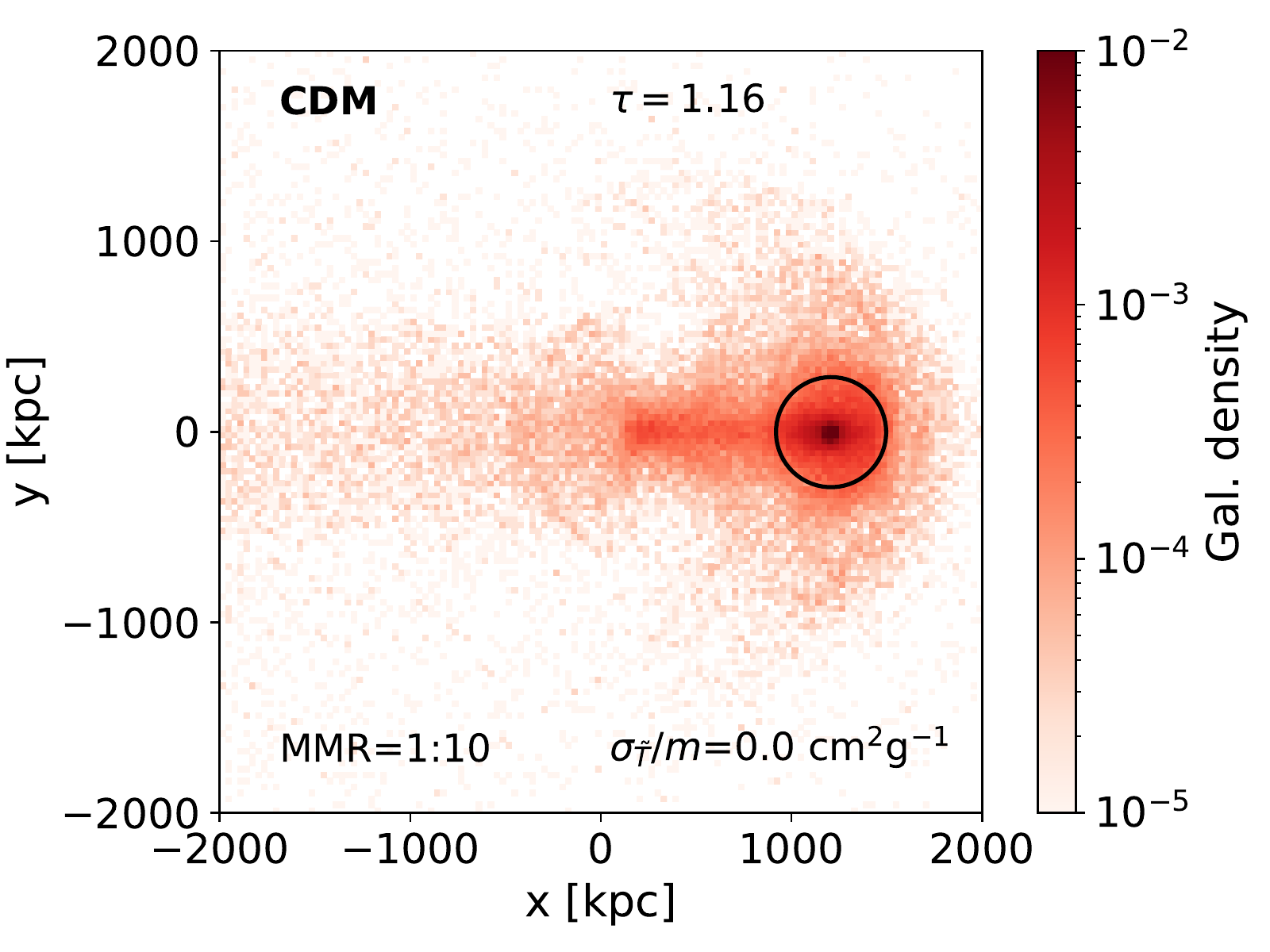}
    \includegraphics[width=\columnwidth]{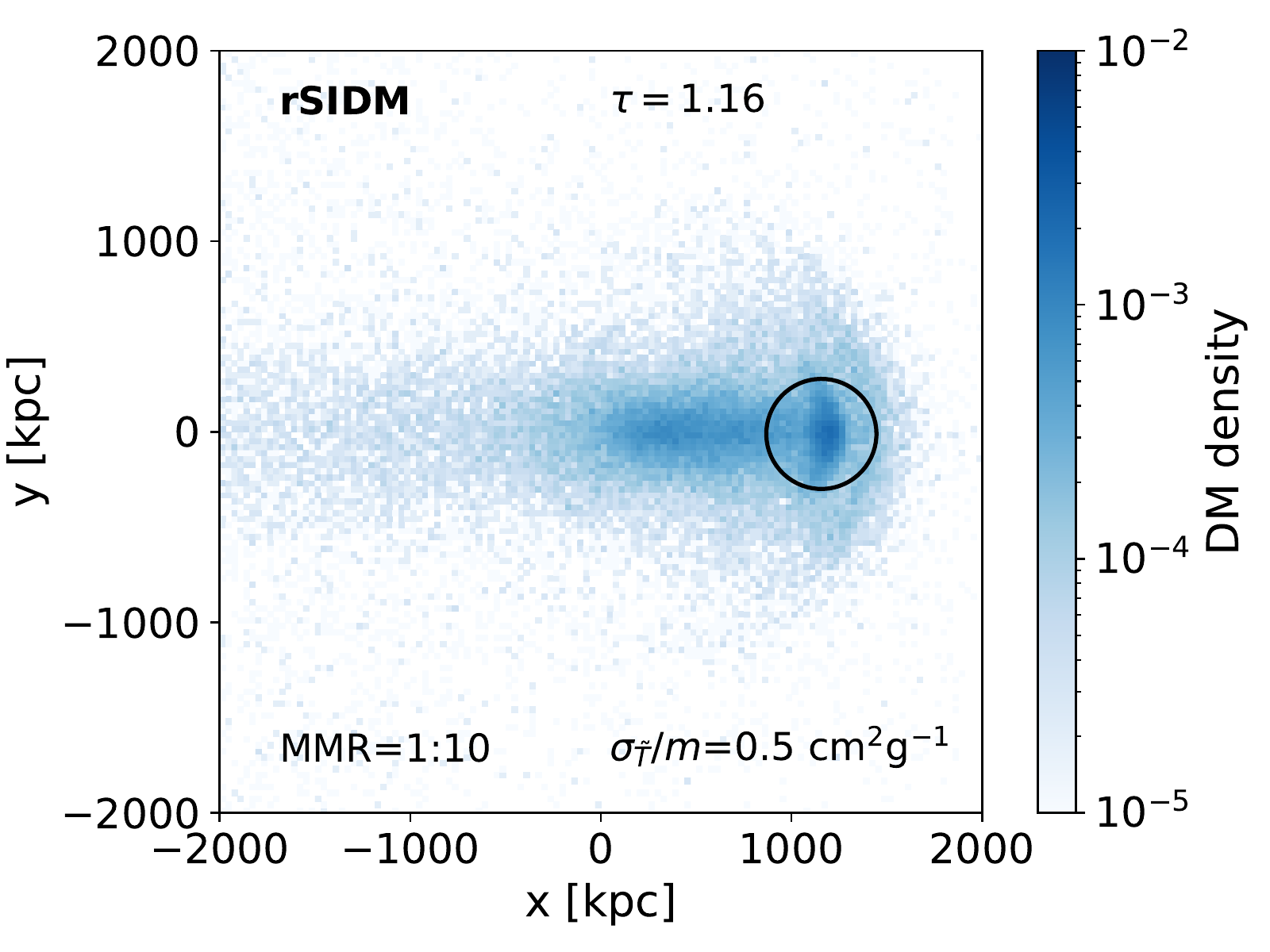}
    \includegraphics[width=\columnwidth]{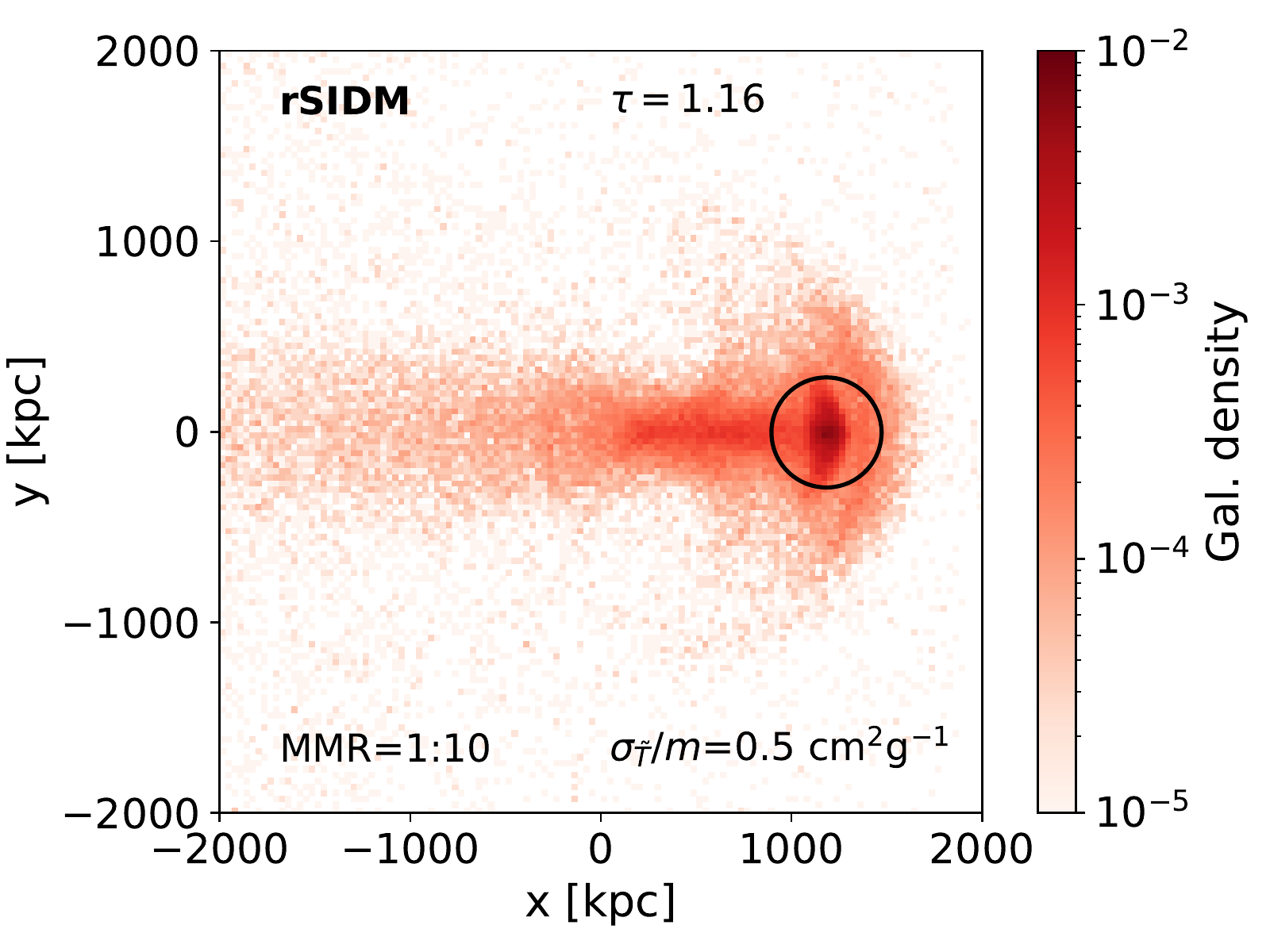}
    \includegraphics[width=\columnwidth]{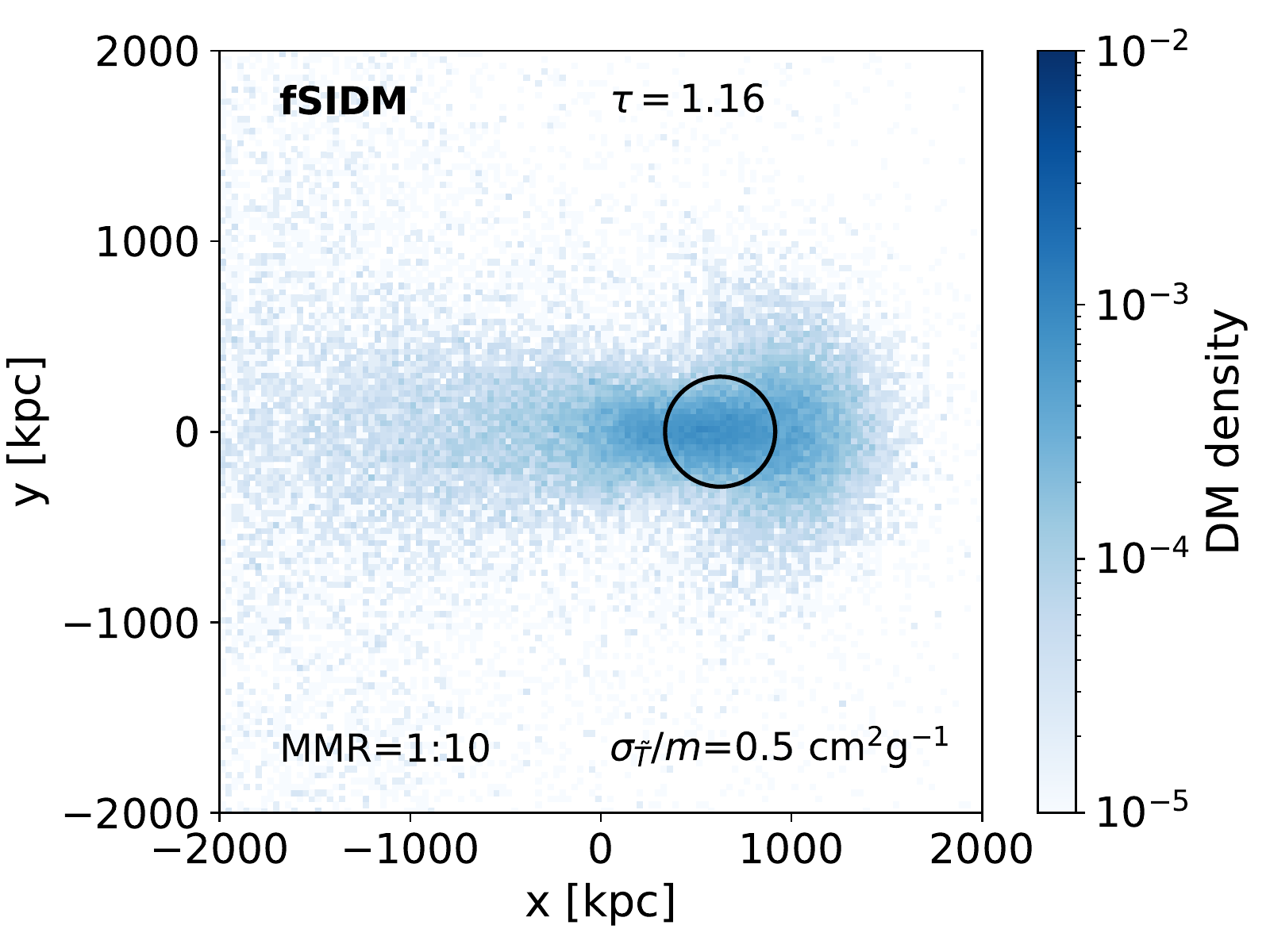}
    \includegraphics[width=\columnwidth]{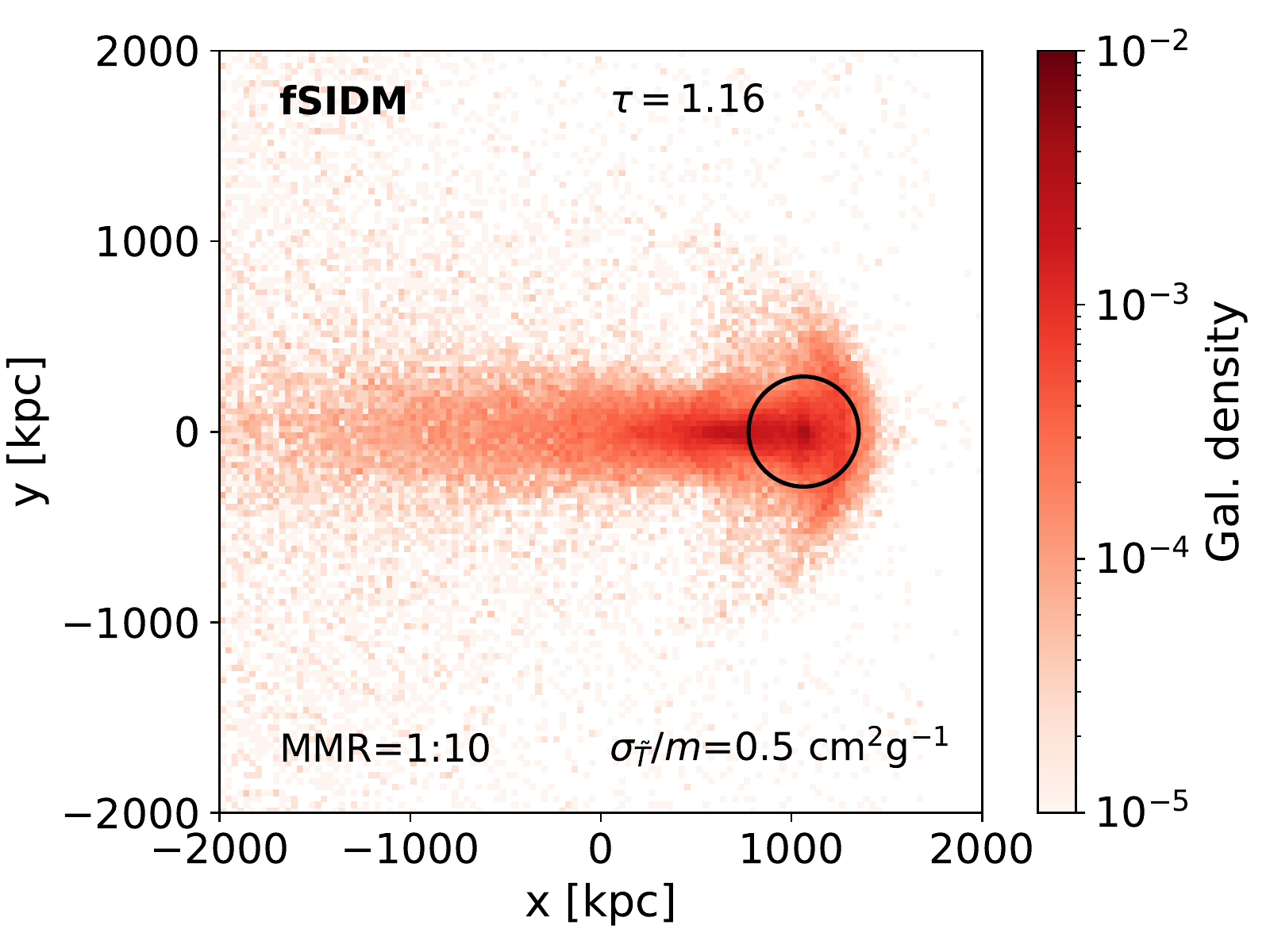}
    \caption{The subhalo's physical density of the DM (left-hand side) and the galactic component (right-hand side) in the merger plane is shown for cluster-scale mergers with an MMR of 1:10.
    The upper panel gives the density for the CDM run, the middle panel for rSIDM, and the bottom panel for fSIDM.
    All panels display the density for $\tau = 1.16$, i.e.\ sometime after the second pericentre passage.
    At this stage, the subhalo is moving in the positive $x$-direction.
    The self-interaction cross-section is $\sigma_\mathrm{\Tilde{T}}/m = 0.5 \, \mathrm{cm}^2 \, \mathrm{g}^{-1}$.
    The black circles are drawn around the potential based peak position of the subhalo and have a radius of twice the initial scale radius.
    Hence, they indicate the area from which the particles for the shape computation are selected.
    However, they are shown even in the case where we considered the peaks as too uncertain for the following analysis (this concerns rSIDM and fSIDM).
    In the supplementary material, we provide the time evolution as a video.
}
    \label{fig:dens_map_R10}
\end{figure*}

Let us start with examining the physical DM and galactic densities in the plane of the merger, where we consider particles within a slice of 100 kpc height.
An illustration of the time evolution of the merger is provided as supplementary material.
For convenience we show both, the density of the two haloes combined as well as only the density of particles which originally belonged to the subhalo to facilitate the physical intuition of the merger process. We also present DM and galactic components separately for clarity. At the first pericentre passage, differences between the different DM models are still very small but they grow over time and become significant at later merger stages, so we will mainly concentrate on these in the following. 
In Fig.~\ref{fig:dens_map_R10}, we show the subhalo density at $\tau = 1.16$, i.e.\ some time after the second pericentre passage.
At these later stages in the evolution we do observe some differences between frequent and rare scatterings. 

For example, the DM densities of the left-hand column show that matter is most concentrated for CDM, less for rSIDM, and least for fSIDM.
The fSIDM subhalo dissolves faster than its rSIDM counterpart thus distinct DM peaks are only detectable for a shorter period of time.
This is related to differences in the gravitational potential, which affect the galaxy particles and creates distributions that differ significantly from each other (see right-hand column of Fig.~\ref{fig:dens_map_R10}).

\begin{figure}
    \includegraphics[width=\columnwidth]{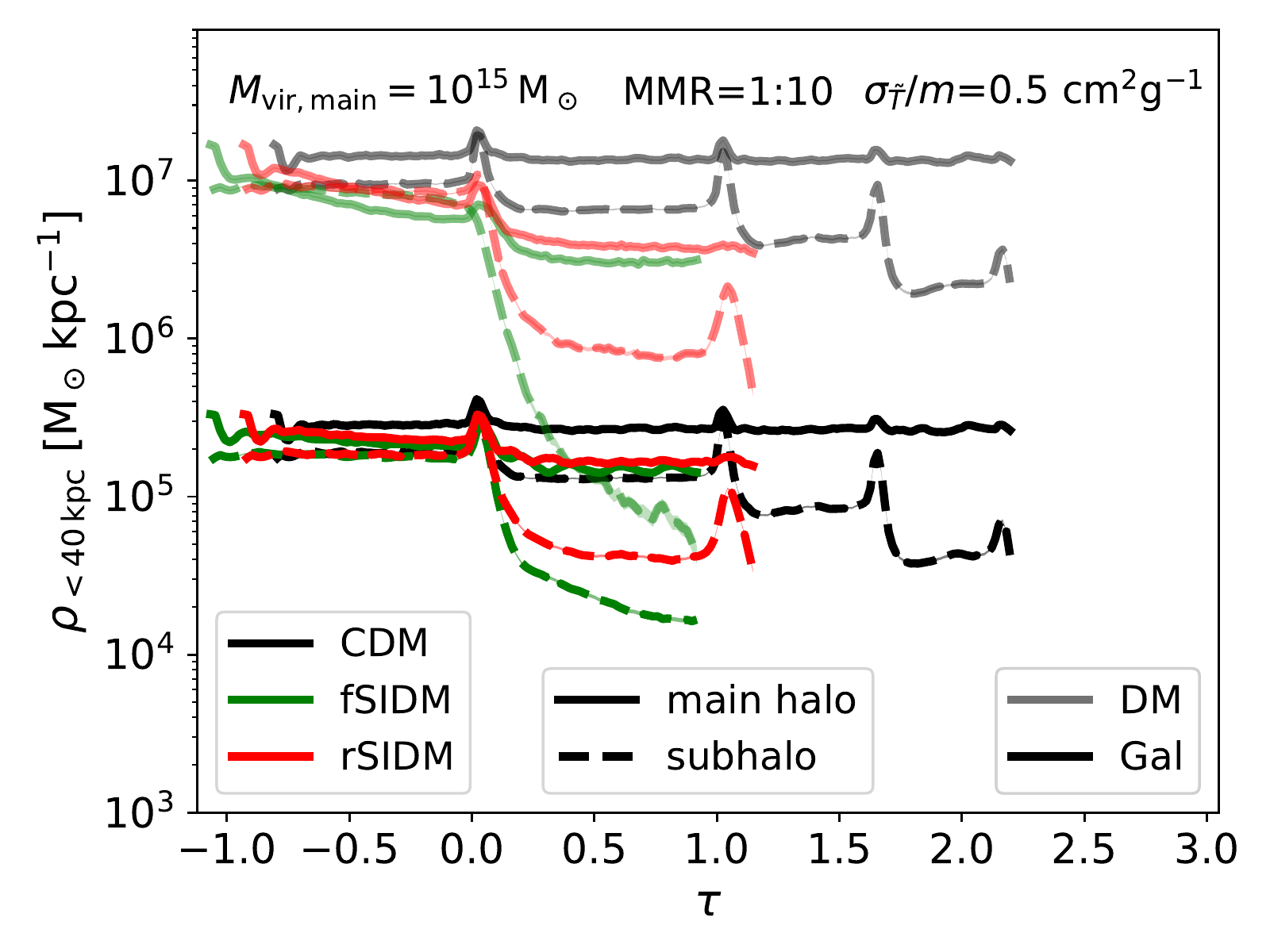}
    \caption{The density at the location of the halo peak is shown as a function of time. The density is computed from the particles within a sphere of $40 \, \mathrm{kpc}$ around the peak. Only the particles which initially belonged to the halo in question were considered for the density computation. The shaded regions display the error. Here, we show the central densities for the same simulations as studied in Fig.~\ref{fig:dens_map_R10}.}
    \label{fig:peak_density}
\end{figure}

The density at the peak position as a function of time is shown in Fig.~\ref{fig:peak_density}. Here, one can see the quantitative differences between the DM models. The central density of the subhalo is more affected than the main halo and galaxies are less affected than the DM. Usually, the density stays constant or is decreasing, except for short periods of density increase that occur subsequent to pericentre passages. Note that we measured the mean density within a sphere that has a radius of $40 \, \mathrm{kpc}$. As the density gradient in the vicinity of the peak position is non-zero, the measured density depends on the chosen radius. However, the results do not qualitatively depend on the selection criterion.

In Fig.~\ref{fig:dens_map_R10}, the shapes of the densest regions for rSIDM and fSIDM look rather different.
The matter distribution for rSIDM appears to be very oblate in the vicinity of the peak for both DM and galaxies.
In contrast, for fSIDM the distribution looks more prolate.
In section~\ref{sec:results_shapes} and \ref{sec:results_comp_shape}, we study the evolution of the halo shape. However, for the fSIDM and rSIDM, runs we do not consider the potential based peaks to be accurate enough to compute the shape at the merger stage we discuss here.

With time, the subhalo particles get caught by the main halo.
For fSIDM, a fraction of stripped particles appear as a dense tail in between the halo peaks (at the left side of the black circle, lower right-hand panel of Fig.\ref{fig:dens_map_R10}). This is less the case for rSIDM.

Besides, there are shell-like features in the galactic distribution.
For rSIDM, there seem to be two shells, the peak belongs to one of them and another one is in front of it.
The fSIDM morphology looks different, there appears to be only one shell which is located in front of the peak.

Based on the morphology, minor mergers seem to be well suited to distinguish rare and frequent self-interactions.
However, in practice, observational limitations could alter the picture. It remains to be seen whether this persists in the presence of baryonic matter and this will be the subject of forthcoming work.
Moreover, we should note that we have only looked at a slice in the merger plane and not a projected two-dimensional density map, which is more relevant from an observational point of view and may look somewhat different due to projection effects. For clarity, we only considered the particles of the subhalo and ignored the main halo which is the dominant component.
But even if the main halo is taken into account, one can recognize differences between DM models as we demonstrate in Appendix~\ref{sec:additional_plots_morphology}.

\subsection{Centre of mass distance and offsets} \label{sec:results_offsets}
\begin{figure}
    \centering
    \includegraphics[width=\columnwidth]{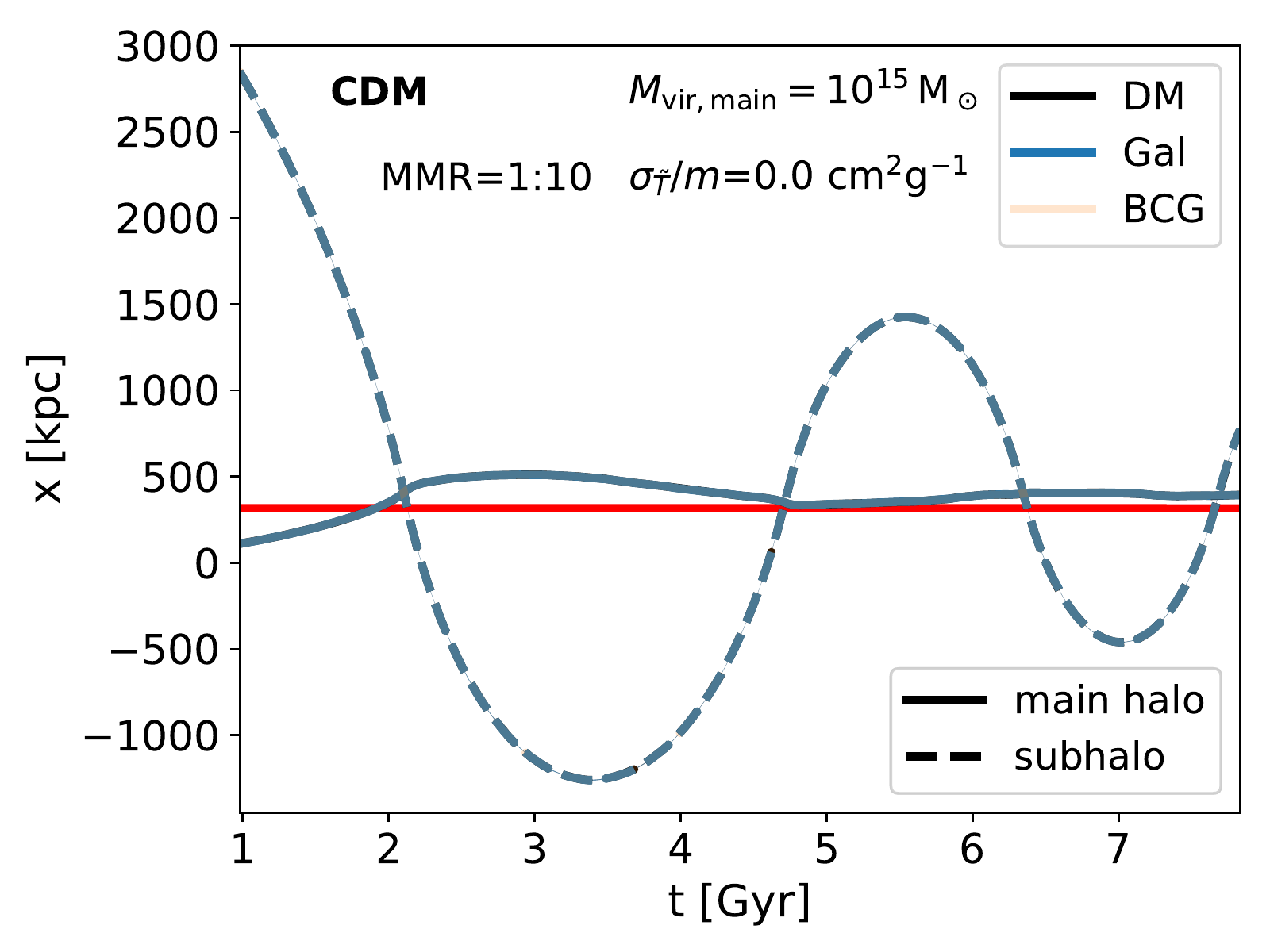}
    \includegraphics[width=\columnwidth]{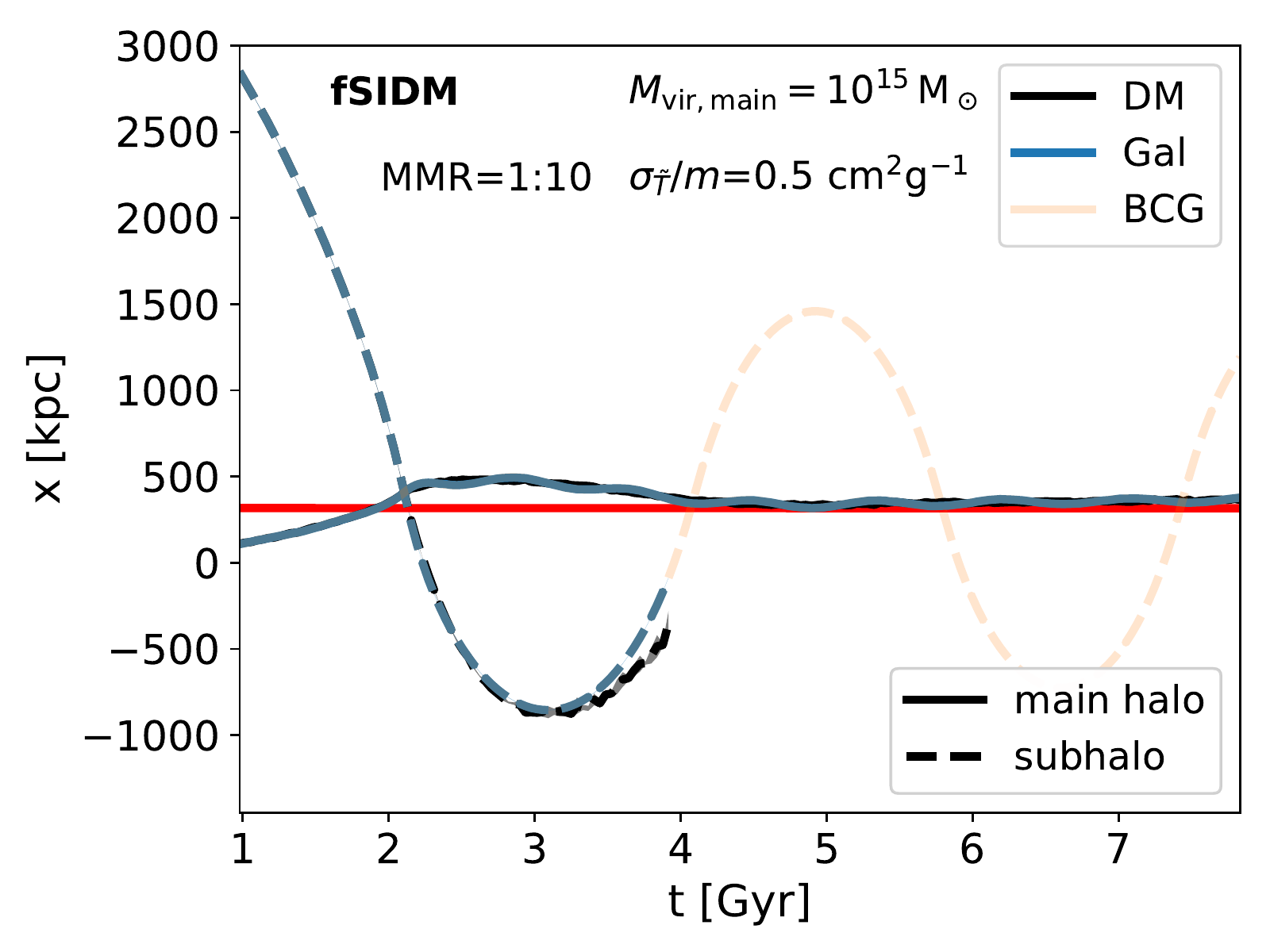}
    \caption{For a 1:10 merger, the peak positions for various components (DM, galaxies, BCGs) are shown.
    The red line indicates the centre of mass of the system.
    The upper panel gives the positions for a simulation with CDM and the lower panel for a simulation with fSIDM and a cross-section of $\sigma_\mathrm{\tilde{T}}/m = 0.5 \, \mathrm{cm}^2 \, \mathrm{g}^{-1}$.
    Peak positions are shown as long as the peak finder provides reasonable results.}
    \label{fig:byc0}
\end{figure}

\begin{figure}
    \centering
    \includegraphics[width=\columnwidth]{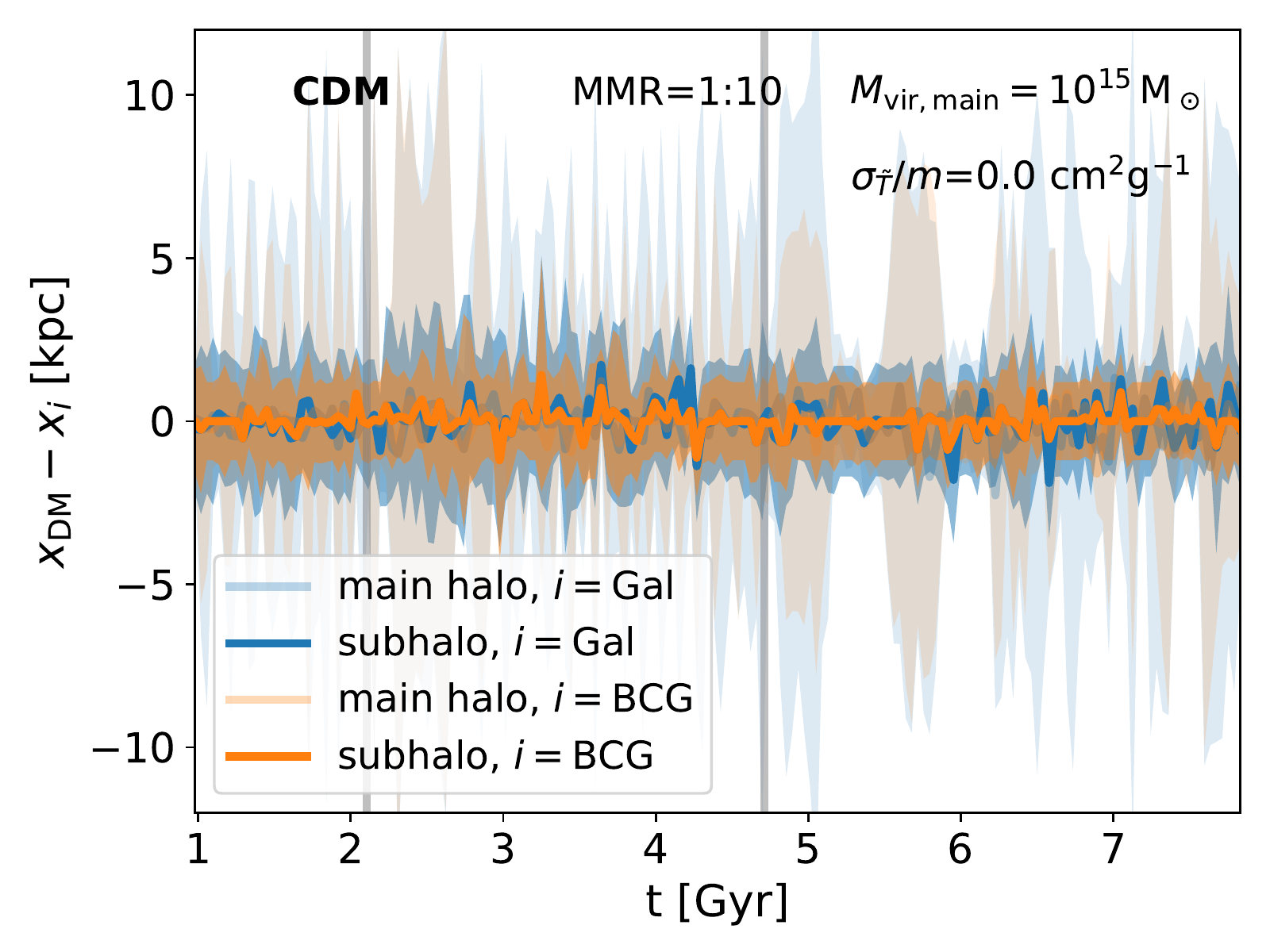}
    \includegraphics[width=\columnwidth]{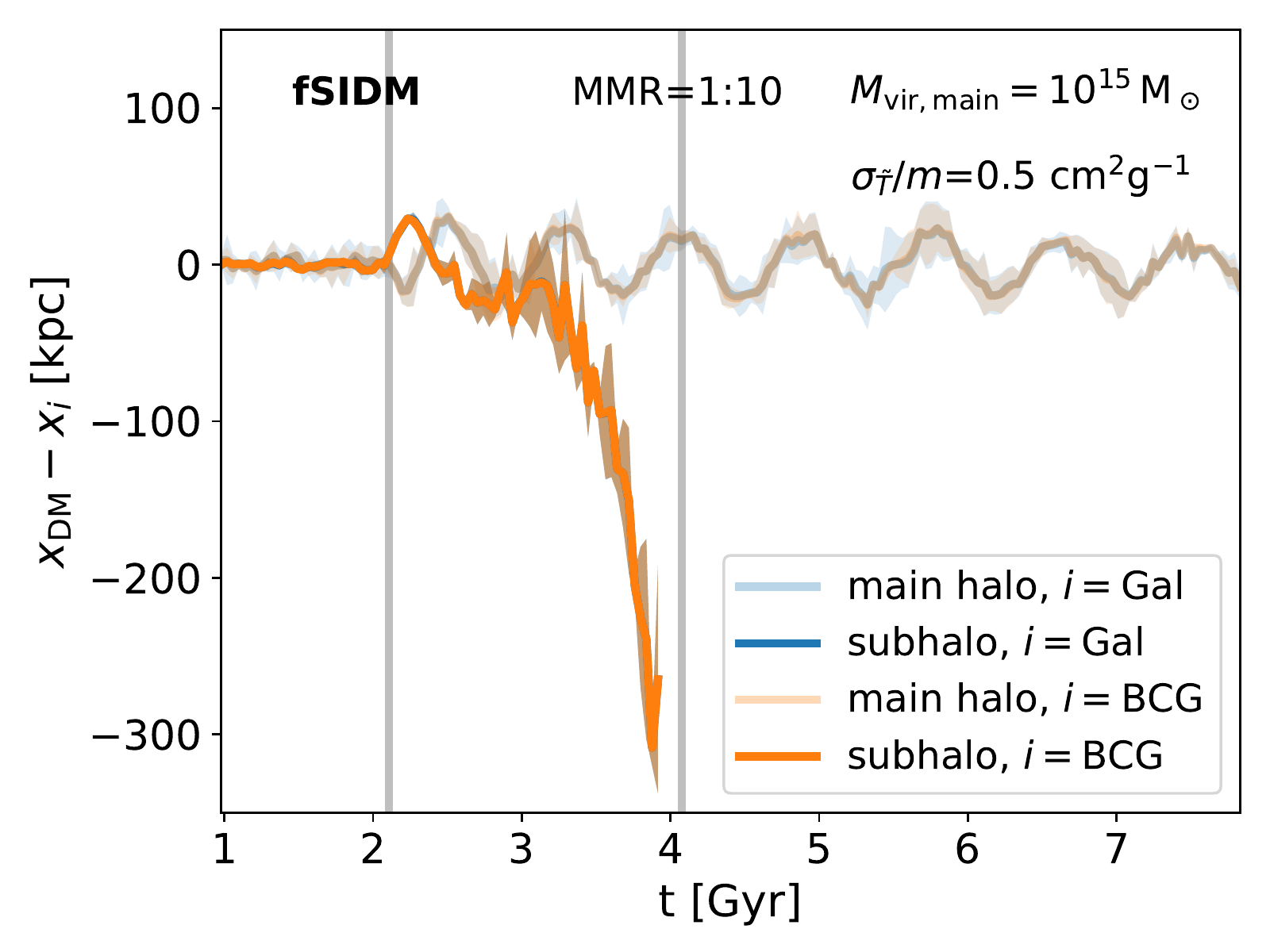}
    \caption{Offsets for the runs shown in Fig.~\ref{fig:byc0}.
    The upper panel displays offsets for the CDM merger and the lower panel for the fSIDM merger. The shaded areas indicate the $1\sigma$ error. The first and second pericentre passage are indicated by the vertical grey lines.}
    \label{fig:offset0}
\end{figure}

Let us now come to a discussion of the peak positions of the different sub-components as well as the inferred offsets for our merger simulations.
In Fig.~\ref{fig:byc0}, we show the positions of the peaks of the various components (DM, galaxies, BCGs) along the merger axis for runs of the 1:10 cluster-scale merger.
For the same simulations, we display the offset in Fig.~\ref{fig:offset0}.

The peaks were determined by using the potential-based peak finder described in Section~\ref{sec:method_peak_find}.
This peak finder has the effect that the peaks of a collisionless component behave very similarly to the BCG positions.
This can be seen when comparing the galaxy peaks with the BCG positions.
For the CDM run, the DM is collisionless and thus all peaks coincide; the vanishing offsets shown in the upper panel of Fig.~\ref{fig:offset0} demonstrate this.
Furthermore, the vanishing DM--galaxy offset demonstrates how small the peak finding error is.
Besides, it seems that the haloes are a little offset from the centre of mass, e.g. the pericentre passage does not coincide with the centre of mass. This is only the case for the unequal-mass mergers and might be caused by the asymmetry of the system. During the infall phase, the haloes are deformed due to tidal forces, which may lead to a shift between the centre of mass and the weighted centre of the two peaks.

For the CDM and fSIDM run, the first pericentre passage occurs after roughly 2.1 Gyr. Self-interactions have the effect of reducing the merger time for the fSIDM run. Also, the second pericentre passage occurs earlier than for CDM.
Another difference between SIDM and CDM shows up in the oscillation of the BCGs in the DM potential.
For CDM, the amplitude decays much faster than for SIDM where the orbital decay is minimal.
This effect exists in unequal-mass mergers as well as in equal-mass mergers as demonstrated in \cite{Kim_2017b}.
At a basic level, this reduction in dynamical friction with SIDM results from the lowered DM densities in the merger remnant compared with in the CDM case (e.g. see Fig. \ref{fig:peak_density}), and the fact that the dynamical friction force is proportional to the background density \citep{Chandrasekhar_1943}. We note however \citep[as also discussed in][]{Kim_2017b} that dynamical friction acting on bodies orbiting in a cored DM distribution is more complicated than the motion through an infinite constant-density background considered by \citet{Chandrasekhar_1943}, and that dynamical friction can vanish almost entirely in such a case \citep{Read_2006}.

Before the merging system reaches equilibrium, the common potential becomes deeper and thus the amplitude of the BCGs oscillation decreases until the DM core of the coalesced halo has formed.
However, the orbits of the BCG could change once the effects of the ICM are considered and the BCG is modelled more realistically.

The offsets for the fSIDM run are large enough such that they can already be identified in the lower panel of Fig.~\ref{fig:byc0}.
These large offsets do not arise close to the first pericentre passage, but between the first apocentre and the second pericentre.
However, we should point out that we have no reliable peak positions for the subhalo at times
later than 4 Gyr.
While in principle a merging system could have its largest offset after the second pericentre passage, the observational identification of separate DM peaks becomes prohibitively difficult at late merger stages as the subhalo dissolves.
In contrast, separate stellar or galactic components could be identified more easily and offsets after coalescence of the DM component associated with core sloshing could provide a signature of SIDM \citep{Kim_2017b}.
In particular, the minimally decaying orbit of the BCGs could be of interest.
We provide further results on core sloshing using multiple peak finding methods in Section~\ref{sec:results_peakfind}.
We also compare offsets of different runs including various merger mass ratios (MMR) and self-interaction cross-sections in Section~\ref{sec:results_comp_offset}.

\subsection{Shapes} \label{sec:results_shapes}

\begin{figure}
    \centering
    \includegraphics[width=\columnwidth]{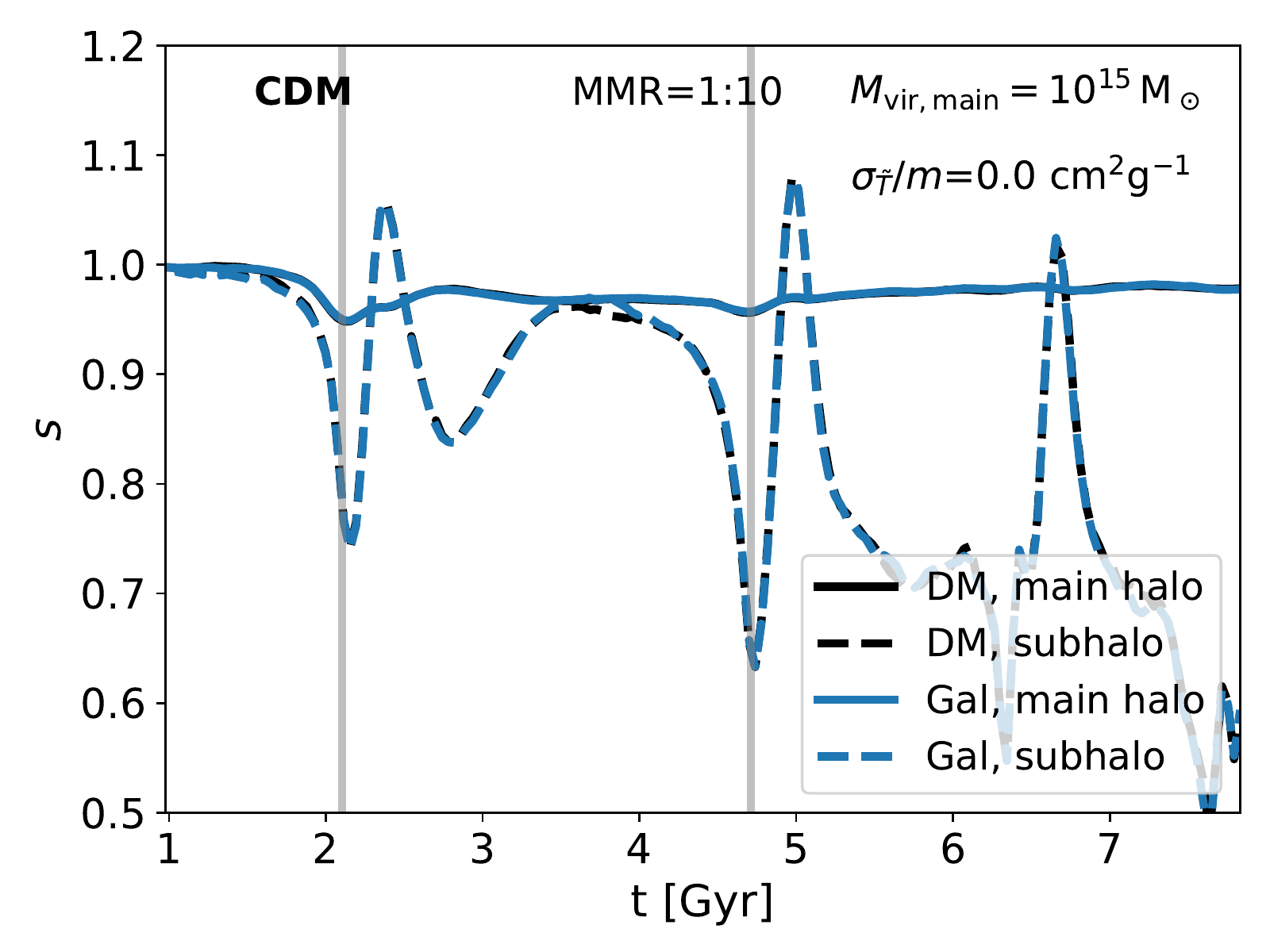}
    \includegraphics[width=\columnwidth]{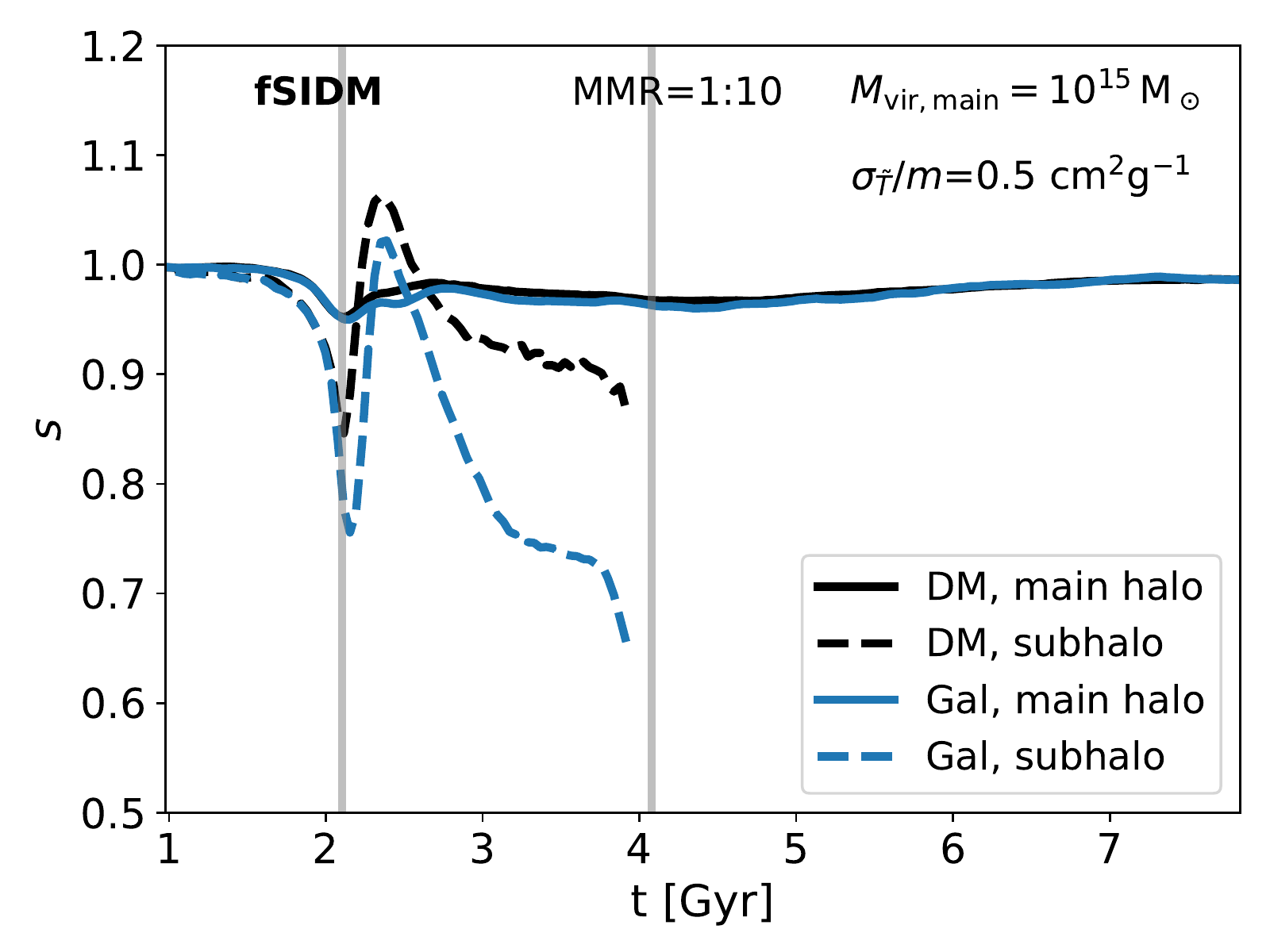}
    \caption{Shapes for the runs shown in Fig.~\ref{fig:byc0}.
    The upper panel displays shapes for the CDM merger and the lower panel for the fSIDM merger. The first and second pericentre passage are indicated by the vertical grey lines.}
    \label{fig:shape0}
\end{figure}

DM self-interactions affect the peak positions of the haloes as well as the higher order moments of the DM distribution.
Here, we focus on the shape variable of the haloes as defined by Eq.~\eqref{eq:shape}.
In contrast to other studies, we pursue a simplified approach by considering all particles within twice the scale radius of the initial NFW profile instead of measuring the shape as a function of radial distance as done in other studies \citep[e.g.][]{Zemp_2011, Peter_2013, Chua_2020, Vargya_2021}.

In Fig.~\ref{fig:shape0}, we show the shape parameter for a 1:10 merger evolved with CDM and fSIDM, the same simulations as shown in Fig.~\ref{fig:byc0} and Fig.~\ref{fig:offset0}.

Initially, the haloes are spherically symmetric ($s = 1$) and subsequently evolve to become more elliptical owing to gravitational interactions with the other halo.
The shape of the main halo indicated by the solid line is only slightly affected by the merger and becomes a little more elliptical, in particular about the pericentre passages.
In contrast, the shape of the subhalo is more strongly affected, for both CDM and fSIDM.
For the CDM merger, we are able to track the peaks for much longer times and thus can compute shapes for later times compared to fSIDM.
As we can see in the upper panel of Fig.~\ref{fig:shape0},
the evolution of the CDM subhalo can be described as follows:
During the infall phase directly before the first pericentre passage, the shape becomes much more prolate due to tidal forces and the size of this distortion depends on the mass ratio.
Close to the core passage, gravity has a different effect, which leads to a more oblate shape.
The halo becomes even more oblate than it has been initially.
But when it climbs out of the potential of the main halo afterwards, then it becomes more prolate due to tidal forces again.
When the separation between the two haloes has grown large enough (about the first apocentre passage, $\sim 3.4$ Gyr) the tidal force can become small compared to the self-gravity of the subhalo.
Hence, the self-gravity makes the subhalo more spherical.
A rough estimate of the tidal radius at the first apocentre passage leads to $536 \, \mathrm{kpc}$ or $3.7 \, r_\mathrm{s}$.
Thus, the particles we selected for the shape computation should be within the tidal radius.
Later on, when the merger is getting close to the second pericentre passage, tidal forces make the halo more prolate again.
The described picture depends strongly on the considered particles. 
If one would take particles beyond twice the scale radius of the initial NFW profile into account, the evolution of the shape would look very different.

In comparison to CDM, the fSIDM subhalo is less prolate about the first pericentre passage and also the maximum in shape shortly after the core passage is more extreme for fSIDM.
Frequent self-interactions transfer energy from the direction of motion to the perpendicular component which contributes to a more oblate halo.
Besides, one can observe that a difference between the galaxies and the DM component arises. This is simply due to the collisionless nature of our galaxies that do not undergo frequent self-interactions.

About the first apocentre passage ($\sim 3.1$  Gyr), when the tidal force becomes less important, the self-gravity can lead to a more spherical halo as we found for CDM.
But for fSIDM the subhalo's gravitational potential is less deep and the apocentre distance is shorter compared to CDM, resulting in a halo that becomes even more prolate.

Going beyond the parameter values assumed here, we compare shapes of different simulations for a variety of MMRs and cross-sections in Section~\ref{sec:results_comp_shape}.

\subsection{Frequent versus rare interactions} \label{sec:results_comp}
\begin{figure}
    \centering
    \includegraphics[width=\columnwidth]{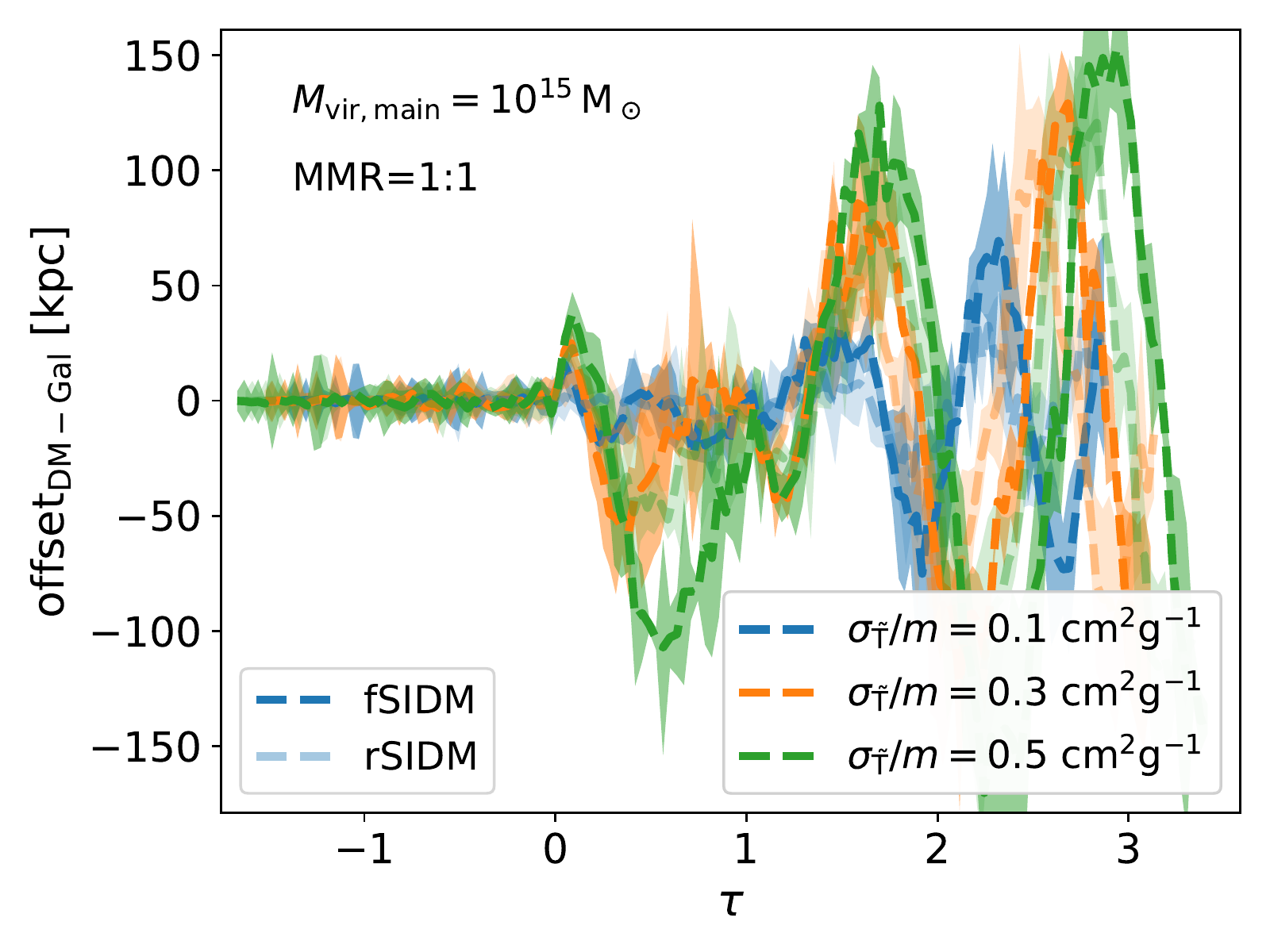}
    \includegraphics[width=\columnwidth]{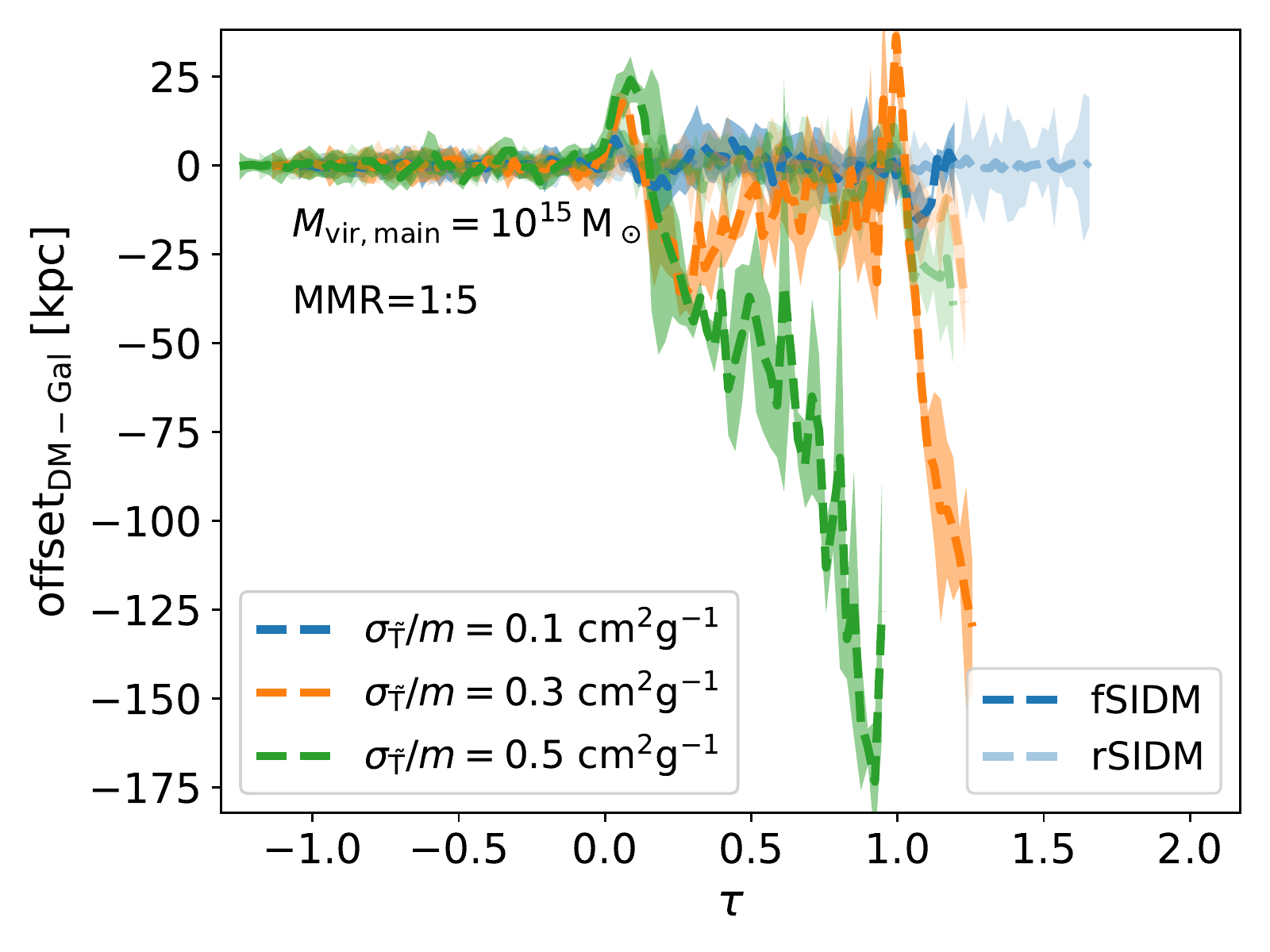}
    \includegraphics[width=\columnwidth]{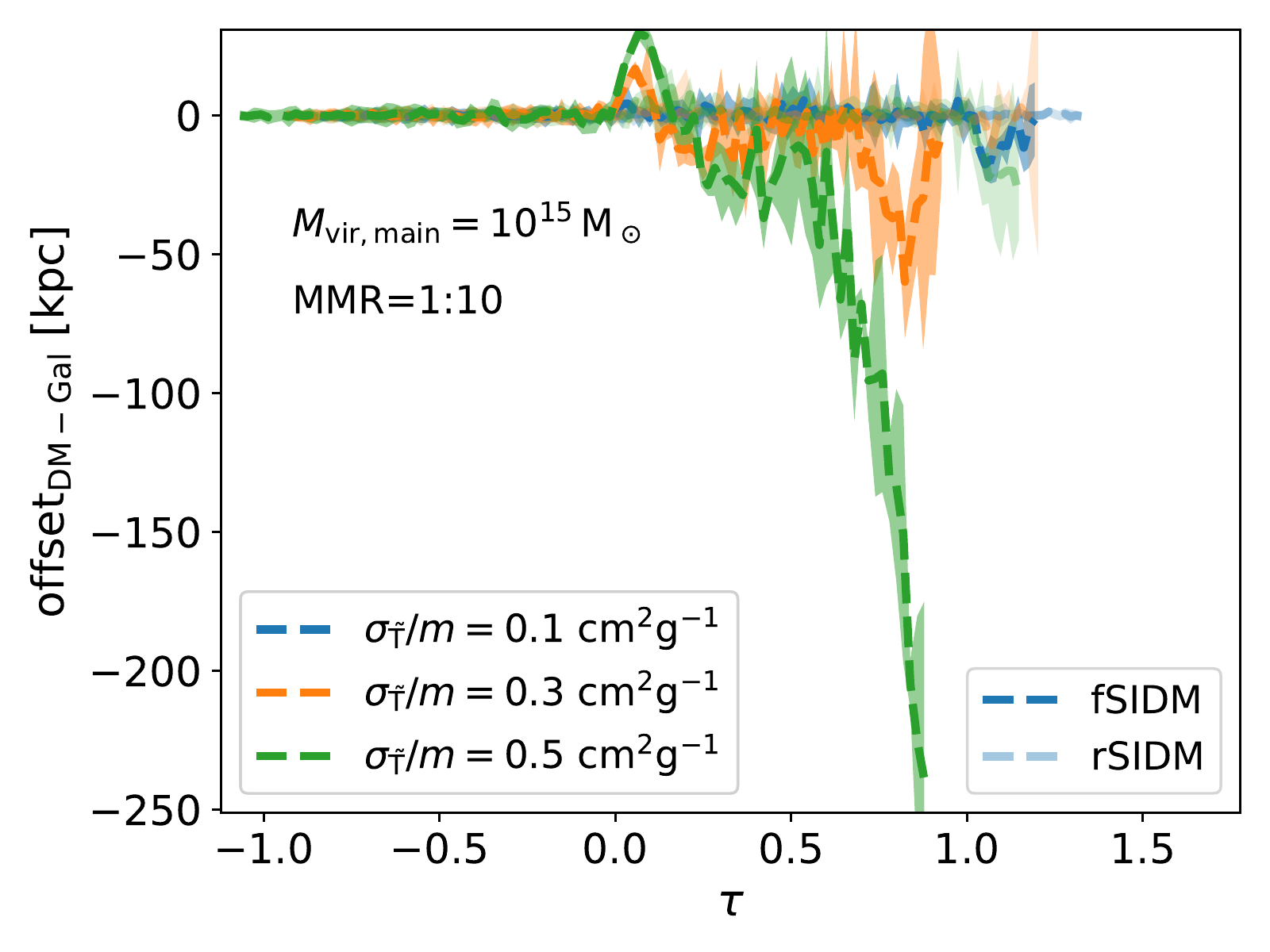}
    \caption{DM-galaxy offsets from merging system with various MMRs (upper row: 1:1, middle row: 1:5, lower row: 1:10) for the subhaloes.
    The results were obtained using the potential based peak finder.
    The offsets are defined according to Eq.~\eqref{eq:offset} and shown for frequent (high opacity) and rare (low opacity) self-interactions as well as for several cross-sections: $\sigma_\mathrm{\Tilde{T}}/m = 0.1 \, \mathrm{cm}^2\,\mathrm{g}^{-1}$ (blue), $\sigma_\mathrm{\Tilde{T}}/m = 0.3 \, \mathrm{cm}^2\,\mathrm{g}^{-1}$ (orange), $\sigma_\mathrm{\Tilde{T}}/m = 0.5 \, \mathrm{cm}^2\,\mathrm{g}^{-1}$ (green).
    The shaded areas indicate the $1\sigma$ error.
    }
    \label{fig:offsets_MMR}
\end{figure}

\begin{figure}
    \centering
    \includegraphics[width=\columnwidth]{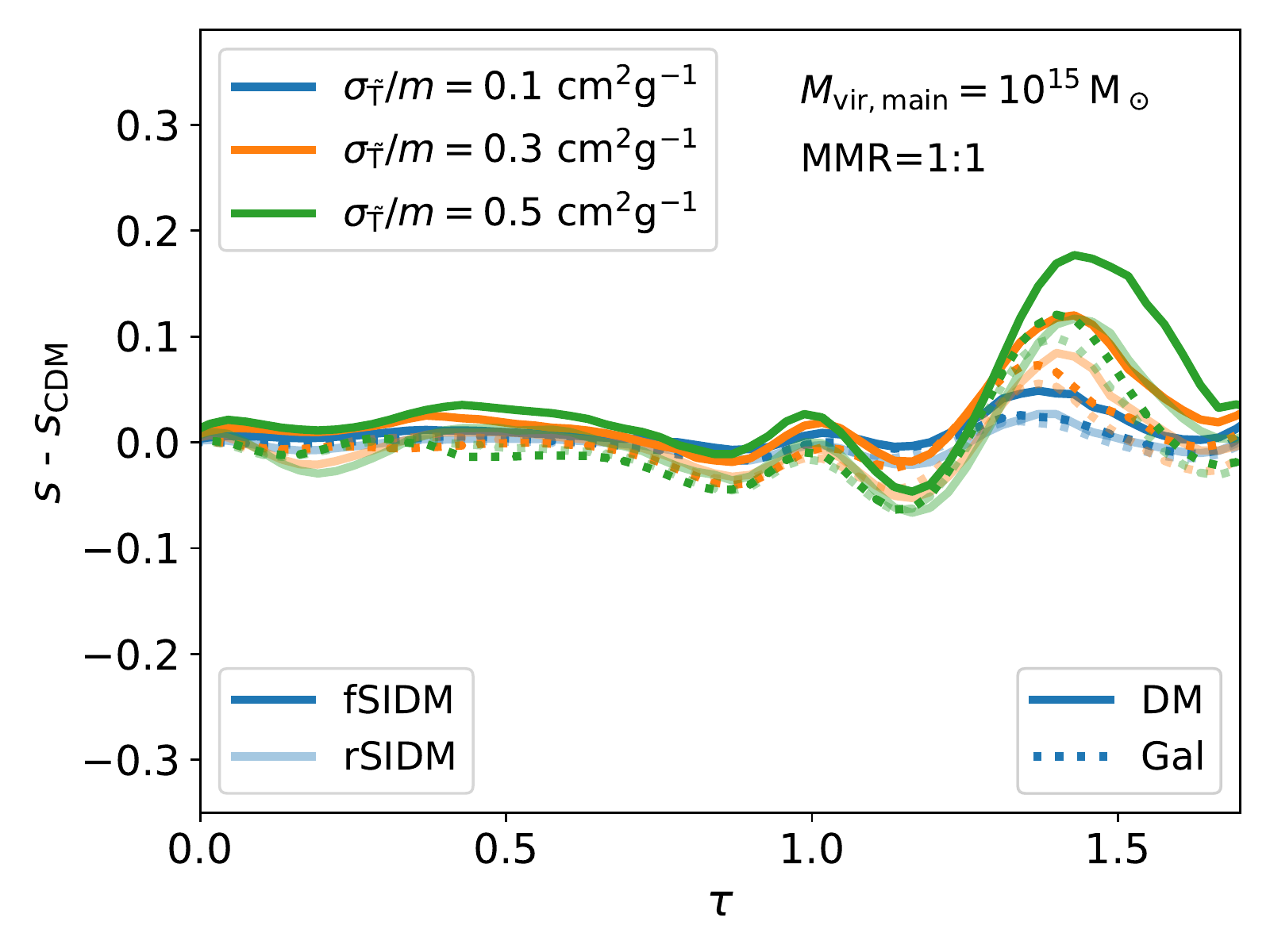}
    \includegraphics[width=\columnwidth]{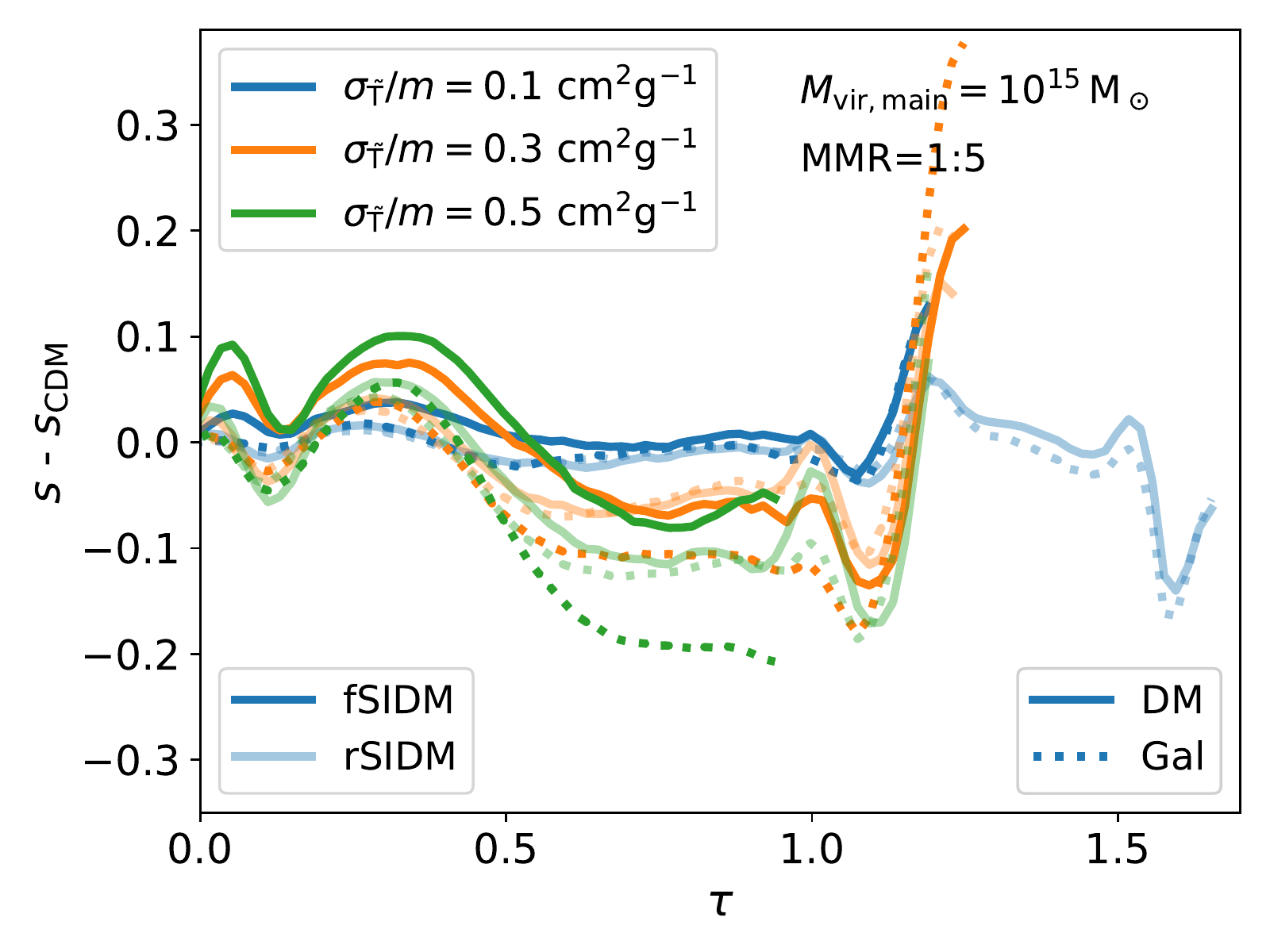}
    \includegraphics[width=\columnwidth]{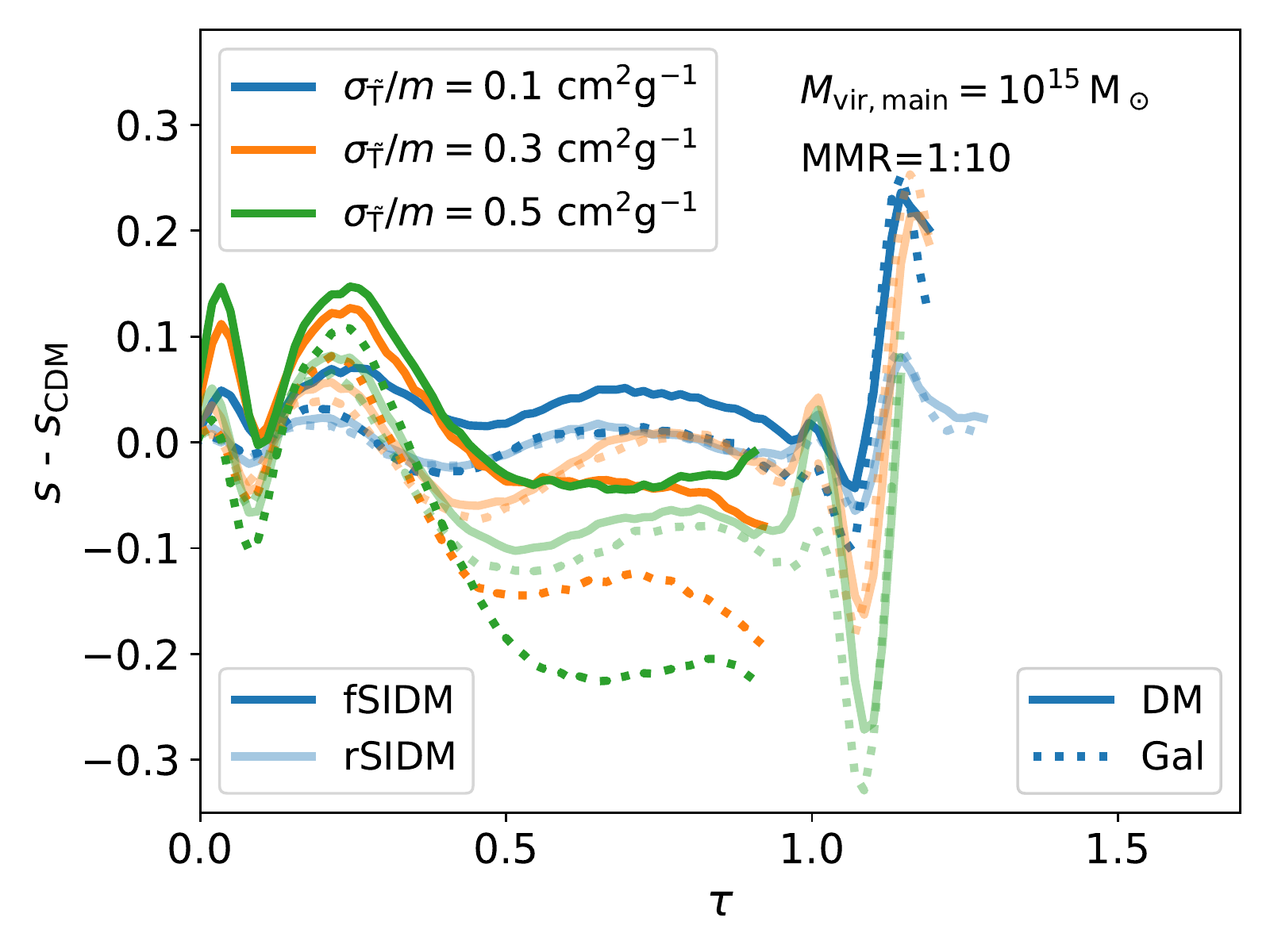}
    \caption{Differences in the shape between SIDM and CDM subhaloes from merging system with various MMRs (upper panel: 1:1, middle panel: 1:5, lower panel: 1:10).
    The shapes are defined according to Eq.~\eqref{eq:shape} and shown for frequent (high opacity) and rare (low opacity) self-interactions as well as for several cross-sections: $\sigma_\mathrm{\Tilde{T}}/m = 0.1 \, \mathrm{cm}^2\,\mathrm{g}^{-1}$ (blue), $\sigma_\mathrm{\Tilde{T}}/m = 0.3 \, \mathrm{cm}^2\,\mathrm{g}^{-1}$ (orange), $\sigma_\mathrm{\Tilde{T}}/m = 0.5 \, \mathrm{cm}^2\,\mathrm{g}^{-1}$ (green).}
    \label{fig:shape_MMR}
\end{figure}

In the following, we compare the effects from rare and frequent self-interactions in mergers for various MMRs.
In Fig.~\ref{fig:offsets_MMR}, we show DM-galaxy offsets for subhaloes employing several cross-sections and in Fig.~\ref{fig:shape_MMR}, we compare the subhalo shapes of SIDM runs to the CDM shapes.
In this section, we concentrate on the cluster-scale simulations. Plots for the galaxy-scale simulations can be found in Appendix~\ref{sec:additional_plots_comp}.

\subsubsection{Offsets} \label{sec:results_comp_offset}

For the equal-mass mergers, we find that the largest offsets occur at late stages, i.e.\ after the second apocentre passage (see Fig.~\ref{fig:offsets_MMR}) of the mergers as the time difference between the pericentre passages of DM and galaxies becomes larger for late times.
This phase shift increases for larger cross-sections, leading to larger offsets.
For unequal-mass merger, we find similar results.
However, in the latter case, we encounter difficulties in determining the peaks at late times due to the evaporating subhalo.

For all mergers, we find the general trend that frequent self-interactions produce larger offsets than rare self-interaction when comparing the same momentum transfer cross-section and the size of offsets increases for larger values of $\sigma_\mathrm{\tilde{T}}/m$ (see Fig.~\ref{fig:offsets_MMR}).
In addition, frequent self-interactions usually lead to somewhat shorter merger times than rare self-interactions. 
We observe that differences between the two cases are maximized for small MMR (i.e.\ larger difference in mass) and large cross-sections.

Furthermore, we find that runs with smaller MMR show larger offsets.
As the subhalo is less massive in this case, its particles are less bound to it.
Therefore, DM self-interactions and tidal forces of the main halo can affect it more, resulting in stronger effects of DM scatterings.
Moreover, for our unequal mass mergers the merger time is longer, i.e.\ the time between two pericentre passages, such that the amplification process as observed in \cite{Fischer_2021} has more time to act on the galactic component.
By amplification, we denote the process that the small initial offsets created by the self-interactions at the time when the system is close to its first pericentre passage evolve to much larger offsets at a later merger phase. Those offsets are caused by the different shapes of the DM gravitational potential acting on the trajectories of the collisionless galaxies/stars.

The magnitude of the offsets for rare and frequent self-interactions is substantially more different for unequal-mass mergers than for equal-mass mergers when considering the time before the second pericentre passage.
In addition, unequal-mass mergers seem to be more sensitive to the cross-section than equal-mass mergers.
For instance, consider the 1:5 merger (middle panel of Fig.~\ref{fig:offsets_MMR}) and compare the fSIDM offset of the runs with $\sigma_\mathrm{\tilde{T}}/m = 0.3 \, \mathrm{cm}^2 \, \mathrm{g}^{-1}$ (orange) and $\sigma_\mathrm{\tilde{T}}/m = 0.5 \, \mathrm{cm}^2 \, \mathrm{g}^{-1}$ (green).
For $\tau \gtrsim 0.4$, the evolution of the offsets is quite different, the smaller cross-section shows decreasing offsets whereas for the larger one the offset continues to grow, implying that the relationship between cross-section and offset size is highly non-linear in general.

To obtain a better understanding of the underlying dynamics, let us now discuss the various effects which enter into the development of the offsets: 
when the drag force decelerates the DM, an offset between galaxies and DM arises. The gravitational pull of the DM halo acts against the offset.
Thus, the smaller the subhalo, the weaker the gravitational attraction and, as a result, larger offsets occur. But it is not as simple as this.
The picture is more complicated as the gravitational attraction depends on the gradient of the potential, which is flattened by the self-interactions and on the offset itself.
Consequently, the first pericentre passage offsets could also be larger for equal-mass mergers depending on the actual mass profile.
The mass profile of galaxy clusters also depends on the ICM, which we did not include in our simulations.
However, the description above is only appropriate for the time about the first pericentre passage.
At a later time, the DM peak overtakes the galaxies (i.e. becomes more distant to the centre of mass) as the galactic component has experienced a larger deceleration due to the offset, i.e. its corresponding DM halo has led to further deceleration.
At the point in time when the DM is overtaking, i.e. the galactic peak is passing the DM peak, the DM gravitational potential is shallower.
As a consequence, the galaxies can escape further and much larger offsets ensue.
The size of these offsets depends on how much shallower the gravitational potential becomes compared to the first pericentre passage.
Lower mass haloes are less gravitationally bound, they dissolve faster due to self-interactions, and thus their gravitational potential becomes shallower and the offsets larger (and at the same time, harder to observe given the dissolving subhalo).
The size of the effective gravitational attraction acting against the offset should depend on the offset and decrease for large enough offsets.
Hence, the growth of large offsets can be accelerated further as observed for the unequal-mass mergers evolved with $\sigma_\mathrm{\Tilde{T}}/m = 0.5 \, \mathrm{cm}^2 \, \mathrm{g}^{-1}$ (middle and lower panel of the left-hand column of Fig.~\ref{fig:offsets_MMR}). 

For equal-mass mergers, we find that rSIDM can show large offsets ($\gtrsim 100 \, \mathrm{kpc}$) only at late times via an accumulated phase shift. 
However, due to shallower density gradients at later times peak finding becomes more difficult, directly impacting the observational prospects of finding large offsets.
However, also the large offsets of fSIDM will, in general, be difficult to observe as we discuss in Sec.~\ref{sec:discussion_tech}.
Nevertheless, the conditions under which an offset of observable size arises are more easily and more often met for fSIDM than for rSIDM.

\subsubsection{Shape} \label{sec:results_comp_shape}

In Fig.~\ref{fig:shape_MMR}, we compare the shapes of SIDM subhaloes to the shapes of the corresponding CDM haloes using the time $\tau$ as given by Eq.~\eqref{eq:time}.
The shapes of the individual mergers are displayed in Appendix~\ref{sec:additional_plots_comp}.
Before the first pericentre passage ($\tau<0$) differences occur only because of the different merger times as we use $\tau$ to match the times of the simulations.
If one used the physical time for the infall phase, any significant difference would vanish and we do not display them in Fig.~\ref{fig:shape_MMR}.

At the first pericentre passage ($\tau=0$), the shapes are almost the same for all cross-sections.
However, for the second pericentre passage ($\tau=1$), this is no longer the case. How much the shapes of SIDM haloes deviate from their CDM counterpart depends strongly on the MMR.
For equal-mass mergers, the differences in the shape parameter between CDM and fSIDM are small before the second pericentre passage and become larger for more unequal halo masses.
As expected, the difference increases with increasing self-interaction cross-section.

Self-interactions can lead to more oblate as well as more prolate shapes compared to CDM. The difference depends on the merger stage and the self-interaction type. For the phase before the first apocentre passage, frequent self-interactions tend to produce haloes that are always more oblate. In contrast, rare self-interaction show also a phase ($\tau \sim$ 0.1--0.2) with a significantly more prolate shape. While both fSIDM and rSIDM lead to a shallower potential, the phase--space distributions are different. Unlike frequent interactions, the isotropic, rare self-interactions do not preferentially transfer the energy from the forward motion to a perpendicular component but can create a tail of back-scattered particles. Hence, the shape of haloes in rSIDM can be more prolate than for fSIDM.

For the unequal-mass mergers, there is a phase at $\tau \sim 0.7$ where the self-interactions -- given the cross-section is large enough -- lead to more elliptical haloes than CDM.
This can be understood in terms of a smaller pericentre distance implying a stronger tidal force and a shallower potential due to the DM scattering as explained in Section~\ref{sec:results_shapes}.
The galactic component can be even more elliptical than the DM as it is not subject to self-interactions.
Thus, the naive picture that self-interactions render haloes always more spherical fails in the case of an unequal-mass merger.

In most cases, fSIDM leads to a more oblate DM halo compared to rSIDM assuming that the same momentum transfer cross-sections are compared.
In addition to the fact that fSIDM is transferring momentum from the direction of motion to a perpendicular component, there could also be differences between frequent and rare self-interactions regarding the efficiency of making haloes more spherical.

Furthermore, we also observe shape differences between the DM component and the galaxies/stars.
The most striking difference can be seen for frequent self-interactions and unequal-mass mergers at around the first apocentre passage.
The galactic component is more prolate than the DM and for the 1:10 merger, this phase lasts remarkably long.
For rSIDM, there is only a very small difference between galaxies and DM shape.
Interestingly, we find that the shape for the galactic component is more prolate for fSIDM than rSIDM.
This occurs due to differences in the gravitational potential, for fSIDM, the particles are less strongly bound and thus more prone to tidal disruption.

Finally, we want to emphasize that our results depend on the particles selected for the shape computation. Particles of the inner or outer halo are affected differently during the merger. Hence, any comparison with simulation data or observations needs to be cautious about the scales on which quantities are measured.

\subsection{Peak finding} \label{sec:results_peakfind}

\begin{figure}
    \centering
    \includegraphics[width=\columnwidth]{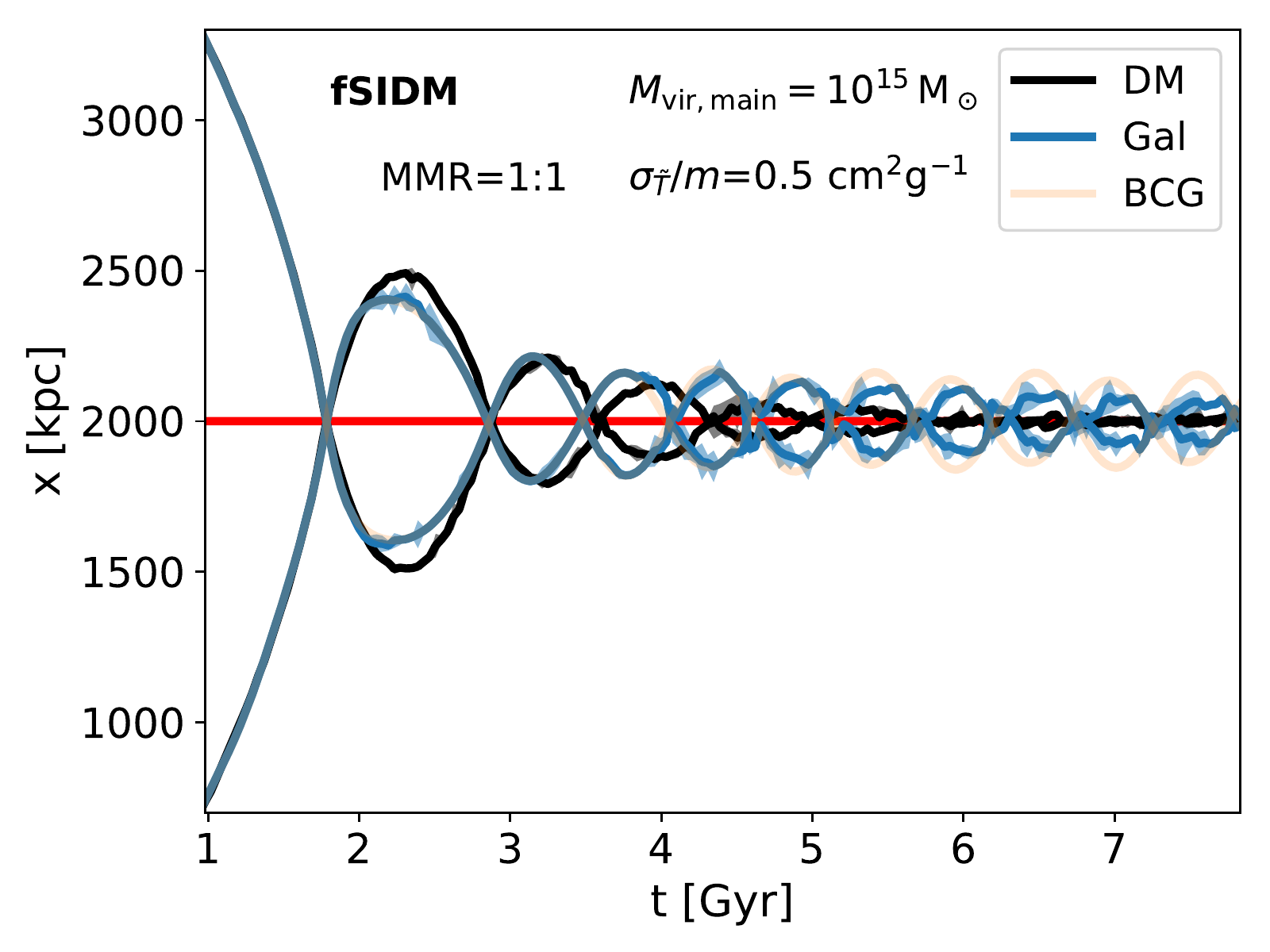}
    \includegraphics[width=\columnwidth]{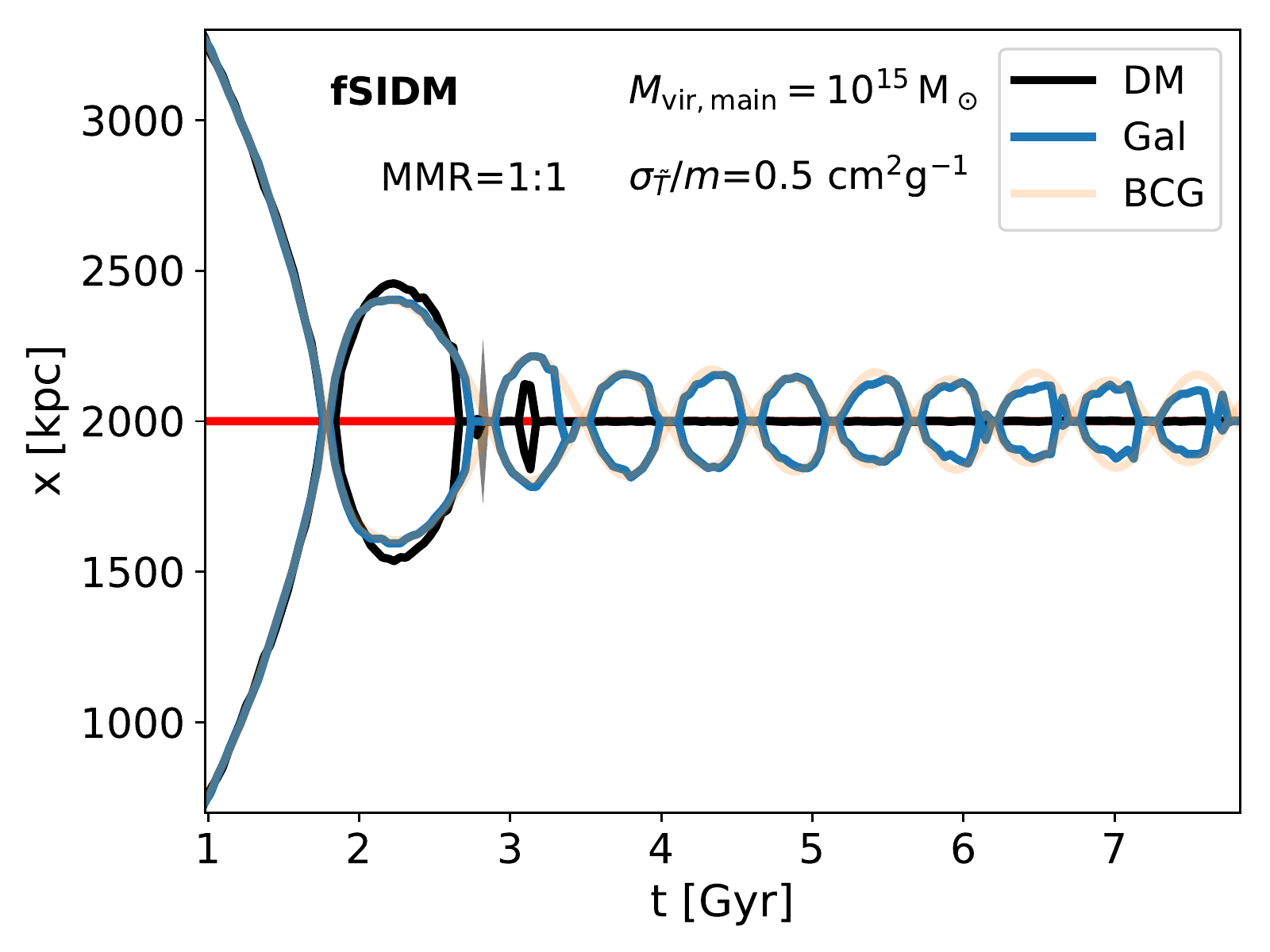}
    \caption{The peak positions for an equal-mass merger are shown as a function of time.
    For the upper panel, the gravitational potential-based peak finder was employed and for the lower one, the one based on isodensity contours.
    DM peaks are indicated in black, galaxy peaks in blue, and the position of BCG particles in orange.
    The red lines indicate the centre of mass of the system.}
    \label{fig:byc_iso}
\end{figure}

\begin{figure}
    \centering
    \includegraphics[width=\columnwidth]{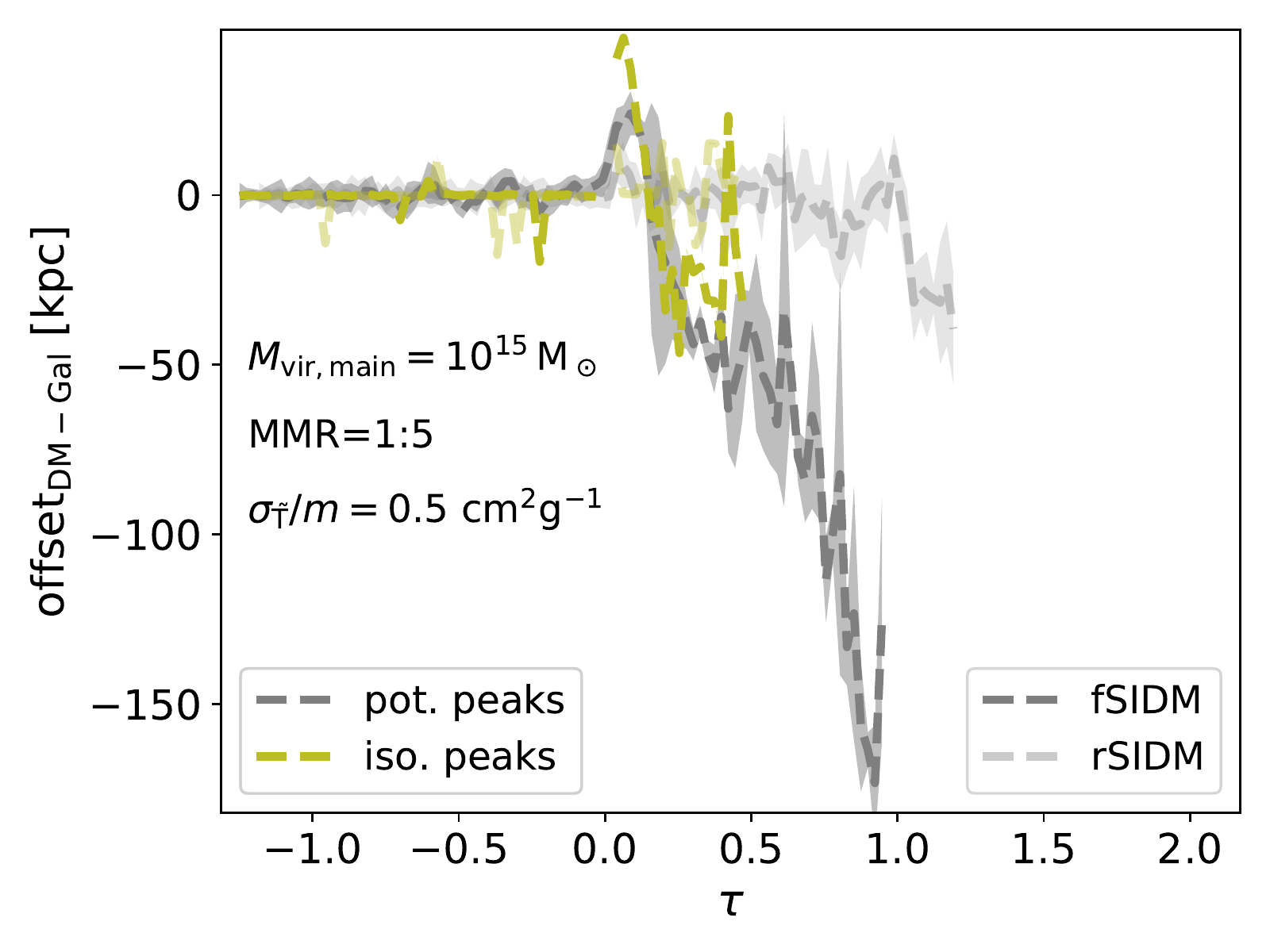}
    \caption{The DM-galaxy offset is shown as a function of time for an unequal mass merger with an MMR of 1:5 of our cluster-scale mergers. The simulation was evolved with a cross-section of $\sigma_\mathrm{\Tilde{T}}/m = 0.5 \, \mathrm{cm}^2 \, \mathrm{g}^{-1}$. We display the offset based on the potential-based peak finder as well as the isodensity contour-based peak finder. The shaded areas indicate the error.}
    \label{fig:peakcomp}
\end{figure}

\begin{figure}
    \centering
    \includegraphics[width=\columnwidth]{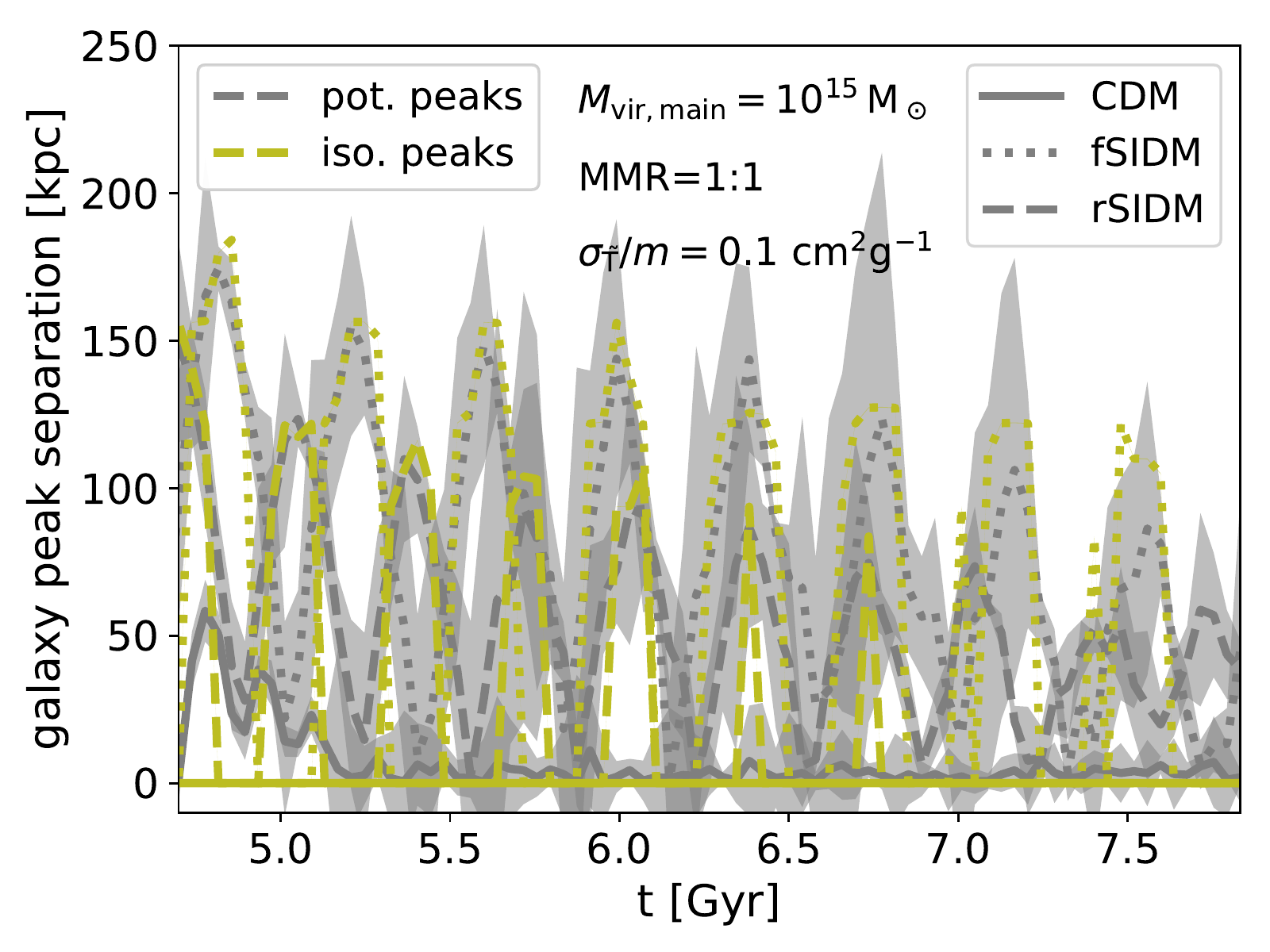}
    \caption{The separation between the peaks of the galactic component is shown as a function of time for an equal-mass cluster-scale merger at late merger stages. The DM component coalesced, but separate peaks for the galaxies can be identified using the potential-based or isodensity contour-based peak finding. The displayed merger was evolved with the lowest cross-section we present in this paper, $\sigma_\mathrm{\Tilde{T}}/m = 0.1 \, \mathrm{cm}^2 \, \mathrm{g}^{-1}$. The shaded areas indicate the error.}
    \label{fig:cphase}
\end{figure}

In Section~\ref{sec:method_peak_find}, we described two peak finding methods.
So far, we have only discussed results relying on the peak finding method based on the gravitational potential.
In this section, we will compare these results to the ones we obtain using the method of isodensity contours and discuss the origin of the resulting differences.

In Fig.~\ref{fig:byc_iso}, we show the peak positions as a function of time for an equal-mass merger evolved with frequent self-interactions and $\sigma_\mathrm{\Tilde{T}}/m = 0.5 \, \mathrm{cm}^2 \, \mathrm{g}^{-1}$.
The upper panel shows the peak position using the peak finder based on the gravitational potential; and the lower panel displays the positions of the peaks based on isodensity contours.
One can recognize two main differences between the peak finding methods:
First, in the lower panel, the peak position around the pericentre passages is biased towards the centre of mass as no separate peaks can be identified for small separations and thus the haloes seem to coalesce earlier.
Secondly, we observe the offsets to be smaller for the isodensity contour peaks at the first apocentre passage. They are about half the size of the potential-based offsets.
This is caused by the projection since self-interactions mainly alter the evolution of the central part of the haloes and the peaks identified in the potential-based method heavily depend on this region, whereas the isodensity contour peaks are due to projection more sensitive to matter in the outer regions of the haloes.

For comparison, the offsets for the two peak finding methods are shown for a 1:5 cluster-scale merger in Fig.\ref{fig:peakcomp}.
The isodensity contour-based offsets are noisier than the potential-based ones, but overall they follow the same trend. Interestingly, also the measured offsets direct after the first pericentre passage appear to be larger in this case. For the isodensity contour peaks, all particles are considered, thus the main halo can influence the position of the subhalo. If the density gradient in the DM component of the subhalo is lower than the one of the galactic component, the DM peak could be more affected by the main halo. Potentially, this could lead to a larger offset measurement.
In contrast to the potential-based offset measurements, the isodensity contour-based ones are not much larger about the first apocentre passage (negative sign) compared to the ones subsequent to the first pericentre passage (positive sign), but they last for a longer time and, as such, could be easier to observe.
But the isodensity method does not provide reasonable offsets at times as late as for the potential-based peaks. However, for comparison with observations, an observationally motivated peak finding strategy should be employed \citep{Robertson_2017a}.

Let us point out that very late merger stages at which the DM haloes already coalesced could nevertheless be of interest from an observational point of view. This is because the presence of self-interactions may lead to distinguishable galactic/stellar components.
In Fig.~\ref{fig:cphase}, we show this late stage for an equal-mass merger because for that MMR the peak finding works best.
The separation between the galaxy peaks for the cluster-scale merger evolved with $\sigma_\mathrm{\Tilde{T}}/m = 0.1 \, \mathrm{cm}^2 \, \mathrm{g}^{-1}$ is shown.
For comparison, we also display the separation for the corresponding CDM merger.
Here, the separation vanishes quickly.
But if self-interactions are present, even if they are rather small, large separations are found with both methods.
For fSIDM, the distance between the galaxy peaks tends to be larger than for rSIDM if the same momentum transfer cross-sections are compared.
In Section~\ref{sec:results_offsets}, we mentioned that a lower density due to self-interactions reduces dynamical friction and thus can lead to core sloshing as previously studied by \cite{Kim_2017b}.
However, it remains to be seen whether this persists in more realistic simulations including the ICM.

\subsection{Phase--space} \label{sec:results_phase_space}

\begin{figure*}
    \centering
    \includegraphics[width=\columnwidth]{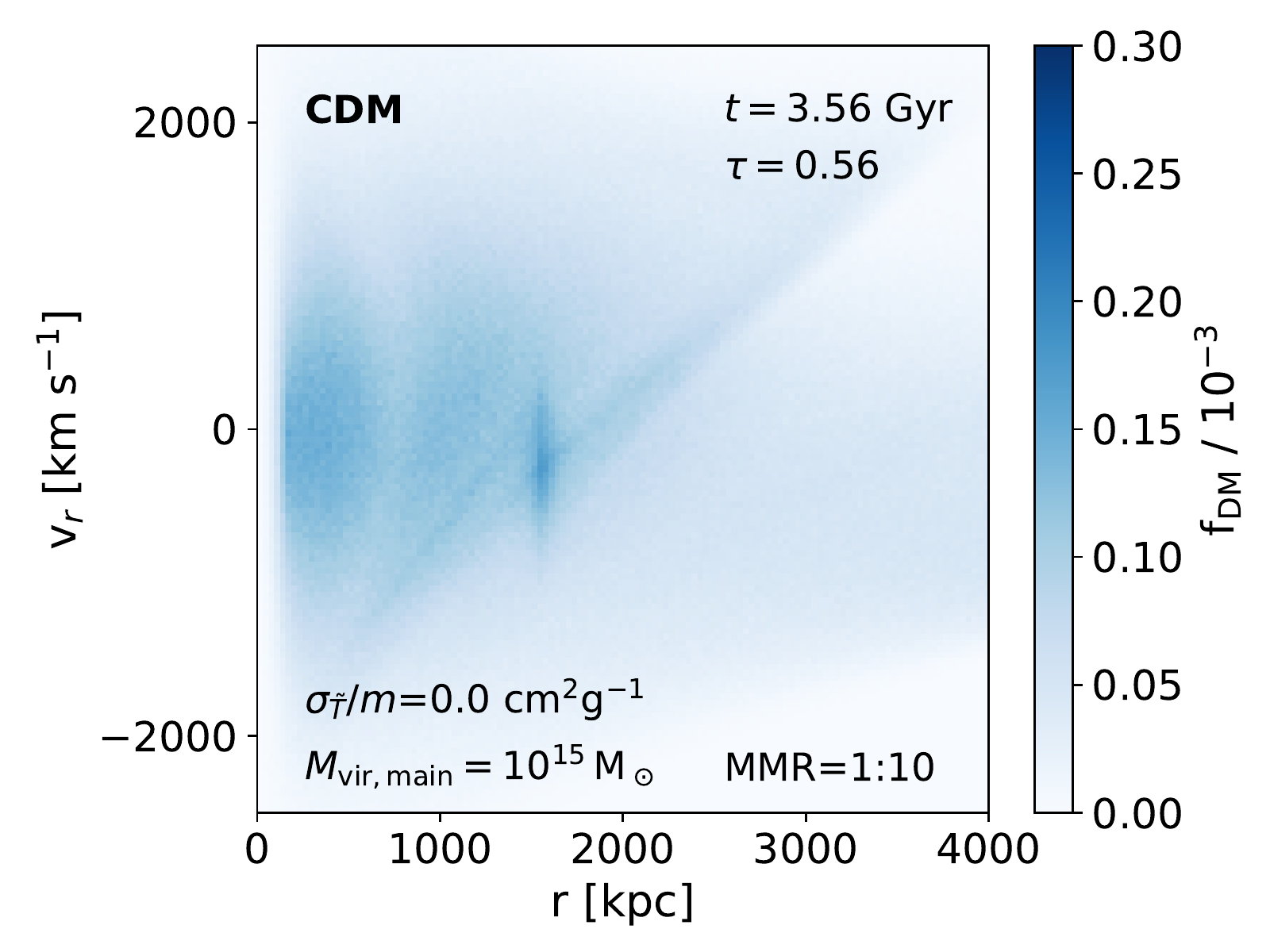}
    \includegraphics[width=\columnwidth]{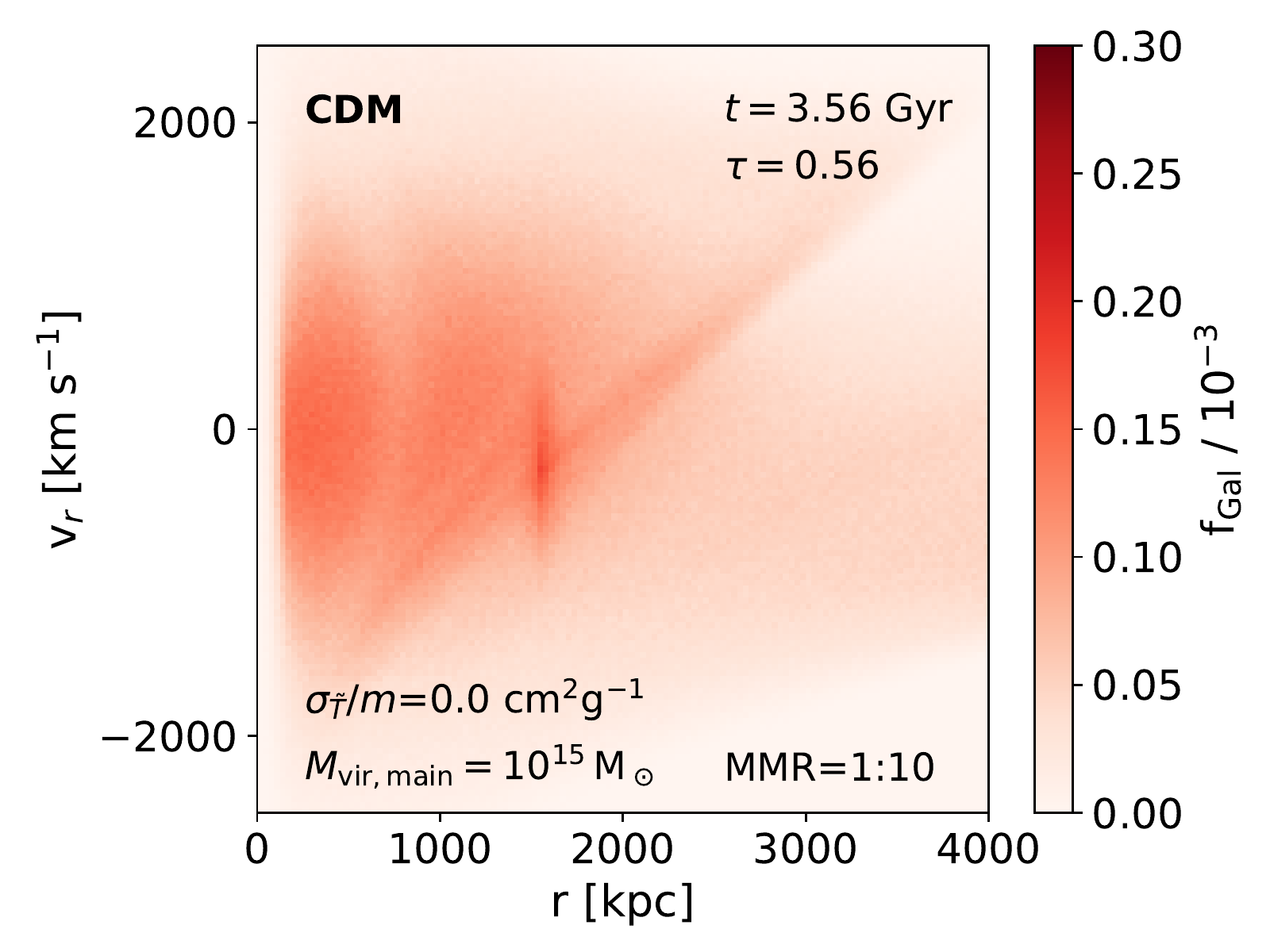}
    \includegraphics[width=\columnwidth]{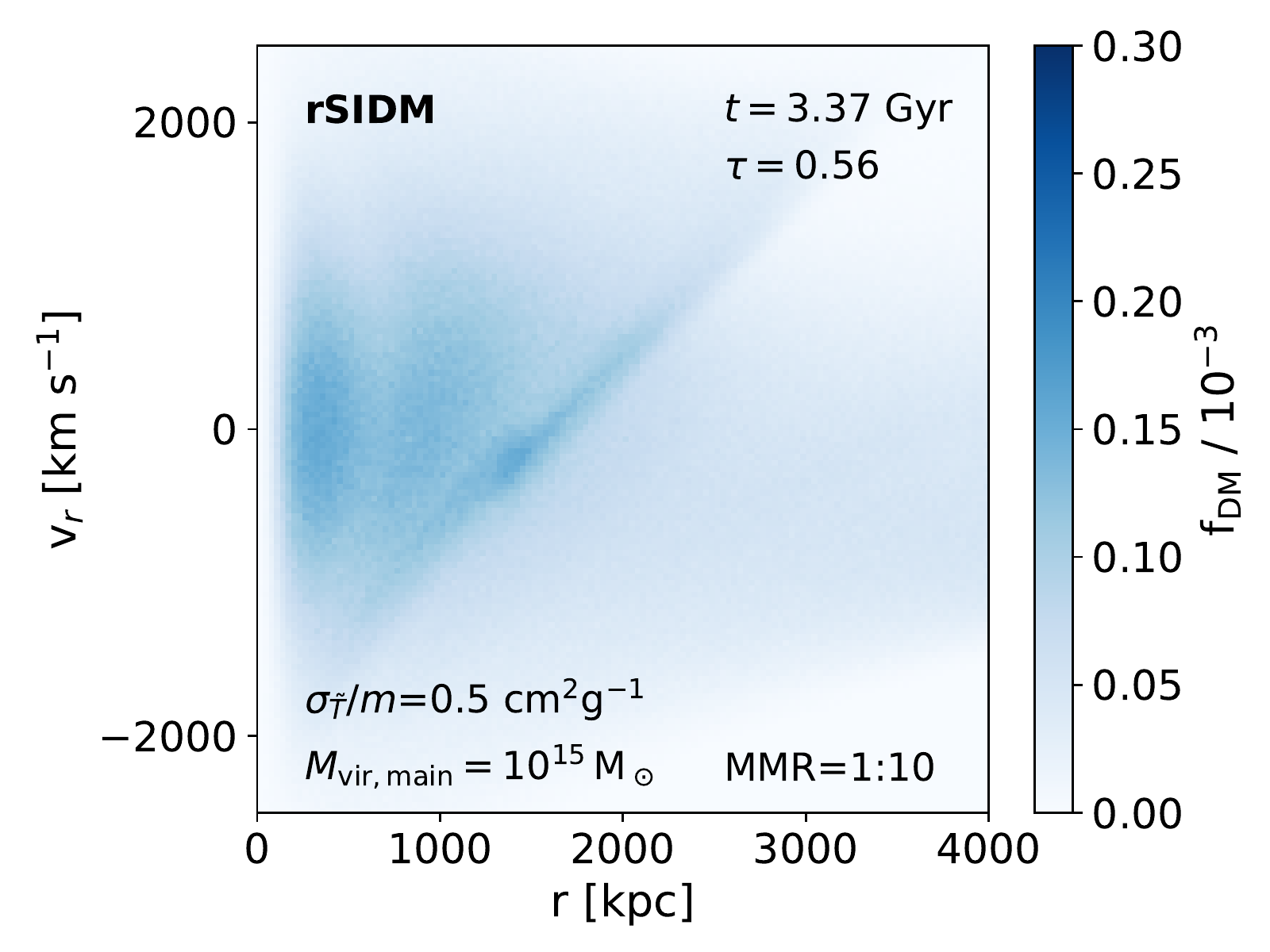}
    \includegraphics[width=\columnwidth]{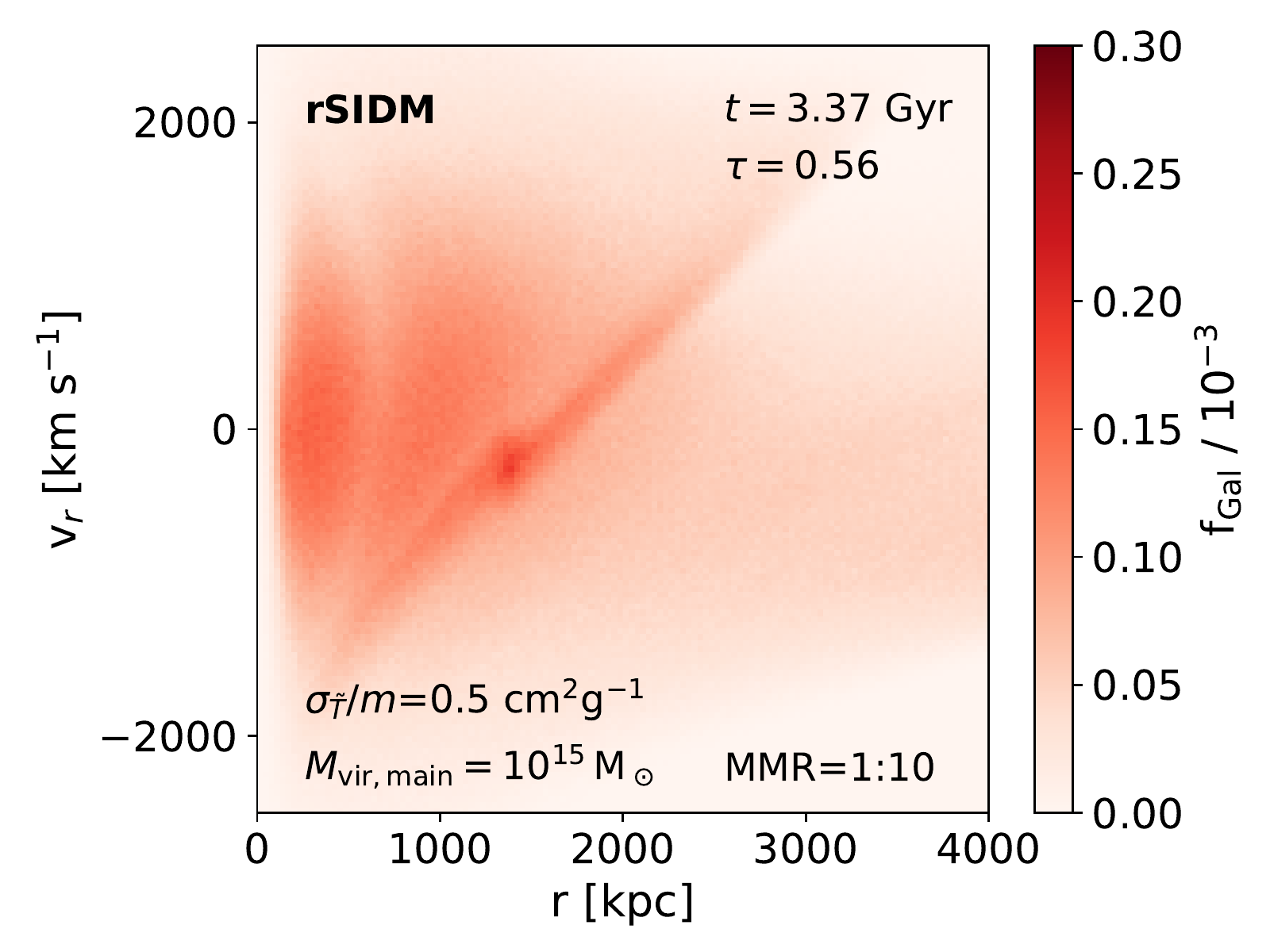}
    \includegraphics[width=\columnwidth]{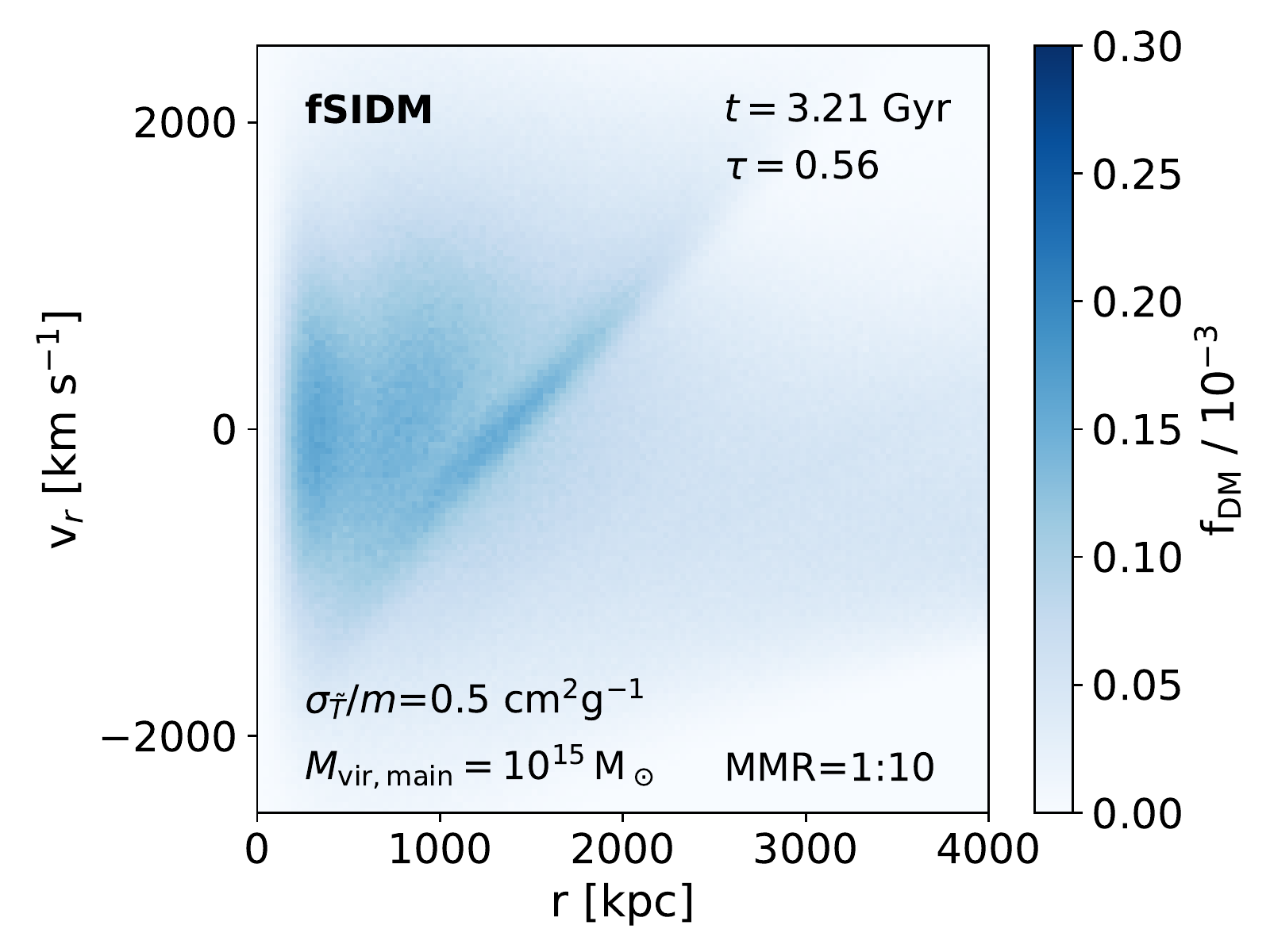}
    \includegraphics[width=\columnwidth]{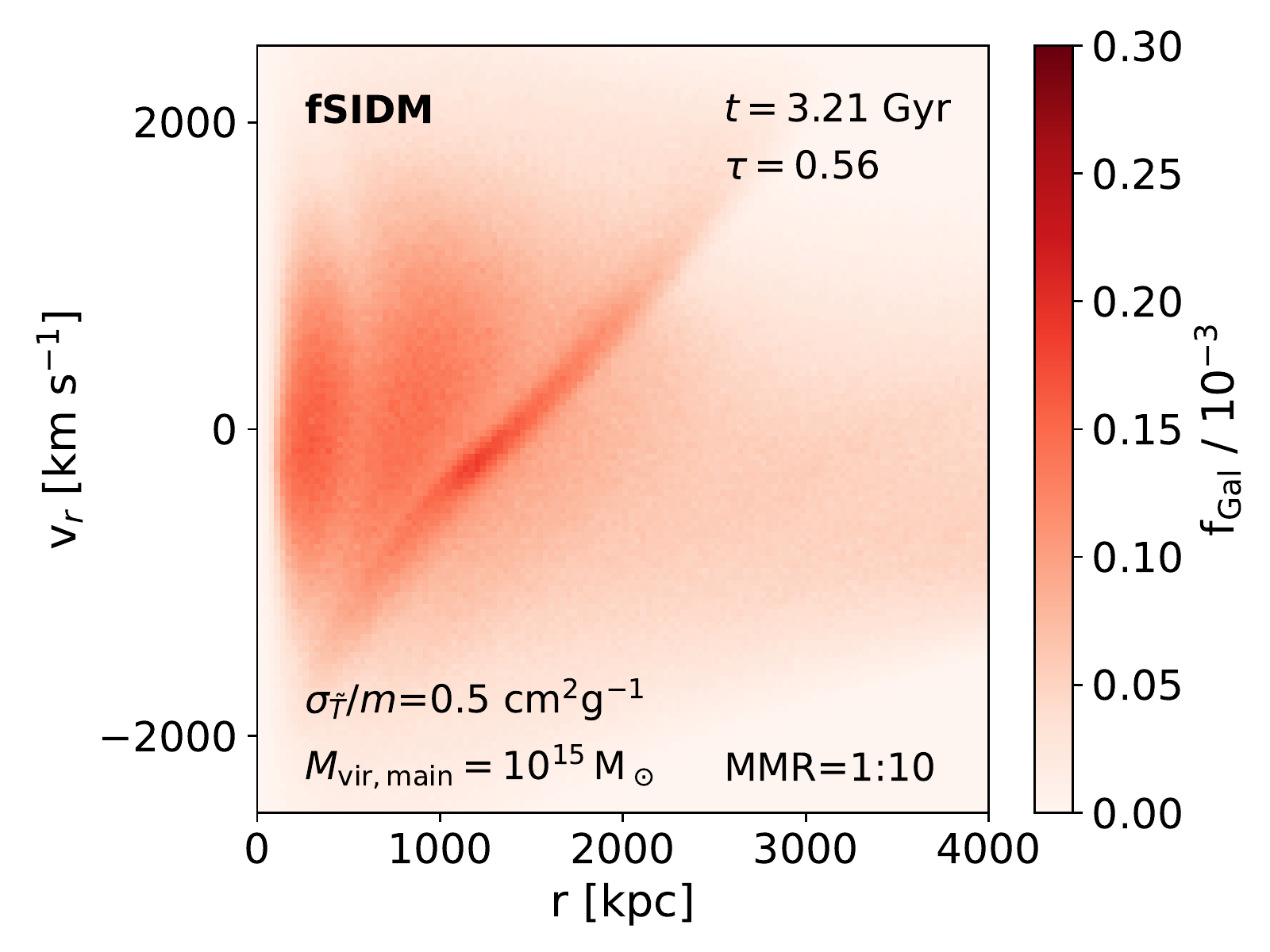}
    \caption{The phase--space distribution of a 1:10 merger evolved with different DM models is shown at $\tau = 0.56$, i.e. close to the first apocentre passage. The left-hand side column displays the DM distribution and on the right-hand side, the smoothed galactic component is shown. The top row gives the phase--space distribution for CDM and below SIDM is shown with a cross-section of $\sigma_\mathrm{\Tilde{T}}/m = 0.5 \, \mathrm{cm}^2 \, \mathrm{g}^{-1}$ for the case of isotropic scattering (middle row) and small-angle scattering (bottom row). }
    \label{fig:phase_space}
\end{figure*}

Finally, we study the phase--space distribution of our 1:10 cluster-scale merger using the same simulation as for the morphology.
For the phase--space distribution, we do not only consider particles of the subhalo but also from the main halo.
In Fig.~\ref{fig:phase_space}, we show the distance to the centre of mass as a function of the radial velocity (with respect to the centre of mass) for $\tau = 0.56$.
These quantities are all computed in 3 d.
We display results for CDM and for rare and frequent self-interactions with a cross-section of $\sigma_\mathrm{\Tilde{T}}/m = 0.5 \, \mathrm{cm}^2 \, \mathrm{g}^{-1}$.
On the left-hand side, we display the DM and on the right-hand side, the smoothed galactic component is shown.

A clear difference between the distributions for the DM models is visible, especially when focusing on the clump of particles at $r \sim 1500 \, \mathrm{kpc}$.
These particles primarily belong to the subhalo and disperse over time depending on the DM physics.
If DM self-interactions are present, the gravitational potential is shallower, which also leads to a faster dispersal of the galactic component.
As we can see in Fig.~\ref{fig:phase_space}, the subhalo clump is strongest for CDM, less strong for rSIDM and the weakest for fSIDM for both DM and galaxies.

\section{Discussion} \label{sec:discussion}

In this section, we first discuss aspects relating to the peak finding and the analysis of our simulation.
Then, we elaborate on the limitations of our model and the physical implications of our results.

\subsection{Technical aspects} \label{sec:discussion_tech}

We compared two peak finding methods, whose results differ significantly.
The one based on the gravitational potential is more robust but less useful when it comes to a potential comparison with observational data, whereas the one based on isodensity contours may be more readily applied to observations.

The isodensity contour method suffers from projection effects and difficulties in the peak identification for small peak separations.
As a result, we found smaller offsets (about half the size for an equal-mass merger) with the isodensity contour method at the first apocentre passage. However, it is possible that the isodensity contour method provides larger offsets close to the pericentre passages as the measured subhalo peak position is affected by the main halo.
Moreover, we failed to identify distinct peaks at an earlier stage of the merging process than with the gravitational potential based method. 
Furthermore, we only investigated projections of the particle distribution perpendicular to the merger axis, whereas in observations, the line of sight and merger axis need not be perpendicular to one another.

However, in real observations, further difficulties arise.
Especially for galaxy clusters, the number of observed galaxies ($\sim 100$--$1000$) is much less than the number of particles ($\sim 9 \times 10^5$--$10^7$) we used for our smoothed galactic cluster component.
Nevertheless, this problem could be allayed by using the BCGs to measure offsets.

We find that observations at a later stage of the merger, rather than around the time of the first pericentre passage, might be more interesting because offsets typically become larger with time. However, the largest offsets might be difficult to observe as the subhalo dissolves rather quickly.
Nevertheless, a late merger phase when the DM haloes coalesce and form a single DM peak may still provide evidence for DM self-interactions as in general two separate galactic or stellar components will be present.
For equal-mass mergers, \cite{Kim_2017b} found at these later merger stages that BCGs and galaxies oscillated around the centre of mass. 
We make a similar observation in our simulations and find that this also persists for unequal mass mergers. Interestingly, the galactic peak separations are found to be generally large even for rather small cross-sections and can also be seen employing the more observationally motivated peak finding method. Hence, these types of observations could be a promising way to provide evidence for DM self-interactions.

Distinguishing between different DM models through the morphologies of the galaxies within merging galaxy clusters may prove difficult owing to the small number of collisionless tracers (galaxies) on these scales.
However, on galaxy scales, stars may provide enough tracers to better estimate the locations and shapes of the collisonless components.
In addition, galaxies might also offer a chance to distinguish stellar components according to their origin using stellar population properties such as metallicity.
Nevertheless, resolution limits of astronomical observations might pose a challenge for such an approach.

\subsection{Physical considerations}

The focus of this paper lies on understanding the different phenomenologies of rare and frequent self-interactions and not a comparison to observational data.
Hence, the dependence of our results on the initial conditions is less of a concern.
For example, we find larger offsets than in \cite{Fischer_2021}, despite the haloes starting with the same virial masses. This is because they start with higher concentration parameters, which for the present study lead to higher central densities and thus larger offsets.
Aside from quantitative differences, some results may change qualitatively as physical processes shaping the evolution of merging systems act on different time-scales. For instance, the merger time-scale differs from the time-scale on which self-interactions isotropize the DM velocity distribution. In particular, this could complicate the evolution of the halo shapes.
Consequently, it would be informative to extract merging systems from cosmological simulations to obtain more realistic results that can be directly compared with observations.
In addition to the study of individual systems, this would also allow us to estimate how frequently offsets of observable size would occur for various SIDM models.
In the literature, observations of fairly large offsets have been claimed but there is also reasonable doubt about them (for references see sec.~\ref{sec:introduction}). However, in the light of large offsets, fSIDM models are particularly interesting as they can explain larger offsets than rSIDM.

We found minor mergers to be interesting in terms of distinguishing rare and frequent self-interactions.
In contrast to equal-mass mergers, they have the advantage of being more abundant in the Universe and thus allow for better statistics than studies based on individual systems.
For fSIDM, the subhalo dissolves faster than for rSIDM, but this statement depends on the matching of the cross-section. In terms of $\sigma_\mathrm{\tilde{T}}$, the constraints for rare and frequent self-interactions would differ from each other. Together with alternative constraints, there could be a chance to distinguish between rSIDM and fSIDM.
However, to derive constraints on the self-interactions with observations would require a stringent observational motivated analysis of the simulations and a more realistic setup as we explain next. Thus, we do not try to derive any constraints on the differential cross-section.

Here, we studied an idealized setup that neglects various physical aspects.
Perhaps, most importantly, we did not include the baryonic matter, i.e.\ the ICM, which contains a significant fraction ($\sim 10\%$) of the cluster mass. The ICM is likely to change the evolution of a merger as it behaves collisionally \citep{Zhang_2016} and is affected by processes such as star formation and feedback.
Moreover, our haloes are totally smooth and do not contain any substructure.
The modelling of the BCGs is also very idealized as they are approximated as collisionless point masses with unrealistically low masses.
Finally, we treated the galaxies in our cluster simulations as collisionless particles, which neglects the fact that they also have a large DM component \citep{Kummer_2018}.
More realistic modelling of the BCGs and the ICM could lead to different results, in particular, the pattern of oscillations of the BCGs and the core sloshing of the galaxies could change.
It is largely unknown how the evolution of a merger subject to self-interactions would change if one improves on the aspects mentioned above. Hence, we do not want to speculate about this but leave it for future studies.

In the present work, we only modelled a constant cross-section and did not consider SIDM with a velocity-dependent cross-section.
A velocity-dependence is natural from a particle physics perspective, especially for light mediator models, which interact frequently \citep[e.g.][]{Buckley_2010, Loeb_2011, Bringmann:2016din}.
In future work, it would be interesting to also investigate models with a velocity dependence, particularly as such models appear to be in better agreement with astrophysical observations than constant cross-sections \citep[e.g.][]{Kaplinghat_2016, Correa_2021, Gilman_2021, Sagunski_2021}.

\section{Summary and Conclusions} \label{sec:conclusion}

In this paper, we have studied idealized equal and unequal-mass mergers undergoing head-on collisions, focusing on the effects arising from DM self-interactions.
In particular, we have investigated galaxy cluster and galaxy mergers and compared simulations with collisionless, rare and frequent self-interacting DM.
In each simulation, we determined the peaks of the different components (DM, galaxies, and a central massive object) for each of our two merging haloes, and measured the offsets between different components as well as the shapes of the different components.
Moreover, we studied the morphology and phase--space distribution of the mergers and compared two peak finding methods.
Our main results from this suite of simulations are as follows:

\begin{itemize}
    \item The morphology of the collisionless particles, i.e.\ galaxies/stars shows significant differences between rSIDM and fSIDM, especially at a later merger phase and in unequal-mass mergers. It is strongly affected by the faster dissolving fSIDM subhalo.

    \item Minor mergers with SIDM can produce large offsets between galactic/stellar and DM peaks.
    Before the second pericentre passage, we found large offsets only for frequent self-interactions.
    
    \item In general, frequent self-interactions tend to produce larger offsets than rare self-interactions if the same momentum transfer cross-sections are compared.
    This is even more extreme in unequal-mass mergers.

    \item Separate galactic/stellar components with a coalesced DM component could provide evidence for DM self-interactions. Core sloshing seems to be most interesting in the case of equal-mass mergers, which has been studied by \cite{Kim_2017b}.
    
    \item For SIDM, the shapes of DM and the galactic/stellar component can differ significantly from collisionless DM. In unequal-mass mergers, self-interactions can lead to more elliptical distributions between the first apocentre and second pericentre passage.
    
    \item We find the merger phase between $\tau \approx 0.25$ (halfway to the first apocentre) and $\tau \approx 1.0$ (second pericentre) to be more interesting in terms of studying effects from self-interactions than the phase near and shortly after the first pericentre passage ($\tau \approx 0.0$).
    This seems to be true for observations as well, as suggested by our results concerning isodensity peaks.
    In general, a late phase of a merging system should be more interesting as the non-linear evolution of the system can amplify differences between DM models.
    
    \item Peak finding is a major challenge in studying effects of SIDM and distinguishing rare and frequent self-interactions.
    In observations, it might prohibit detecting the very large offsets at late merger stages.
    The positions obtained by various methods can differ a lot with respect to one another, which makes it important to analyse observations and simulations in a similar manner.
\end{itemize}

In this paper, we performed a parameter study on idealized mergers in order to understand the dominant physical effects and to develop an intuition about the best indicators of DM interaction properties.
A detailed comparison with observations will require more realistic simulations that include baryonic matter as well as more realistic substructure in the haloes. This is the subject of forthcoming work. 

\section*{Acknowledgements}
We thank Felix Kahlhoefer for useful comments and suggestions.
This work is funded by the Deutsche Forschungsgemeinschaft (DFG, German Research Foundation) under Germany's Excellence Strategy -- EXC 2121 ``Quantum Universe'' --  390833306, Germany’s Excellence Strategy -- EXC-2094 ``Origins'' -- 390783311, and the Emmy Noether Grant No.\ KA 4662/1-1. AR is supported by the European Research Council's Horizon 2020 project `EWC' (award AMD-776247- 6).
Preprint number: DESY-21-116

Software:
numpy \citep{NumPy},
matplotlib \citep{Matplotlib},
scipy \citep{SciPy}

\section*{Data Availability}
The data underlying this article will be shared on reasonable request to the corresponding author.



\bibliographystyle{mnras}
\bibliography{references.bib}

\begin{thebibliography}{}
\makeatletter
\relax
\def\mn@urlcharsother{\let\do\@makeother \do\$\do\&\do\#\do\^\do\_\do\%\do\~}
\def\mn@doi{\begingroup\mn@urlcharsother \@ifnextchar [ {\mn@doi@}
  {\mn@doi@[]}}
\def\mn@doi@[#1]#2{\def\@tempa{#1}\ifx\@tempa\@empty \href
  {http://dx.doi.org/#2} {doi:#2}\else \href {http://dx.doi.org/#2} {#1}\fi
  \endgroup}
\def\mn@eprint#1#2{\mn@eprint@#1:#2::\@nil}
\def\mn@eprint@arXiv#1{\href {http://arxiv.org/abs/#1} {{\tt arXiv:#1}}}
\def\mn@eprint@dblp#1{\href {http://dblp.uni-trier.de/rec/bibtex/#1.xml}
  {dblp:#1}}
\def\mn@eprint@#1:#2:#3:#4\@nil{\def\@tempa {#1}\def\@tempb {#2}\def\@tempc
  {#3}\ifx \@tempc \@empty \let \@tempc \@tempb \let \@tempb \@tempa \fi \ifx
  \@tempb \@empty \def\@tempb {arXiv}\fi \@ifundefined
  {mn@eprint@\@tempb}{\@tempb:\@tempc}{\expandafter \expandafter \csname
  mn@eprint@\@tempb\endcsname \expandafter{\@tempc}}}

\bibitem[\protect\citeauthoryear{Ackerman, Buckley, Carroll  \&
  Kamionkowski}{Ackerman et~al.}{2009}]{Ackermann_2009}
Ackerman L.,  Buckley M.~R.,  Carroll S.~M.,   Kamionkowski M.,  2009, \mn@doi
  [Phys. Rev. D] {10.1103/PhysRevD.79.023519}, 79, 023519

\bibitem[\protect\citeauthoryear{Allgood, Flores, Primack, Kravtsov, Wechsler,
  Faltenbacher  \& Bullock}{Allgood et~al.}{2006}]{Allgood_2006}
Allgood B.,  Flores R.~A.,  Primack J.~R.,  Kravtsov A.~V.,  Wechsler R.~H.,
  Faltenbacher A.,   Bullock J.~S.,  2006, \mn@doi [Monthly Notices of the
  Royal Astronomical Society] {10.1111/j.1365-2966.2006.10094.x}, 367, 1781

\bibitem[\protect\citeauthoryear{Banerjee, Adhikari, Dalal, More  \&
  Kravtsov}{Banerjee et~al.}{2020}]{Banerjee_2020}
Banerjee A.,  Adhikari S.,  Dalal N.,  More S.,   Kravtsov A.,  2020, \mn@doi
  [Journal of Cosmology and Astroparticle Physics]
  {10.1088/1475-7516/2020/02/024}, 2020, 024

\bibitem[\protect\citeauthoryear{Berezhiani, Dolgov  \& Mohapatra}{Berezhiani
  et~al.}{1996}]{Berezhiani_1996}
Berezhiani Z.,  Dolgov A.,   Mohapatra R.,  1996, \mn@doi [Physics Letters B]
  {https://doi.org/10.1016/0370-2693(96)00219-5}, 375, 26

\bibitem[\protect\citeauthoryear{Bett}{Bett}{2012}]{Bett_2012}
Bett P.,  2012, \mn@doi [Monthly Notices of the Royal Astronomical Society]
  {10.1111/j.1365-2966.2011.20258.x}, 420, 3303

\bibitem[\protect\citeauthoryear{{Blinnikov} \& {Khlopov}}{{Blinnikov} \&
  {Khlopov}}{1983}]{Blinnikov_1983}
{Blinnikov} S.~I.,  {Khlopov} M.~Y.,  1983, \sovast, \href
  {https://ui.adsabs.harvard.edu/abs/1983SvA....27..371B} {27, 371}

\bibitem[\protect\citeauthoryear{Boddy, Kaplinghat, Kwa  \& Peter}{Boddy
  et~al.}{2016}]{Boddy_2016}
Boddy K.~K.,  Kaplinghat M.,  Kwa A.,   Peter A. H.~G.,  2016, \mn@doi [Phys.
  Rev. D] {10.1103/PhysRevD.94.123017}, 94, 123017

\bibitem[\protect\citeauthoryear{Boylan-Kolchin, Springel, White, Jenkins  \&
  Lemson}{Boylan-Kolchin et~al.}{2009}]{Boylan-Kolchin_2009}
Boylan-Kolchin M.,  Springel V.,  White S. D.~M.,  Jenkins A.,   Lemson G.,
  2009, \mn@doi [Monthly Notices of the Royal Astronomical Society]
  {10.1111/j.1365-2966.2009.15191.x}, 398, 1150

\bibitem[\protect\citeauthoryear{{Brada{\v{c}}}, {Allen}, {Treu}, {Ebeling},
  {Massey}, {Morris}, {von der Linden}  \& {Applegate}}{{Brada{\v{c}}}
  et~al.}{2008}]{Bradac_2008}
{Brada{\v{c}}} M.,  {Allen} S.~W.,  {Treu} T.,  {Ebeling} H.,  {Massey} R.,
  {Morris} R.~G.,  {von der Linden} A.,   {Applegate} D.,  2008, \mn@doi [\apj]
  {10.1086/591246}, \href
  {https://ui.adsabs.harvard.edu/abs/2008ApJ...687..959B} {687, 959}

\bibitem[\protect\citeauthoryear{Bringmann, Kahlhoefer, Schmidt-Hoberg  \&
  Walia}{Bringmann et~al.}{2017}]{Bringmann:2016din}
Bringmann T.,  Kahlhoefer F.,  Schmidt-Hoberg K.,   Walia P.,  2017, \mn@doi
  [Phys. Rev. Lett.] {10.1103/PhysRevLett.118.141802}, 118, 141802

\bibitem[\protect\citeauthoryear{Buckley \& Fox}{Buckley \&
  Fox}{2010}]{Buckley_2010}
Buckley M.~R.,  Fox P.~J.,  2010, \mn@doi [Phys. Rev. D]
  {10.1103/PhysRevD.81.083522}, 81, 083522

\bibitem[\protect\citeauthoryear{{Bullock} \& {Boylan-Kolchin}}{{Bullock} \&
  {Boylan-Kolchin}}{2017}]{Bullock_2017}
{Bullock} J.~S.,  {Boylan-Kolchin} M.,  2017, \mn@doi [\araa]
  {10.1146/annurev-astro-091916-055313}, \href
  {https://ui.adsabs.harvard.edu/abs/2017ARA&A..55..343B} {55, 343}

\bibitem[\protect\citeauthoryear{{Carlson}, {Machacek}  \& {Hall}}{{Carlson}
  et~al.}{1992}]{Carlson_1992}
{Carlson} E.~D.,  {Machacek} M.~E.,   {Hall} L.~J.,  1992, \mn@doi [\apj]
  {10.1086/171833}, \href
  {https://ui.adsabs.harvard.edu/abs/1992ApJ...398...43C} {398, 43}

\bibitem[\protect\citeauthoryear{{Chandrasekhar}}{{Chandrasekhar}}{1943}]{Chandrasekhar_1943}
{Chandrasekhar} S.,  1943, \mn@doi [\apj] {10.1086/144517}, \href
  {https://ui.adsabs.harvard.edu/abs/1943ApJ....97..255C} {97, 255}

\bibitem[\protect\citeauthoryear{Chua, Dibert, Vogelsberger  \& Zavala}{Chua
  et~al.}{2020}]{Chua_2020}
Chua K. T.~E.,  Dibert K.,  Vogelsberger M.,   Zavala J.,  2020, \mn@doi
  [Monthly Notices of the Royal Astronomical Society] {10.1093/mnras/staa3315},
  500, 1531

\bibitem[\protect\citeauthoryear{Cline, Liu  \& Xue}{Cline
  et~al.}{2012}]{Cline_2012}
Cline J.~M.,  Liu Z.,   Xue W.,  2012, \mn@doi [Phys. Rev. D]
  {10.1103/PhysRevD.85.101302}, 85, 101302

\bibitem[\protect\citeauthoryear{Colin, Avila-Reese, Valenzuela  \&
  Firmani}{Colin et~al.}{2002}]{Colin_2002}
Colin P.,  Avila-Reese V.,  Valenzuela O.,   Firmani C.,  2002, \mn@doi [The
  Astrophysical Journal] {10.1086/344259}, 581, 777

\bibitem[\protect\citeauthoryear{Correa}{Correa}{2021}]{Correa_2021}
Correa C.~A.,  2021, \mn@doi [Monthly Notices of the Royal Astronomical
  Society] {10.1093/mnras/stab506}, 503, 920

\bibitem[\protect\citeauthoryear{Cyr-Racine \& Sigurdson}{Cyr-Racine \&
  Sigurdson}{2013}]{Cyr-Racine_2013}
Cyr-Racine F.-Y.,  Sigurdson K.,  2013, \mn@doi [Phys. Rev. D]
  {10.1103/PhysRevD.87.103515}, 87, 103515

\bibitem[\protect\citeauthoryear{{Dawson}}{{Dawson}}{2013}]{Dawson_2013}
{Dawson} W.~A.,  2013, PhD thesis, Univ. California

\bibitem[\protect\citeauthoryear{{Dawson} et~al.,}{{Dawson}
  et~al.}{2012}]{Dawson_2012}
{Dawson} W.~A.,  et~al., 2012, \mn@doi [\apjl] {10.1088/2041-8205/747/2/L42},
  \href {https://ui.adsabs.harvard.edu/abs/2012ApJ...747L..42D} {747, L42}

\bibitem[\protect\citeauthoryear{Despali, Sparre, Vegetti, Vogelsberger, Zavala
   \& Marinacci}{Despali et~al.}{2019}]{Despali_2019}
Despali G.,  Sparre M.,  Vegetti S.,  Vogelsberger M.,  Zavala J.,   Marinacci
  F.,  2019, \mn@doi [Monthly Notices of the Royal Astronomical Society]
  {10.1093/mnras/stz273}, 484, 4563–4573

\bibitem[\protect\citeauthoryear{{Dodelson} \& {Widrow}}{{Dodelson} \&
  {Widrow}}{1994}]{Dodelson_1994}
{Dodelson} S.,  {Widrow} L.~M.,  1994, \mn@doi [\prl]
  {10.1103/PhysRevLett.72.17}, \href
  {https://ui.adsabs.harvard.edu/abs/1994PhRvL..72...17D} {72, 17}

\bibitem[\protect\citeauthoryear{Donnert}{Donnert}{2014}]{Donnert_2014}
Donnert J. M.~F.,  2014, \mn@doi [Monthly Notices of the Royal Astronomical
  Society] {10.1093/mnras/stt2291}, 438, 1971

\bibitem[\protect\citeauthoryear{Donnert, Beck, Dolag  \& Röttgering}{Donnert
  et~al.}{2017}]{Donnert_2017}
Donnert J. M.~F.,  Beck A.~M.,  Dolag K.,   Röttgering H. J.~A.,  2017,
  \mn@doi [Monthly Notices of the Royal Astronomical Society]
  {10.1093/mnras/stx1819}, 471, 4587

\bibitem[\protect\citeauthoryear{Doubrawa, Machado, Laganá, Lima Neto,
  Monteiro-Oliveira  \& Cypriano}{Doubrawa et~al.}{2020}]{Doubrawa_2020}
Doubrawa L.,  Machado R. E.~G.,  Laganá T.~F.,  Lima Neto G.~B.,
  Monteiro-Oliveira R.,   Cypriano E.~S.,  2020, \mn@doi [Monthly Notices of
  the Royal Astronomical Society] {10.1093/mnras/staa1051}, 495, 2022

\bibitem[\protect\citeauthoryear{{Dutton} \& {Macci{\`o}}}{{Dutton} \&
  {Macci{\`o}}}{2014}]{Dutton_2014}
{Dutton} A.~A.,  {Macci{\`o}} A.~V.,  2014, \mn@doi [\mnras]
  {10.1093/mnras/stu742}, \href
  {https://ui.adsabs.harvard.edu/abs/2014MNRAS.441.3359D} {441, 3359}

\bibitem[\protect\citeauthoryear{{Dyer} \& {Ip}}{{Dyer} \&
  {Ip}}{1993}]{Dyer_1993}
{Dyer} C.~C.,  {Ip} P. S.~S.,  1993, \mn@doi [\apj] {10.1086/172641}, \href
  {https://ui.adsabs.harvard.edu/abs/1993ApJ...409...60D} {409, 60}

\bibitem[\protect\citeauthoryear{{Essig}, {McDermott}, {Yu}  \&
  {Zhong}}{{Essig} et~al.}{2019}]{Essig_2019}
{Essig} R.,  {McDermott} S.~D.,  {Yu} H.-B.,   {Zhong} Y.-M.,  2019, \mn@doi
  [\prl] {10.1103/PhysRevLett.123.121102}, \href
  {https://ui.adsabs.harvard.edu/abs/2019PhRvL.123l1102E} {123, 121102}

\bibitem[\protect\citeauthoryear{Feng, Kaplinghat, Tu  \& Yu}{Feng
  et~al.}{2009}]{Feng_2009}
Feng J.~L.,  Kaplinghat M.,  Tu H.,   Yu H.-B.,  2009, \mn@doi [Journal of
  Cosmology and Astroparticle Physics] {10.1088/1475-7516/2009/07/004}, 2009,
  004

\bibitem[\protect\citeauthoryear{Fischer, Brüggen, Schmidt-Hoberg, Dolag,
  Kahlhoefer, Ragagnin  \& Robertson}{Fischer et~al.}{2021}]{Fischer_2021}
Fischer M.~S.,  Brüggen M.,  Schmidt-Hoberg K.,  Dolag K.,  Kahlhoefer F.,
  Ragagnin A.,   Robertson A.,  2021, \mn@doi [Monthly Notices of the Royal
  Astronomical Society] {10.1093/mnras/stab1198}, 505, 851

\bibitem[\protect\citeauthoryear{Foot}{Foot}{2004}]{Foot_2004}
Foot R.,  2004, \mn@doi [Int. J. Mod. Phys. D] {10.1142/S0218271804006449}, 13,
  2161

\bibitem[\protect\citeauthoryear{Foot \& Vagnozzi}{Foot \&
  Vagnozzi}{2015}]{Foot_2015}
Foot R.,  Vagnozzi S.,  2015, \mn@doi [Phys. Rev. D]
  {10.1103/PhysRevD.91.023512}, 91, 023512

\bibitem[\protect\citeauthoryear{Frandsen, Sarkar  \& Schmidt-Hoberg}{Frandsen
  et~al.}{2011}]{Frandsen_2011}
Frandsen M.~T.,  Sarkar S.,   Schmidt-Hoberg K.,  2011, \mn@doi [Phys. Rev. D]
  {10.1103/PhysRevD.84.051703}, 84, 051703

\bibitem[\protect\citeauthoryear{Gilman, Bovy, Treu, Nierenberg, Birrer, Benson
   \& Sameie}{Gilman et~al.}{2021}]{Gilman_2021}
Gilman D.,  Bovy J.,  Treu T.,  Nierenberg A.,  Birrer S.,  Benson A.,   Sameie
  O.,  2021, \mn@doi [Monthly Notices of the Royal Astronomical Society]
  {10.1093/mnras/stab2335}, 507, 2432

\bibitem[\protect\citeauthoryear{Harris et~al.,}{Harris et~al.}{2020}]{NumPy}
Harris C.~R.,  et~al., 2020, \mn@doi [Nature] {10.1038/s41586-020-2649-2}, 585,
  357

\bibitem[\protect\citeauthoryear{{Harvey}, {Robertson}, {Massey}  \&
  {Kneib}}{{Harvey} et~al.}{2017}]{Harvey_2017}
{Harvey} D.,  {Robertson} A.,  {Massey} R.,   {Kneib} J.-P.,  2017, \mn@doi
  [\mnras] {10.1093/mnras/stw2671}, \href
  {https://ui.adsabs.harvard.edu/abs/2017MNRAS.464.3991H} {464, 3991}

\bibitem[\protect\citeauthoryear{Hopkins et~al.,}{Hopkins
  et~al.}{2018}]{Hopkins_2018}
Hopkins P.~F.,  et~al., 2018, \mn@doi [Monthly Notices of the Royal
  Astronomical Society] {10.1093/mnras/sty1690}, 480, 800

\bibitem[\protect\citeauthoryear{{Hu}, {Barkana}  \& {Gruzinov}}{{Hu}
  et~al.}{2000}]{Hu_2000}
{Hu} W.,  {Barkana} R.,   {Gruzinov} A.,  2000, \mn@doi [\prl]
  {10.1103/PhysRevLett.85.1158}, \href
  {https://ui.adsabs.harvard.edu/abs/2000PhRvL..85.1158H} {85, 1158}

\bibitem[\protect\citeauthoryear{{Hunter}}{{Hunter}}{2007}]{Matplotlib}
{Hunter} J.~D.,  2007, Comput. Sci. Engg., 9, 90

\bibitem[\protect\citeauthoryear{Huo, Yu  \& Zhong}{Huo
  et~al.}{2020}]{Huo_2019}
Huo R.,  Yu H.-B.,   Zhong Y.-M.,  2020, \mn@doi [Journal of Cosmology and
  Astroparticle Physics] {10.1088/1475-7516/2020/06/051}, 2020, 051

\bibitem[\protect\citeauthoryear{{Jee}, {Hughes}, {Menanteau}, {Sif{\'o}n},
  {Mandelbaum}, {Barrientos}, {Infante}  \& {Ng}}{{Jee}
  et~al.}{2014}]{Jee_2014}
{Jee} M.~J.,  {Hughes} J.~P.,  {Menanteau} F.,  {Sif{\'o}n} C.,  {Mandelbaum}
  R.,  {Barrientos} L.~F.,  {Infante} L.,   {Ng} K.~Y.,  2014, \mn@doi [\apj]
  {10.1088/0004-637X/785/1/20}, \href
  {https://ui.adsabs.harvard.edu/abs/2014ApJ...785...20J} {785, 20}

\bibitem[\protect\citeauthoryear{{Jee} et~al.,}{{Jee} et~al.}{2015}]{Jee_2015}
{Jee} M.~J.,  et~al., 2015, \mn@doi [\apj] {10.1088/0004-637X/802/1/46}, \href
  {https://ui.adsabs.harvard.edu/abs/2015ApJ...802...46J} {802, 46}

\bibitem[\protect\citeauthoryear{{Kahlhoefer}, {Schmidt-Hoberg}, {Frandsen}  \&
  {Sarkar}}{{Kahlhoefer} et~al.}{2014}]{Kahlhoefer_2014}
{Kahlhoefer} F.,  {Schmidt-Hoberg} K.,  {Frandsen} M.~T.,   {Sarkar} S.,  2014,
  \mn@doi [\mnras] {10.1093/mnras/stt2097}, \href
  {https://ui.adsabs.harvard.edu/abs/2014MNRAS.437.2865K} {437, 2865}

\bibitem[\protect\citeauthoryear{Kahlhoefer, Schmidt-Hoberg  \&
  Wild}{Kahlhoefer et~al.}{2017}]{Kahlhoefer:2017umn}
Kahlhoefer F.,  Schmidt-Hoberg K.,   Wild S.,  2017, \mn@doi [JCAP]
  {10.1088/1475-7516/2017/08/003}, 08, 003

\bibitem[\protect\citeauthoryear{Kaplan, Krnjaic, Rehermann  \& Wells}{Kaplan
  et~al.}{2010}]{Kaplan_2010}
Kaplan D.~E.,  Krnjaic G.~Z.,  Rehermann K.~R.,   Wells C.~M.,  2010, \mn@doi
  [Journal of Cosmology and Astroparticle Physics]
  {10.1088/1475-7516/2010/05/021}, 2010, 021

\bibitem[\protect\citeauthoryear{Kaplinghat, Tulin  \& Yu}{Kaplinghat
  et~al.}{2016}]{Kaplinghat_2016}
Kaplinghat M.,  Tulin S.,   Yu H.-B.,  2016, \mn@doi [Phys. Rev. Lett.]
  {10.1103/PhysRevLett.116.041302}, 116, 041302

\bibitem[\protect\citeauthoryear{{Kim}, {Peter}  \& {Wittman}}{{Kim}
  et~al.}{2017}]{Kim_2017b}
{Kim} S.~Y.,  {Peter} A. H.~G.,   {Wittman} D.,  2017, \mn@doi [\mnras]
  {10.1093/mnras/stx896}, \href
  {https://ui.adsabs.harvard.edu/abs/2017MNRAS.469.1414K} {469, 1414}

\bibitem[\protect\citeauthoryear{Klypin, Trujillo-Gomez  \& Primack}{Klypin
  et~al.}{2011}]{Klypin_2011}
Klypin A.~A.,  Trujillo-Gomez S.,   Primack J.,  2011, \mn@doi [The
  Astrophysical Journal] {10.1088/0004-637x/740/2/102}, 740, 102

\bibitem[\protect\citeauthoryear{{Kolb}, {Seckel}  \& {Turner}}{{Kolb}
  et~al.}{1985}]{Kolb_1985}
{Kolb} E.~W.,  {Seckel} D.,   {Turner} M.~S.,  1985, \mn@doi [\nat]
  {10.1038/314415a0}, \href
  {https://ui.adsabs.harvard.edu/abs/1985Natur.314..415K} {314, 415}

\bibitem[\protect\citeauthoryear{{Kummer}, {Kahlhoefer}  \&
  {Schmidt-Hoberg}}{{Kummer} et~al.}{2018}]{Kummer_2018}
{Kummer} J.,  {Kahlhoefer} F.,   {Schmidt-Hoberg} K.,  2018, \mn@doi [\mnras]
  {10.1093/mnras/stx2715}, \href
  {https://ui.adsabs.harvard.edu/abs/2018MNRAS.474..388K} {474, 388}

\bibitem[\protect\citeauthoryear{Kummer, Br\"uggen, Dolag, Kahlhoefer  \&
  Schmidt-Hoberg}{Kummer et~al.}{2019}]{Kummer:2019yrb}
Kummer J.,  Br\"uggen M.,  Dolag K.,  Kahlhoefer F.,   Schmidt-Hoberg K.,
  2019, \mn@doi [MNRAS] {10.1093/mnras/stz1261}, 487, 354

\bibitem[\protect\citeauthoryear{Kusenko \& Steinhardt}{Kusenko \&
  Steinhardt}{2001}]{Kusenko_2001}
Kusenko A.,  Steinhardt P.~J.,  2001, \mn@doi [Phys. Rev. Lett.]
  {10.1103/PhysRevLett.87.141301}, 87, 141301

\bibitem[\protect\citeauthoryear{Lacey \& Cole}{Lacey \&
  Cole}{1993}]{Lacey_1993}
Lacey C.,  Cole S.,  1993, \mn@doi [Monthly Notices of the Royal Astronomical
  Society] {10.1093/mnras/262.3.627}, 262, 627

\bibitem[\protect\citeauthoryear{Lage \& Farrar}{Lage \&
  Farrar}{2014}]{Lage_2014}
Lage C.,  Farrar G.,  2014, \mn@doi [The Astrophysical Journal]
  {10.1088/0004-637x/787/2/144}, 787, 144

\bibitem[\protect\citeauthoryear{Loeb \& Weiner}{Loeb \&
  Weiner}{2011}]{Loeb_2011}
Loeb A.,  Weiner N.,  2011, \mn@doi [Phys. Rev. Lett.]
  {10.1103/PhysRevLett.106.171302}, 106, 171302

\bibitem[\protect\citeauthoryear{Machado \& Lima~Neto}{Machado \&
  Lima~Neto}{2015}]{Machado_2015a}
Machado R. E.~G.,  Lima~Neto G.~B.,  2015, \mn@doi [Monthly Notices of the
  Royal Astronomical Society] {10.1093/mnras/stu2669}, 447, 2915–2924

\bibitem[\protect\citeauthoryear{Machado, Monteiro-Oliveira, Lima~Neto  \&
  Cypriano}{Machado et~al.}{2015}]{Machado_2015b}
Machado R. E.~G.,  Monteiro-Oliveira R.,  Lima~Neto G.~B.,   Cypriano E.~S.,
  2015, \mn@doi [Monthly Notices of the Royal Astronomical Society]
  {10.1093/mnras/stv1162}, 451, 3309

\bibitem[\protect\citeauthoryear{Mastropietro \& Burkert}{Mastropietro \&
  Burkert}{2008}]{Mastropietro_2008}
Mastropietro C.,  Burkert A.,  2008, \mn@doi [Monthly Notices of the Royal
  Astronomical Society] {10.1111/j.1365-2966.2008.13626.x}, 389, 967

\bibitem[\protect\citeauthoryear{Mohapatra, Nussinov  \& Teplitz}{Mohapatra
  et~al.}{2002}]{Mohapatra_2002}
Mohapatra R.~N.,  Nussinov S.,   Teplitz V.~L.,  2002, \mn@doi [Phys. Rev. D]
  {10.1103/PhysRevD.66.063002}, 66, 063002

\bibitem[\protect\citeauthoryear{Molnar \& Broadhurst}{Molnar \&
  Broadhurst}{2015}]{Molnar_2015}
Molnar S.~M.,  Broadhurst T.,  2015, \mn@doi [The Astrophysical Journal]
  {10.1088/0004-637x/800/1/37}, 800, 37

\bibitem[\protect\citeauthoryear{Molnar \& Broadhurst}{Molnar \&
  Broadhurst}{2017}]{Molnar_2017}
Molnar S.~M.,  Broadhurst T.,  2017, \mn@doi [The Astrophysical Journal]
  {10.3847/1538-4357/aa70a3}, 841, 46

\bibitem[\protect\citeauthoryear{Molnar \& Broadhurst}{Molnar \&
  Broadhurst}{2018}]{Molnar_2018}
Molnar S.~M.,  Broadhurst T.,  2018, \mn@doi [The Astrophysical Journal]
  {10.3847/1538-4357/aad04c}, 862, 112

\bibitem[\protect\citeauthoryear{Monteiro-Oliveira, Cypriano, Machado,
  Lima~Neto, Ribeiro, Sodré  \& Dupke}{Monteiro-Oliveira
  et~al.}{2016}]{Monteiro-Oliveira_2017}
Monteiro-Oliveira R.,  Cypriano E.~S.,  Machado R. E.~G.,  Lima~Neto G.~B.,
  Ribeiro A. L.~B.,  Sodré L. J.,   Dupke R.,  2016, \mn@doi [Monthly Notices
  of the Royal Astronomical Society] {10.1093/mnras/stw3238}, 466, 2614

\bibitem[\protect\citeauthoryear{Moura, Machado  \& Monteiro-Oliveira}{Moura
  et~al.}{2020}]{Moura_2020}
Moura M.~T.,  Machado R. E.~G.,   Monteiro-Oliveira R.,  2020, \mn@doi [Monthly
  Notices of the Royal Astronomical Society] {10.1093/mnras/staa3399}, 500,
  1858

\bibitem[\protect\citeauthoryear{Nadler, Banerjee, Adhikari, Mao  \&
  Wechsler}{Nadler et~al.}{2020}]{Nadler_2020}
Nadler E.~O.,  Banerjee A.,  Adhikari S.,  Mao Y.-Y.,   Wechsler R.~H.,  2020,
  \mn@doi [The Astrophysical Journal] {10.3847/1538-4357/ab94b0}, 896, 112

\bibitem[\protect\citeauthoryear{{Navarro}, {Frenk}  \& {White}}{{Navarro}
  et~al.}{1996}]{Navarro_1996}
{Navarro} J.~F.,  {Frenk} C.~S.,   {White} S. D.~M.,  1996, \mn@doi [\apj]
  {10.1086/177173}, \href
  {https://ui.adsabs.harvard.edu/abs/1996ApJ...462..563N} {462, 563}

\bibitem[\protect\citeauthoryear{{Peel}, {Lanusse}  \& {Starck}}{{Peel}
  et~al.}{2017}]{Peel_2017}
{Peel} A.,  {Lanusse} F.,   {Starck} J.-L.,  2017, \mn@doi [\apj]
  {10.3847/1538-4357/aa850d}, \href
  {https://ui.adsabs.harvard.edu/abs/2017ApJ...847...23P} {847, 23}

\bibitem[\protect\citeauthoryear{{Peter}, {Rocha}, {Bullock}  \&
  {Kaplinghat}}{{Peter} et~al.}{2013}]{Peter_2013}
{Peter} A. H.~G.,  {Rocha} M.,  {Bullock} J.~S.,   {Kaplinghat} M.,  2013,
  \mn@doi [\mnras] {10.1093/mnras/sts535}, \href
  {https://ui.adsabs.harvard.edu/abs/2013MNRAS.430..105P} {430, 105}

\bibitem[\protect\citeauthoryear{Pillepich et~al.,}{Pillepich
  et~al.}{2017}]{Pillepich_2017}
Pillepich A.,  et~al., 2017, \mn@doi [Monthly Notices of the Royal Astronomical
  Society] {10.1093/mnras/stx2656}, 473, 4077

\bibitem[\protect\citeauthoryear{Poole, Fardal, Babul, McCarthy, Quinn  \&
  Wadsley}{Poole et~al.}{2006}]{Pool_2006}
Poole G.~B.,  Fardal M.~A.,  Babul A.,  McCarthy I.~G.,  Quinn T.,   Wadsley
  J.,  2006, \mn@doi [Monthly Notices of the Royal Astronomical Society]
  {10.1111/j.1365-2966.2006.10916.x}, 373, 881

\bibitem[\protect\citeauthoryear{{Power}, {Navarro}, {Jenkins}, {Frenk},
  {White}, {Springel}, {Stadel}  \& {Quinn}}{{Power} et~al.}{2003}]{Power_2003}
{Power} C.,  {Navarro} J.~F.,  {Jenkins} A.,  {Frenk} C.~S.,  {White} S.~D.~M.,
   {Springel} V.,  {Stadel} J.,   {Quinn} T.,  2003, \mn@doi [\mnras]
  {10.1046/j.1365-8711.2003.05925.x}, \href
  {https://ui.adsabs.harvard.edu/abs/2003MNRAS.338...14P} {338, 14}

\bibitem[\protect\citeauthoryear{{Randall}, {Markevitch}, {Clowe}, {Gonzalez}
  \& {Brada{\v{c}}}}{{Randall} et~al.}{2008}]{Randall_2008}
{Randall} S.~W.,  {Markevitch} M.,  {Clowe} D.,  {Gonzalez} A.~H.,
  {Brada{\v{c}}} M.,  2008, \mn@doi [\apj] {10.1086/587859}, \href
  {https://ui.adsabs.harvard.edu/abs/2008ApJ...679.1173R} {679, 1173}

\bibitem[\protect\citeauthoryear{Read, Goerdt, Moore, Pontzen, Stadel  \&
  Lake}{Read et~al.}{2006}]{Read_2006}
Read J.~I.,  Goerdt T.,  Moore B.,  Pontzen A.~P.,  Stadel J.,   Lake G.,
  2006, \mn@doi [Monthly Notices of the Royal Astronomical Society]
  {10.1111/j.1365-2966.2006.11022.x}, 373, 1451

\bibitem[\protect\citeauthoryear{{Robertson}, {Massey}  \& {Eke}}{{Robertson}
  et~al.}{2017a}]{Robertson_2017a}
{Robertson} A.,  {Massey} R.,   {Eke} V.,  2017a, \mn@doi [\mnras]
  {10.1093/mnras/stw2670}, \href
  {https://ui.adsabs.harvard.edu/abs/2017MNRAS.465..569R} {465, 569}

\bibitem[\protect\citeauthoryear{{Robertson}, {Massey}  \& {Eke}}{{Robertson}
  et~al.}{2017b}]{Robertson_2017b}
{Robertson} A.,  {Massey} R.,   {Eke} V.,  2017b, \mn@doi [\mnras]
  {10.1093/mnras/stx463}, \href
  {https://ui.adsabs.harvard.edu/abs/2017MNRAS.467.4719R} {467, 4719}

\bibitem[\protect\citeauthoryear{Robertson, Massey, Eke, Schaye  \&
  Theuns}{Robertson et~al.}{2020}]{Robertson_2020}
Robertson A.,  Massey R.,  Eke V.,  Schaye J.,   Theuns T.,  2020, \mn@doi
  [Monthly Notices of the Royal Astronomical Society] {10.1093/mnras/staa3954},
  501, 4610–4634

\bibitem[\protect\citeauthoryear{Rocha, Peter, Bullock, Kaplinghat,
  Garrison-Kimmel, O\~{n}orbe  \& Moustakas}{Rocha et~al.}{2013}]{Rocha_2013}
Rocha M.,  Peter A. H.~G.,  Bullock J.~S.,  Kaplinghat M.,  Garrison-Kimmel S.,
   O\~{n}orbe J.,   Moustakas L.~A.,  2013, \mn@doi [Monthly Notices of the
  Royal Astronomical Society] {10.1093/mnras/sts514}, 430, 81

\bibitem[\protect\citeauthoryear{Sagunski, Gad-Nasr, Colquhoun, Robertson  \&
  Tulin}{Sagunski et~al.}{2021}]{Sagunski_2021}
Sagunski L.,  Gad-Nasr S.,  Colquhoun B.,  Robertson A.,   Tulin S.,  2021,
  \mn@doi [Journal of Cosmology and Astroparticle Physics]
  {10.1088/1475-7516/2021/01/024}, 2021, 024

\bibitem[\protect\citeauthoryear{Sameie et~al.,}{Sameie
  et~al.}{2021}]{Sameie_2021}
Sameie O.,  et~al., 2021, \mn@doi [Monthly Notices of the Royal Astronomical
  Society] {10.1093/mnras/stab2173}, 507, 720

\bibitem[\protect\citeauthoryear{Shen, Hopkins, Necib, Jiang, Boylan-Kolchin
  \& Wetzel}{Shen et~al.}{2021}]{Shen_2021}
Shen X.,  Hopkins P.~F.,  Necib L.,  Jiang F.,  Boylan-Kolchin M.,   Wetzel A.,
   2021, \mn@doi [Monthly Notices of the Royal Astronomical Society]
  {10.1093/mnras/stab2042}, 506, 4421

\bibitem[\protect\citeauthoryear{Spergel \& Steinhardt}{Spergel \&
  Steinhardt}{2000}]{Spergel_2000}
Spergel D.~N.,  Steinhardt P.~J.,  2000, \mn@doi [Physical Review Letters]
  {10.1103/physrevlett.84.3760}, 84, 3760

\bibitem[\protect\citeauthoryear{{Springel}}{{Springel}}{2005}]{gadget2}
{Springel} V.,  2005, \mn@doi [\mnras] {10.1111/j.1365-2966.2005.09655.x},
  \href {https://ui.adsabs.harvard.edu/abs/2005MNRAS.364.1105S} {364, 1105}

\bibitem[\protect\citeauthoryear{Springel \& Farrar}{Springel \&
  Farrar}{2007}]{Springel_2007}
Springel V.,  Farrar G.~R.,  2007, \mn@doi [Monthly Notices of the Royal
  Astronomical Society] {10.1111/j.1365-2966.2007.12159.x}, 380, 911

\bibitem[\protect\citeauthoryear{Springel et~al.,}{Springel
  et~al.}{2005}]{Springel_2005n}
Springel V.,  et~al., 2005, \mn@doi [Nature] {10.1038/nature03597}, 435,
  629–636

\bibitem[\protect\citeauthoryear{Springel, Frenk  \& White}{Springel
  et~al.}{2006}]{Springel_2006}
Springel V.,  Frenk C.~S.,   White S. D.~M.,  2006, \mn@doi [Nature]
  {10.1038/nature04805}, 440, 1137–1144

\bibitem[\protect\citeauthoryear{{Taylor}, {Massey}, {Jauzac}, {Courbin},
  {Harvey}, {Joseph}  \& {Robertson}}{{Taylor} et~al.}{2017}]{Taylor_2017}
{Taylor} P.,  {Massey} R.,  {Jauzac} M.,  {Courbin} F.,  {Harvey} D.,  {Joseph}
  R.,   {Robertson} A.,  2017, \mn@doi [\mnras] {10.1093/mnras/stx855}, \href
  {https://ui.adsabs.harvard.edu/abs/2017MNRAS.468.5004T} {468, 5004}

\bibitem[\protect\citeauthoryear{{Tulin} \& {Yu}}{{Tulin} \&
  {Yu}}{2018}]{Tulin_2018}
{Tulin} S.,  {Yu} H.-B.,  2018, \mn@doi [\physrep]
  {10.1016/j.physrep.2017.11.004}, \href
  {https://ui.adsabs.harvard.edu/abs/2018PhR...730....1T} {730, 1}

\bibitem[\protect\citeauthoryear{Tulin, Yu  \& Zurek}{Tulin
  et~al.}{2013}]{Tulin_2013a}
Tulin S.,  Yu H.-B.,   Zurek K.~M.,  2013, \mn@doi [Phys. Rev. D]
  {10.1103/PhysRevD.87.115007}, 87, 115007

\bibitem[\protect\citeauthoryear{Turner, Lovell, Zavala  \&
  Vogelsberger}{Turner et~al.}{2021}]{Turner_2021}
Turner H.~C.,  Lovell M.~R.,  Zavala J.,   Vogelsberger M.,  2021, \mn@doi
  [Monthly Notices of the Royal Astronomical Society] {10.1093/mnras/stab1725},
  505, 5327

\bibitem[\protect\citeauthoryear{Vargya, Sanderson, Sameie, Boylan-Kolchin,
  Hopkins, Wetzel  \& Graus}{Vargya et~al.}{2021}]{Vargya_2021}
Vargya D.,  Sanderson R.,  Sameie O.,  Boylan-Kolchin M.,  Hopkins P.~F.,
  Wetzel A.,   Graus A.,  2021, Shapes of Milky-Way-Mass Galaxies with
  Self-Interacting Dark Matter (\mn@eprint {arXiv} {2104.14069})

\bibitem[\protect\citeauthoryear{Vega-Ferrero, Dana, Diego, Yepes, Cui  \&
  Meneghetti}{Vega-Ferrero et~al.}{2020}]{Vega-Ferrero_2020}
Vega-Ferrero J.,  Dana J.~M.,  Diego J.~M.,  Yepes G.,  Cui W.,   Meneghetti
  M.,  2020, \mn@doi [Monthly Notices of the Royal Astronomical Society]
  {10.1093/mnras/staa3235}, 500, 247

\bibitem[\protect\citeauthoryear{Virtanen et~al.,}{Virtanen
  et~al.}{2020}]{SciPy}
Virtanen P.,  et~al., 2020, \mn@doi [Nature Meth.] {10.1038/s41592-019-0686-2},
  \href {https://rdcu.be/b08Wh} {17, 261}

\bibitem[\protect\citeauthoryear{{Vogelsberger} \& {Zavala}}{{Vogelsberger} \&
  {Zavala}}{2013}]{Vogelsberger_2013}
{Vogelsberger} M.,  {Zavala} J.,  2013, \mn@doi [\mnras]
  {10.1093/mnras/sts712}, \href
  {https://ui.adsabs.harvard.edu/abs/2013MNRAS.430.1722V} {430, 1722}

\bibitem[\protect\citeauthoryear{Vogelsberger, Zavala  \& Loeb}{Vogelsberger
  et~al.}{2012}]{Vogelsberger_2012}
Vogelsberger M.,  Zavala J.,   Loeb A.,  2012, \mn@doi [Monthly Notices of the
  Royal Astronomical Society] {10.1111/j.1365-2966.2012.21182.x}, 423, 3740

\bibitem[\protect\citeauthoryear{{Vogelsberger}, {Zavala}, {Simpson}  \&
  {Jenkins}}{{Vogelsberger} et~al.}{2014}]{Vogelsberger_2014}
{Vogelsberger} M.,  {Zavala} J.,  {Simpson} C.,   {Jenkins} A.,  2014, \mn@doi
  [\mnras] {10.1093/mnras/stu1713}, \href
  {https://ui.adsabs.harvard.edu/abs/2014MNRAS.444.3684V} {444, 3684}

\bibitem[\protect\citeauthoryear{{Wittman}, {Golovich}  \& {Dawson}}{{Wittman}
  et~al.}{2018}]{Wittman_2018}
{Wittman} D.,  {Golovich} N.,   {Dawson} W.~A.,  2018, \mn@doi [\apj]
  {10.3847/1538-4357/aaee77}, \href
  {https://ui.adsabs.harvard.edu/abs/2018ApJ...869..104W} {869, 104}

\bibitem[\protect\citeauthoryear{Zemp, Gnedin, Gnedin  \& Kravtsov}{Zemp
  et~al.}{2011}]{Zemp_2011}
Zemp M.,  Gnedin O.~Y.,  Gnedin N.~Y.,   Kravtsov A.~V.,  2011, \mn@doi [ApJSS]
  {10.1088/0067-0049/197/2/30}, 197, 30

\bibitem[\protect\citeauthoryear{Zhang, Yu  \& Lu}{Zhang
  et~al.}{2015}]{Zhang_2015}
Zhang C.,  Yu Q.,   Lu Y.,  2015, \mn@doi [The Astrophysical Journal]
  {10.1088/0004-637x/813/2/129}, 813, 129

\bibitem[\protect\citeauthoryear{Zhang, Yu  \& Lu}{Zhang
  et~al.}{2016}]{Zhang_2016}
Zhang C.,  Yu Q.,   Lu Y.,  2016, \mn@doi [The Astrophysical Journal]
  {10.3847/0004-637x/820/2/85}, 820, 85

\bibitem[\protect\citeauthoryear{ZuHone}{ZuHone}{2011}]{ZuHone_2011}
ZuHone J.~A.,  2011, \mn@doi [The Astrophysical Journal]
  {10.1088/0004-637x/728/1/54}, 728, 54

\bibitem[\protect\citeauthoryear{van~den Aarssen, Bringmann  \&
  Pfrommer}{van~den Aarssen et~al.}{2012}]{van_den_Aarssen_2012}
van~den Aarssen L.~G.,  Bringmann T.,   Pfrommer C.,  2012, \mn@doi [Phys. Rev.
  Lett.] {10.1103/PhysRevLett.109.231301}, 109, 231301

\makeatother
\end{thebibliography}




\appendix

\section{Relabelling particles for rSIDM} \label{sec:rSIDMa_comp}

\begin{figure}
    \centering
    \includegraphics[width=\columnwidth]{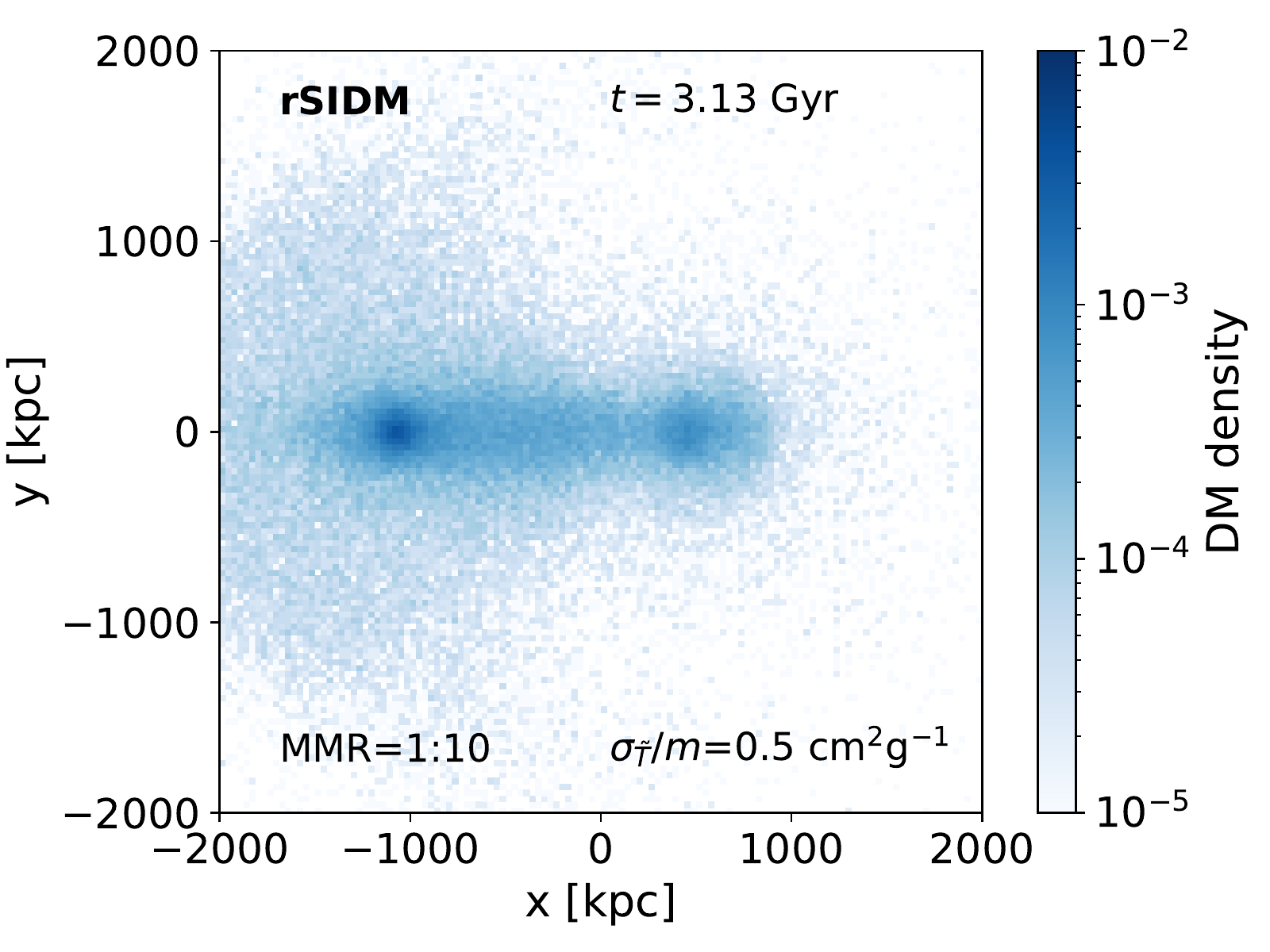}
    \includegraphics[width=\columnwidth]{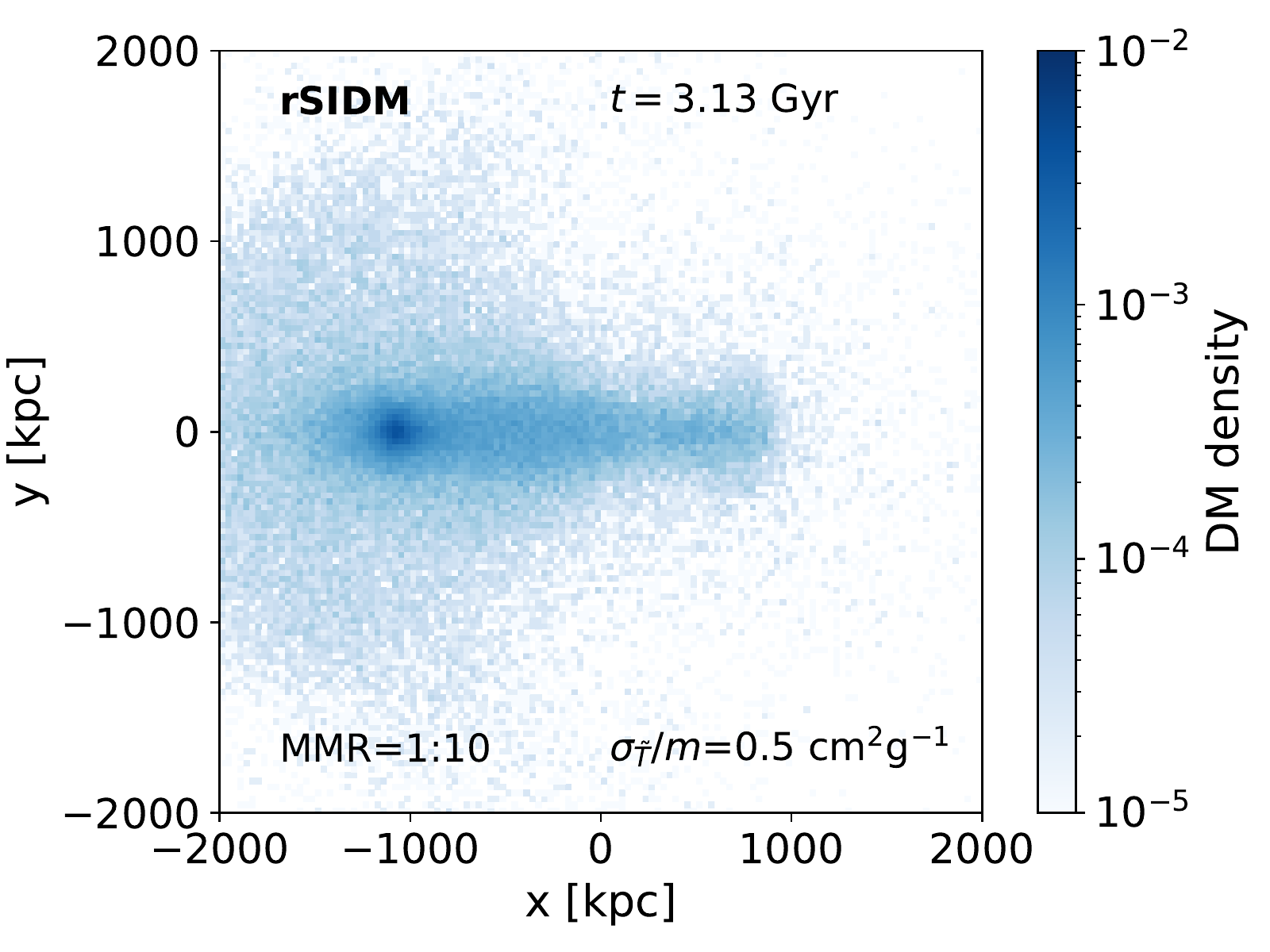}
    \caption{The physical density of the subhalo of a merging system in the plane of the merger is shown at about the first apocentre passage.
    The system is a 1:10 merger simulated with rare self-interactions ($\sigma_\mathrm{\Tilde{T}}/m = 0.5\,\mathrm{cm}^2\,\mathrm{g}^{-1}$).
    The simulation of the upper panel allowed for scattering angles larger than 90°, whereas the lower panel does not.}
    \label{fig:dens_maps_R10}
\end{figure}

For rare self-interactions, large scattering angles are common.
When two particles (one from each merging halo) scatter, a scattering angle larger than 90° could be interpreted as an exchange of particles.
If so, it would make sense to switch the labels of the corresponding particles.
We implemented this by limiting the maximum scattering angle to 90°.
In Fig.~\ref{fig:dens_maps_R10}, we show the density of the subhalo of a 1:10 merger for both rSIDM versions at about the first apocentre passage.
In the upper panel with and in the lower one without relabelling.
One can see that this affects the formation of a second peak at the position of the main halo.

The relabelling procedure suppresses the second peak which in consequence leads to less elliptical haloes.
However, how much the shape is affected depends on the particles which are considered in the inertia tensor calculation.
If only the matter within twice the scale radius of the initial NFW profile is taken into account, the effects from the relabelling procedure are negligible, but if all particles are considered, it can make a significant difference.

\section{Morphology} \label{sec:additional_plots_morphology}

\begin{figure*}
    \centering
    \includegraphics[width=\columnwidth]{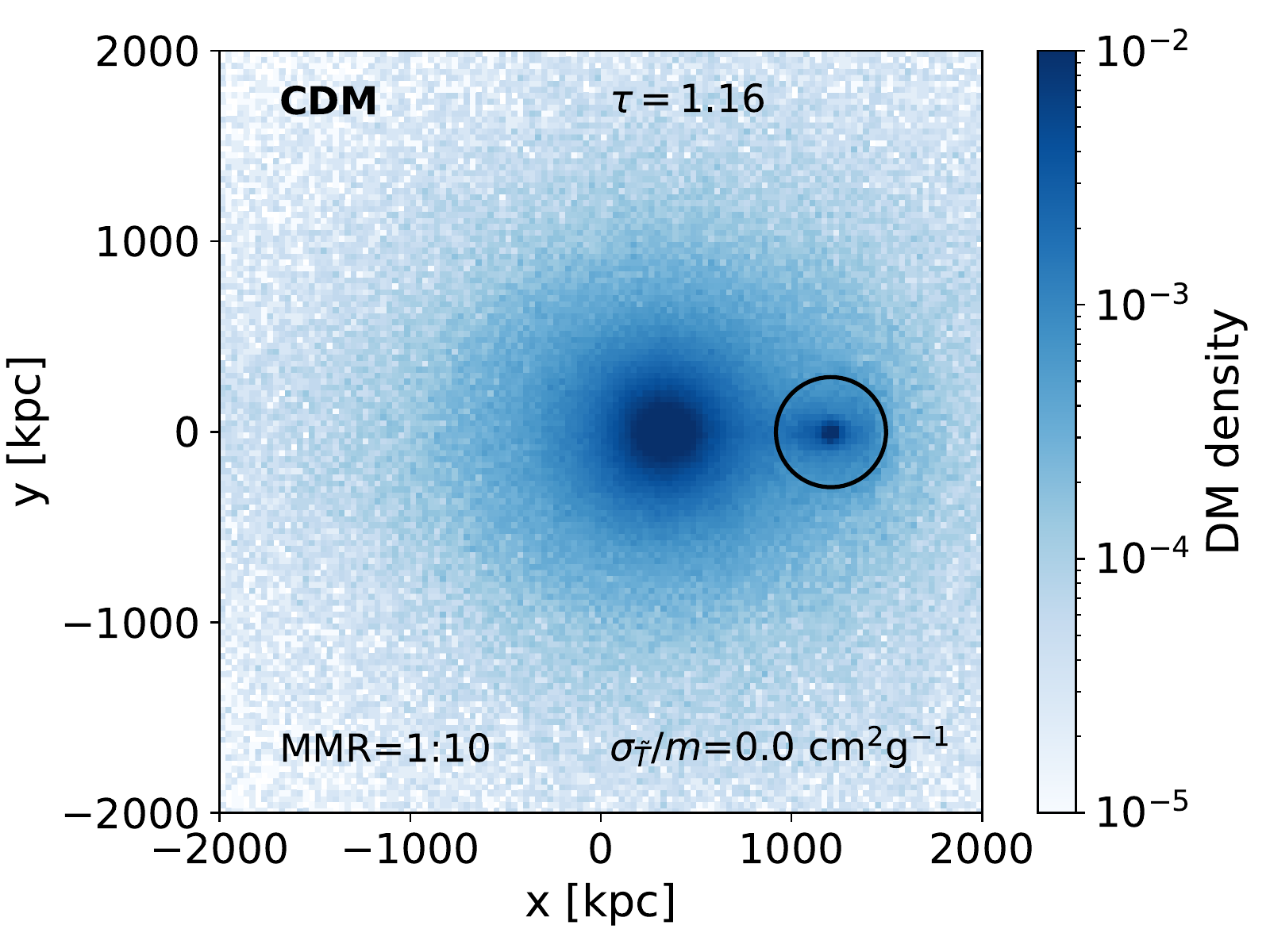}
    \includegraphics[width=\columnwidth]{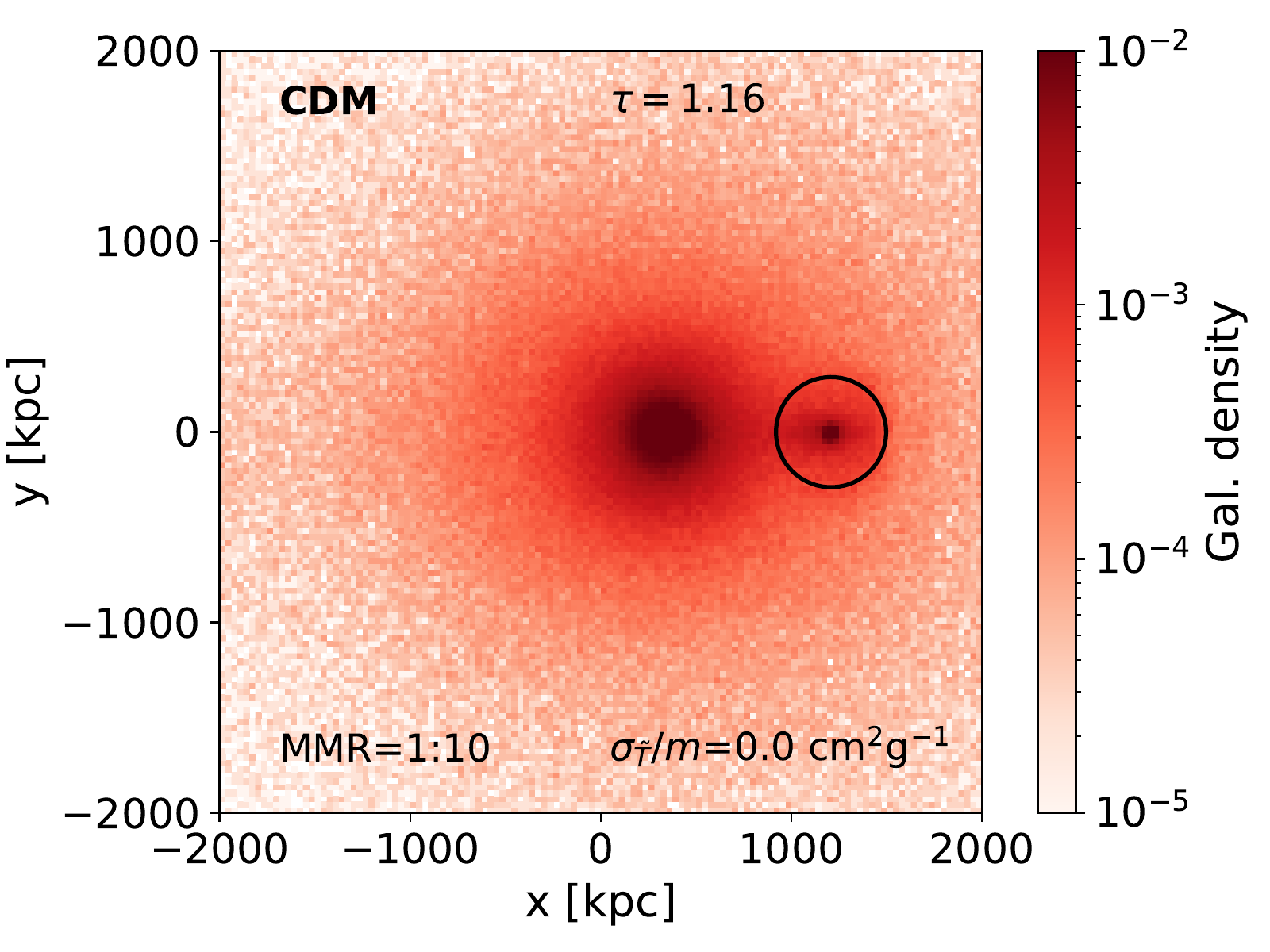}
    \includegraphics[width=\columnwidth]{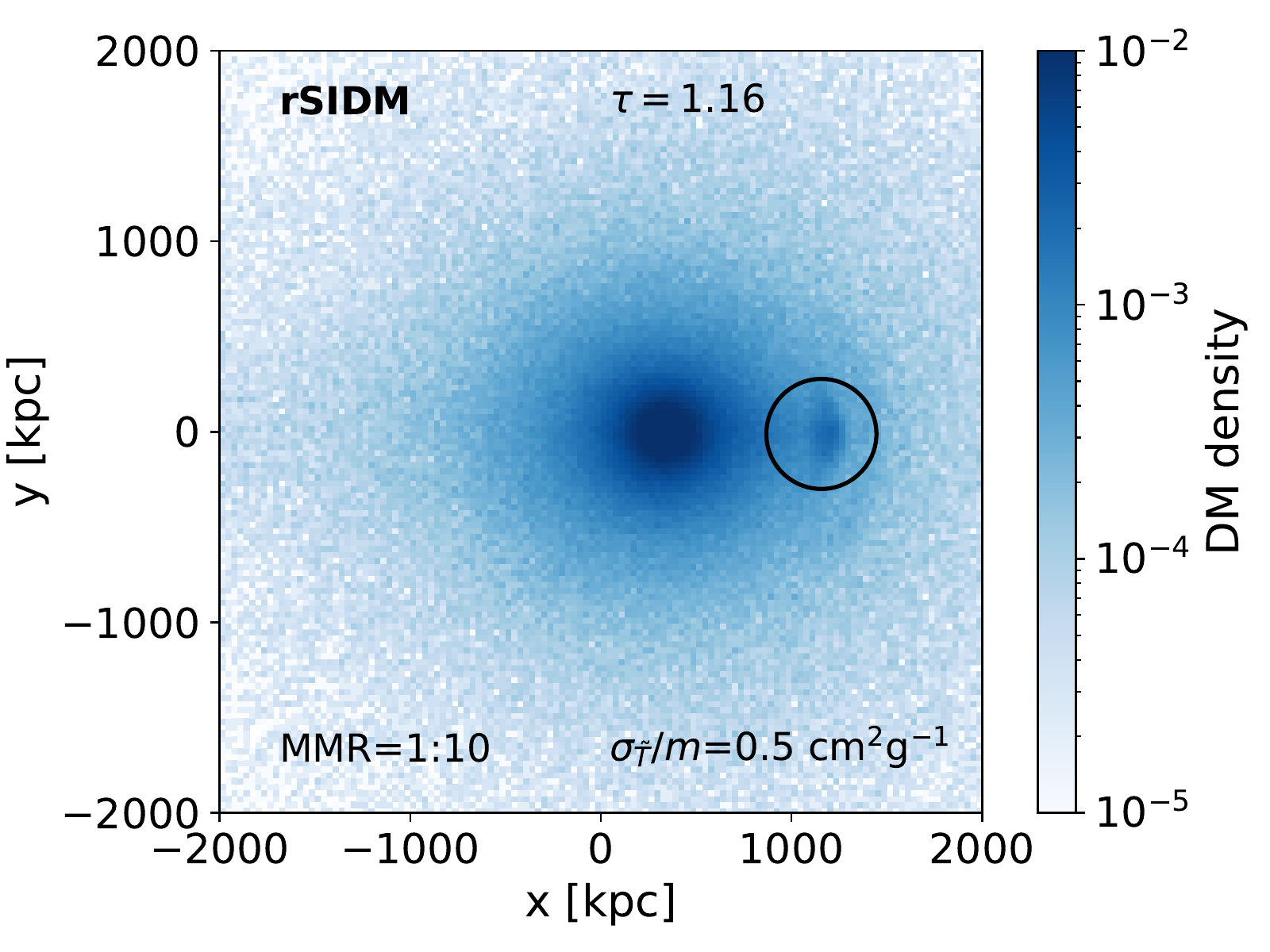}
    \includegraphics[width=\columnwidth]{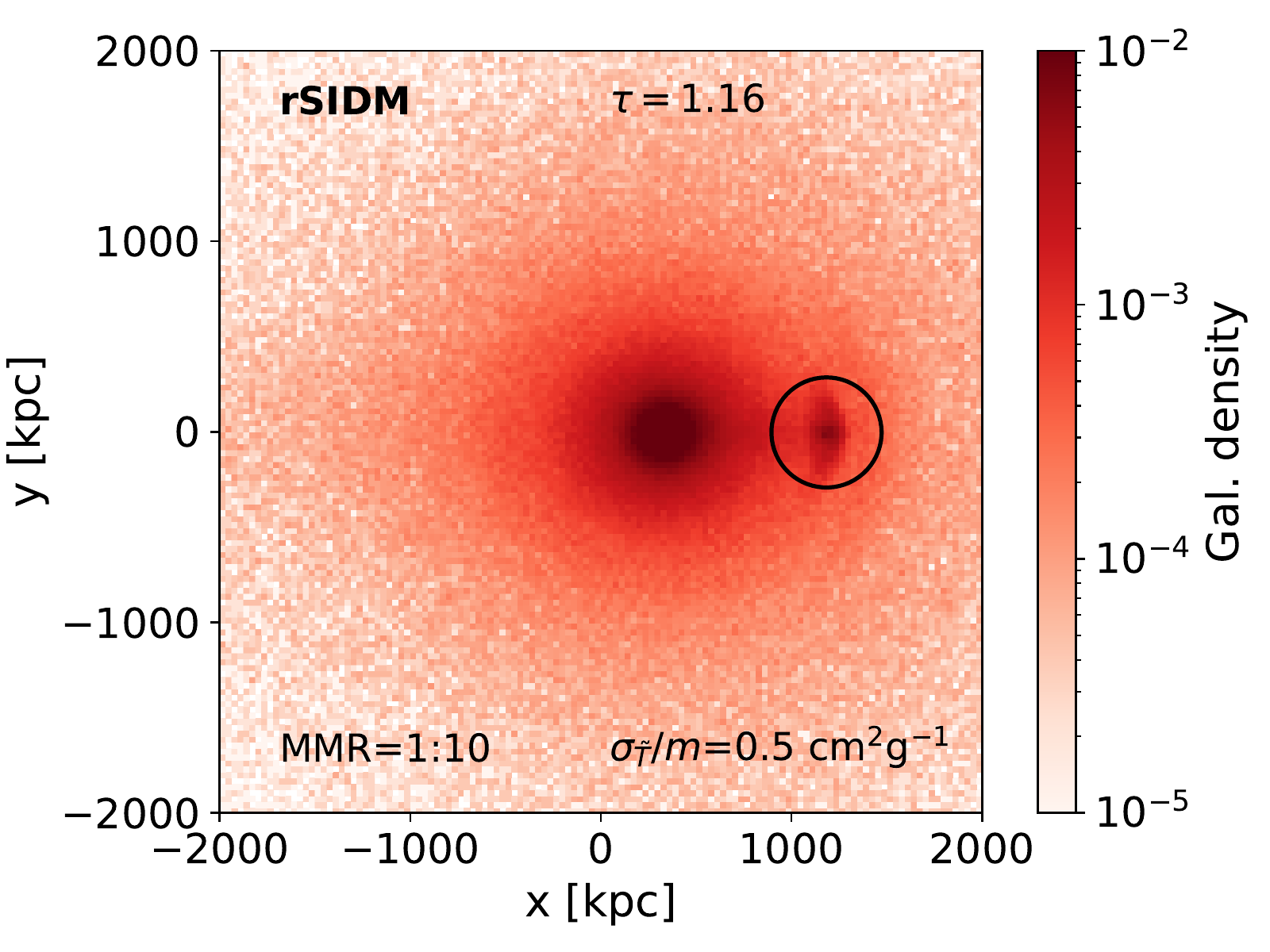}
    \includegraphics[width=\columnwidth]{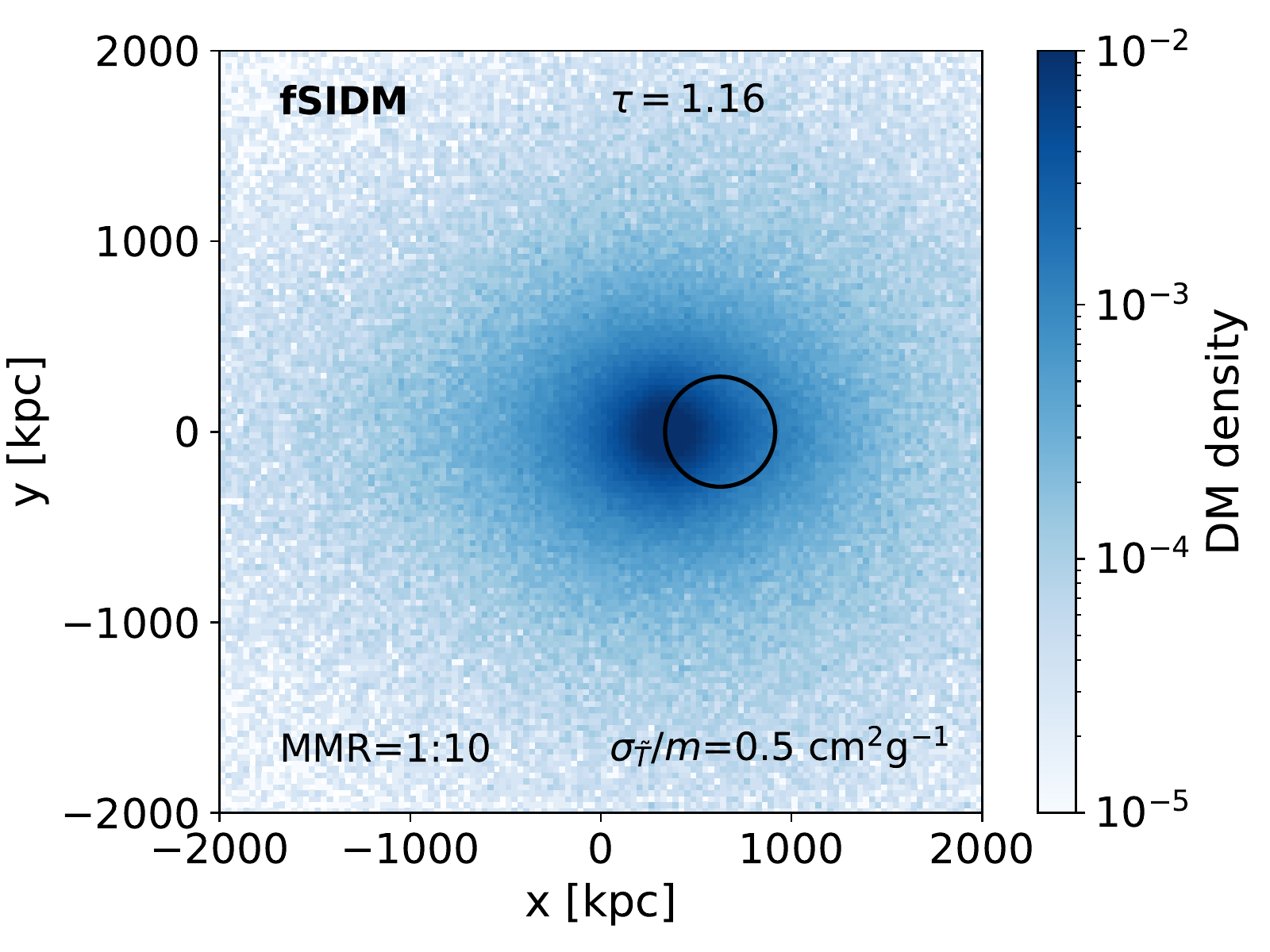}
    \includegraphics[width=\columnwidth]{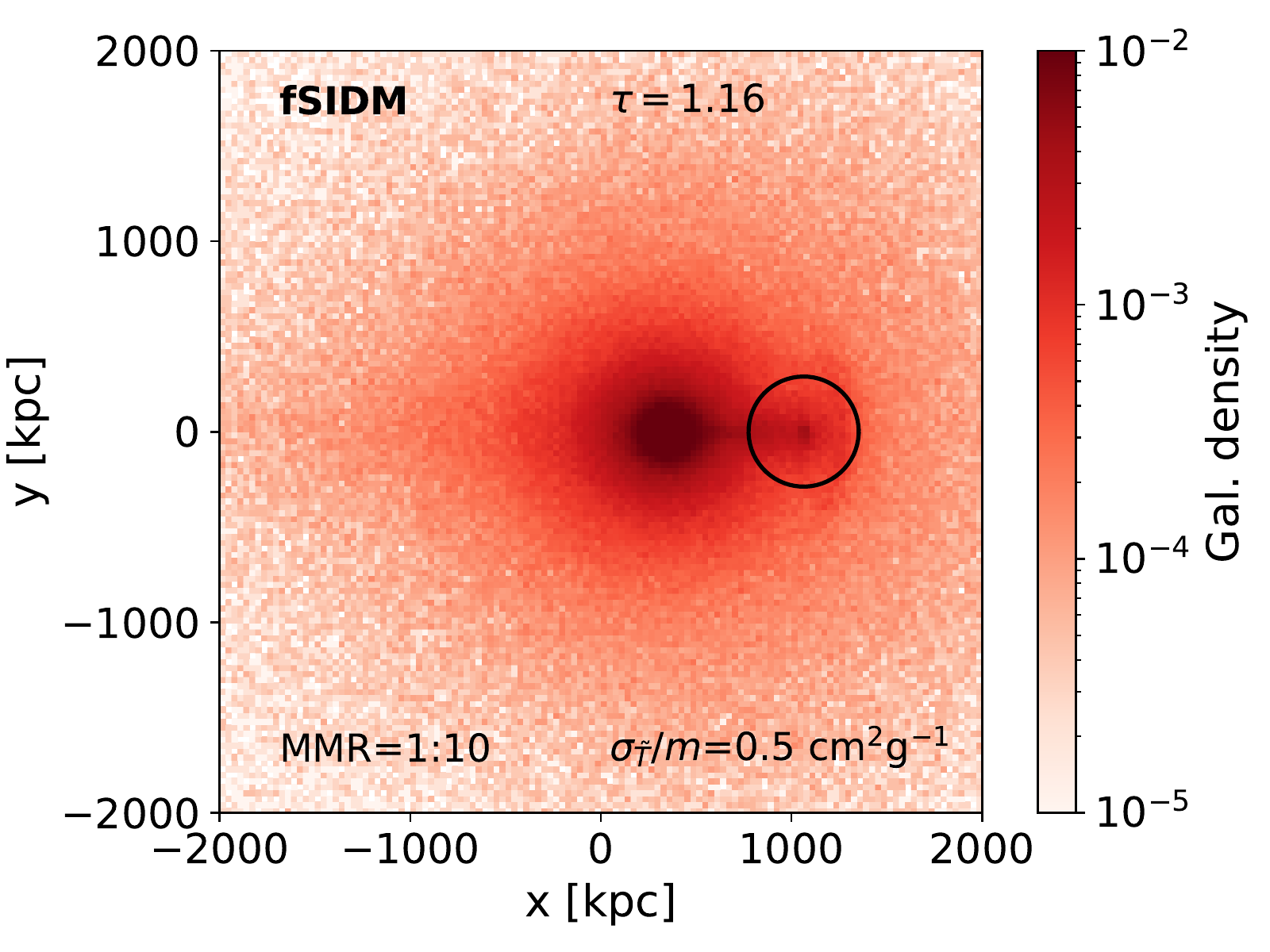}
    \caption{The same as in Fig.~\ref{fig:dens_map_R10}, but considering both haloes. In the supplementary material, we provide the time evolution as a video.}
    \label{fig:dens_map_R10_2}
\end{figure*}

In Fig.~\ref{fig:dens_map_R10_2}, we show the same plots as in Fig.~\ref{fig:dens_map_R10} but this time including the main halo.
Due to the density contribution of the main halo which has 10 times the mass of the subhalo, it becomes more difficult to identify morphological features.
Nevertheless, differences between the DM models are clearly visible.

\section{Peak position, Offset, and Shape} \label{sec:additional_plots_comp}

\begin{figure}
    \centering
    \includegraphics[width=\columnwidth]{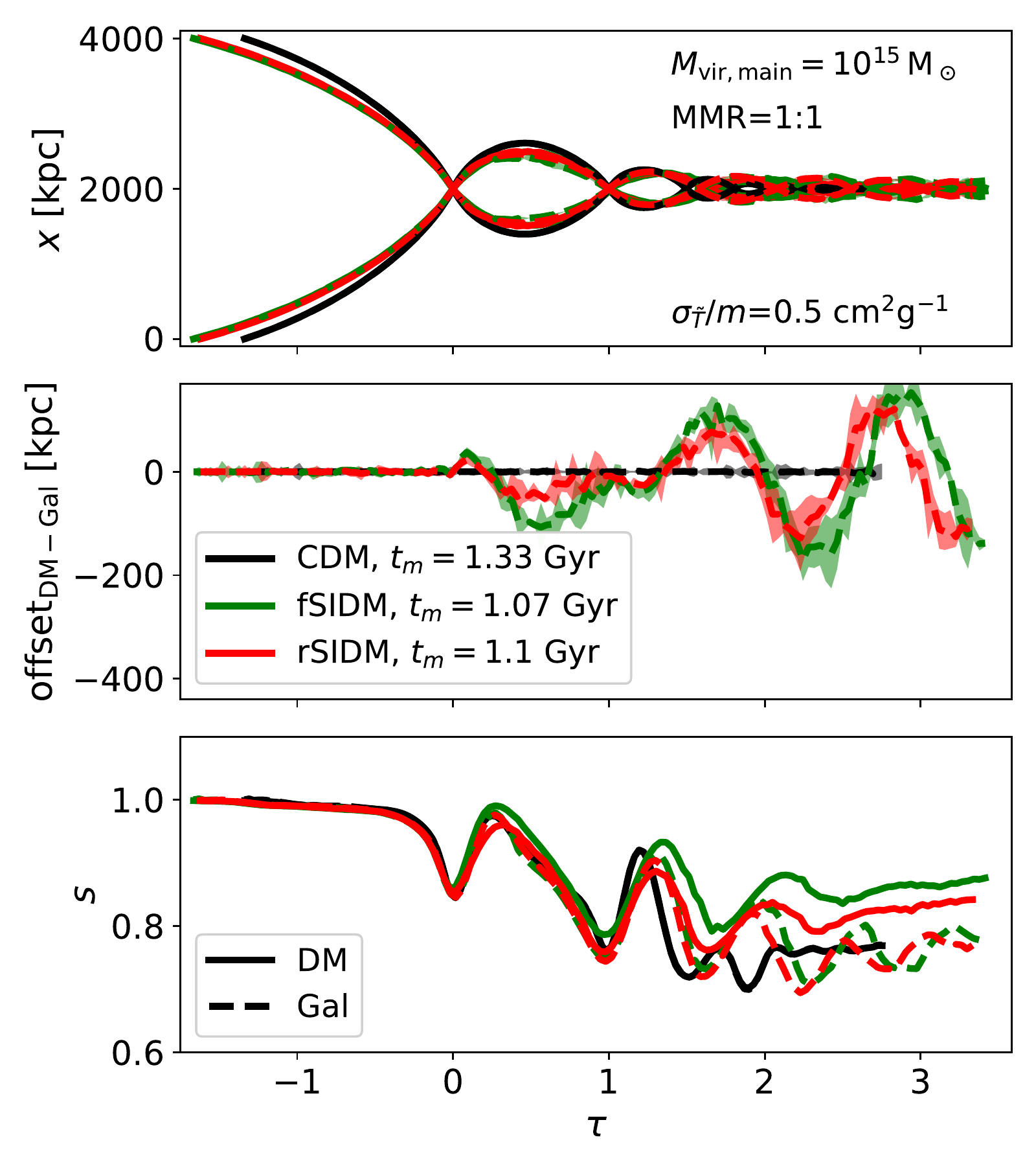}
    \caption{Peak position along the merger axis (top panel), the DM-galaxy offset (middle panel), and the shape (bottom panel) for an equal mass merger as a function of time.
    Results for CDM are shown in black, for rSIDM in red, and for fSIDM in green. The SIDM runs were conducted with a cross-section of $\sigma_\mathrm{\tilde{T}}/m = 0.5 \, \mathrm{cm}^2 \, \mathrm{g}^{-1}$.
    Offsets and shapes are shown only for the subhalo.
    The DM component is indicated by a solid line and the galaxies by a dashed line.}
    \label{fig:main_R1_S5}
\end{figure}

\begin{figure}
    \centering
    \includegraphics[width=\columnwidth]{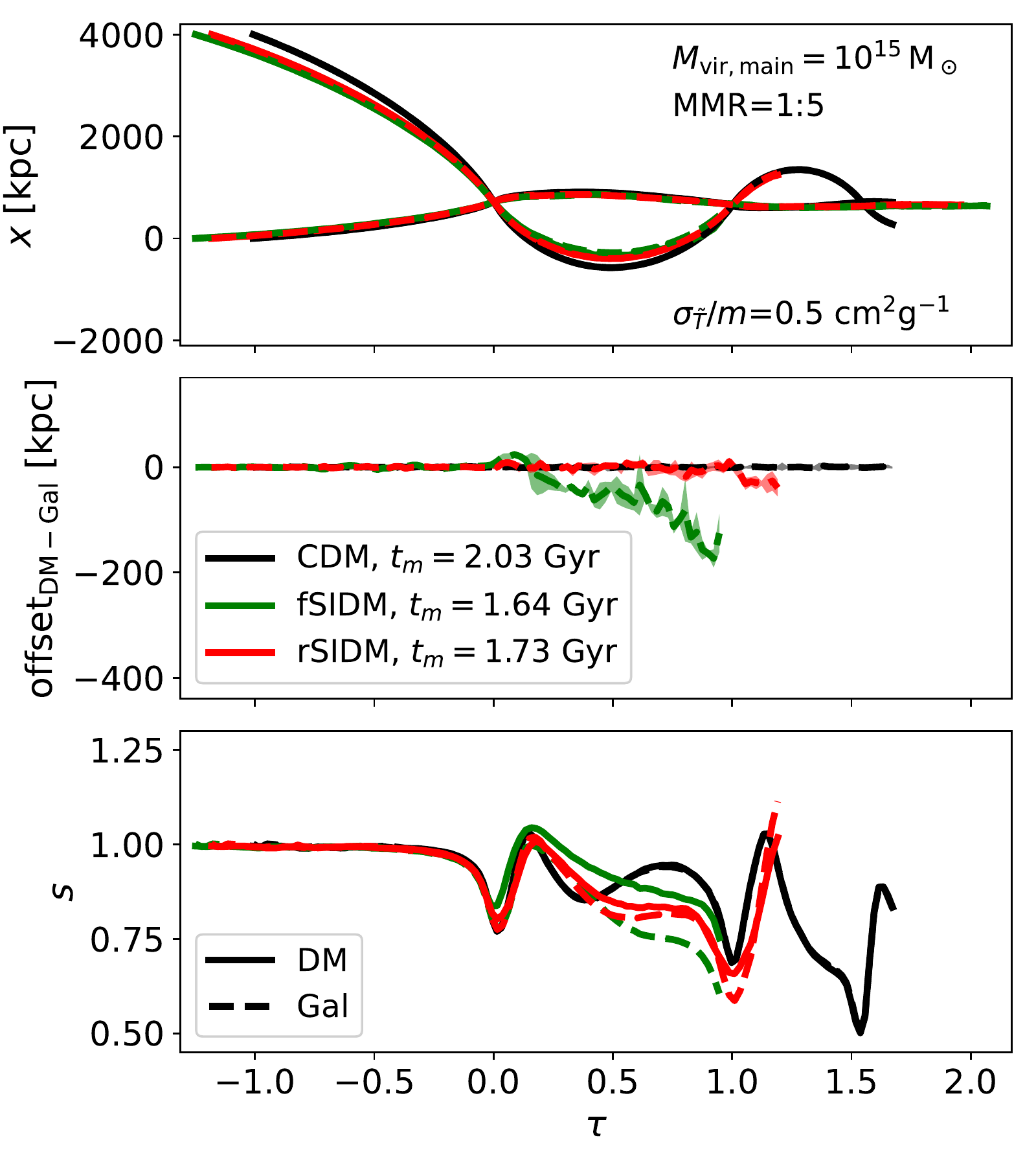}
    \caption{The same as in Fig.~\ref{fig:main_R1_S5}, but for an MMR of 1:5.}
    \label{fig:main_R5_S5}
\end{figure}

\begin{figure}
    \centering
    \includegraphics[width=\columnwidth]{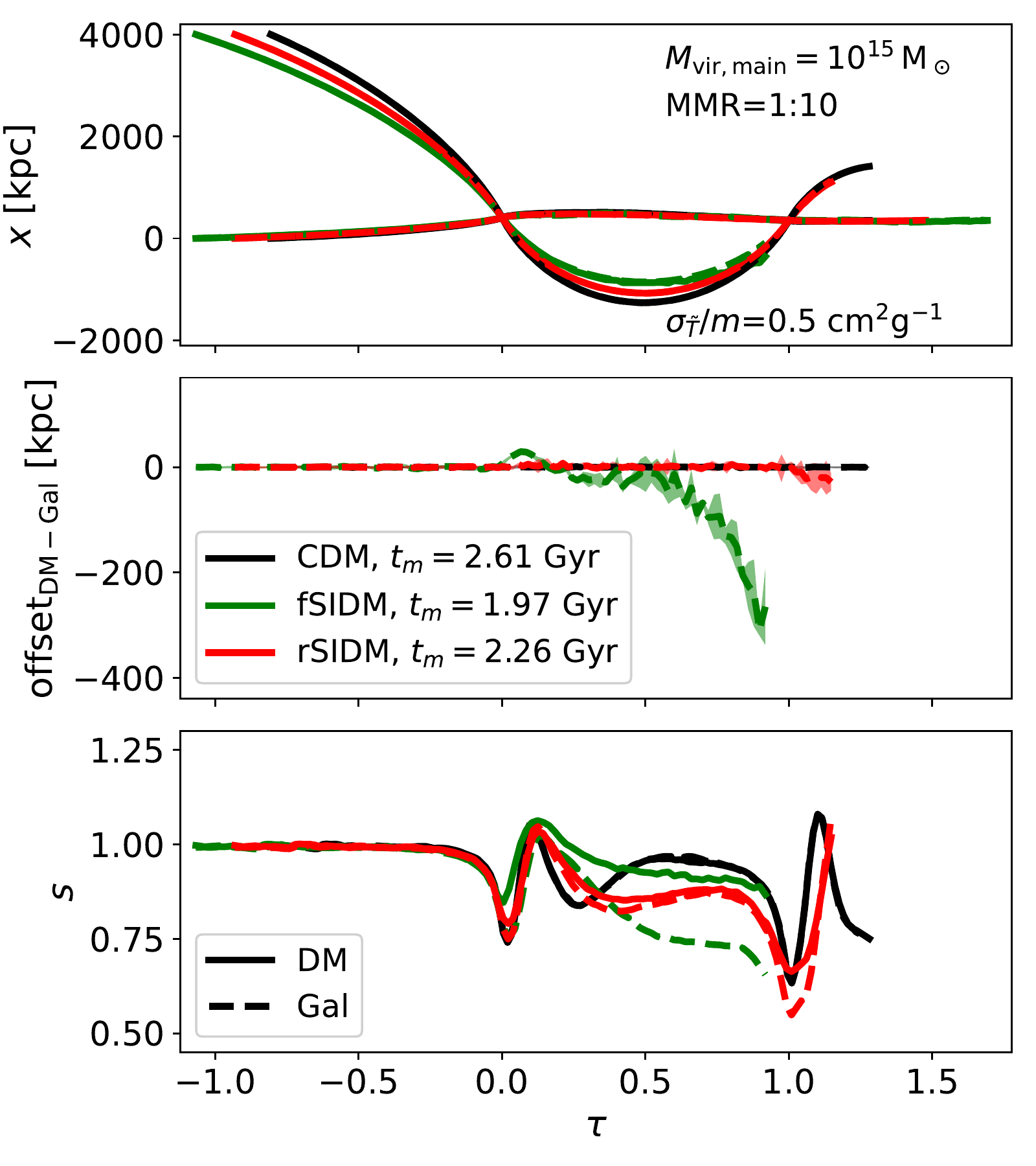}
    \caption{The same as in Fig.~\ref{fig:main_R1_S5}, but for an MMR of 1:10.}
    \label{fig:main_R10_S5}
\end{figure}

\begin{figure}
    \centering
    \includegraphics[width=\columnwidth]{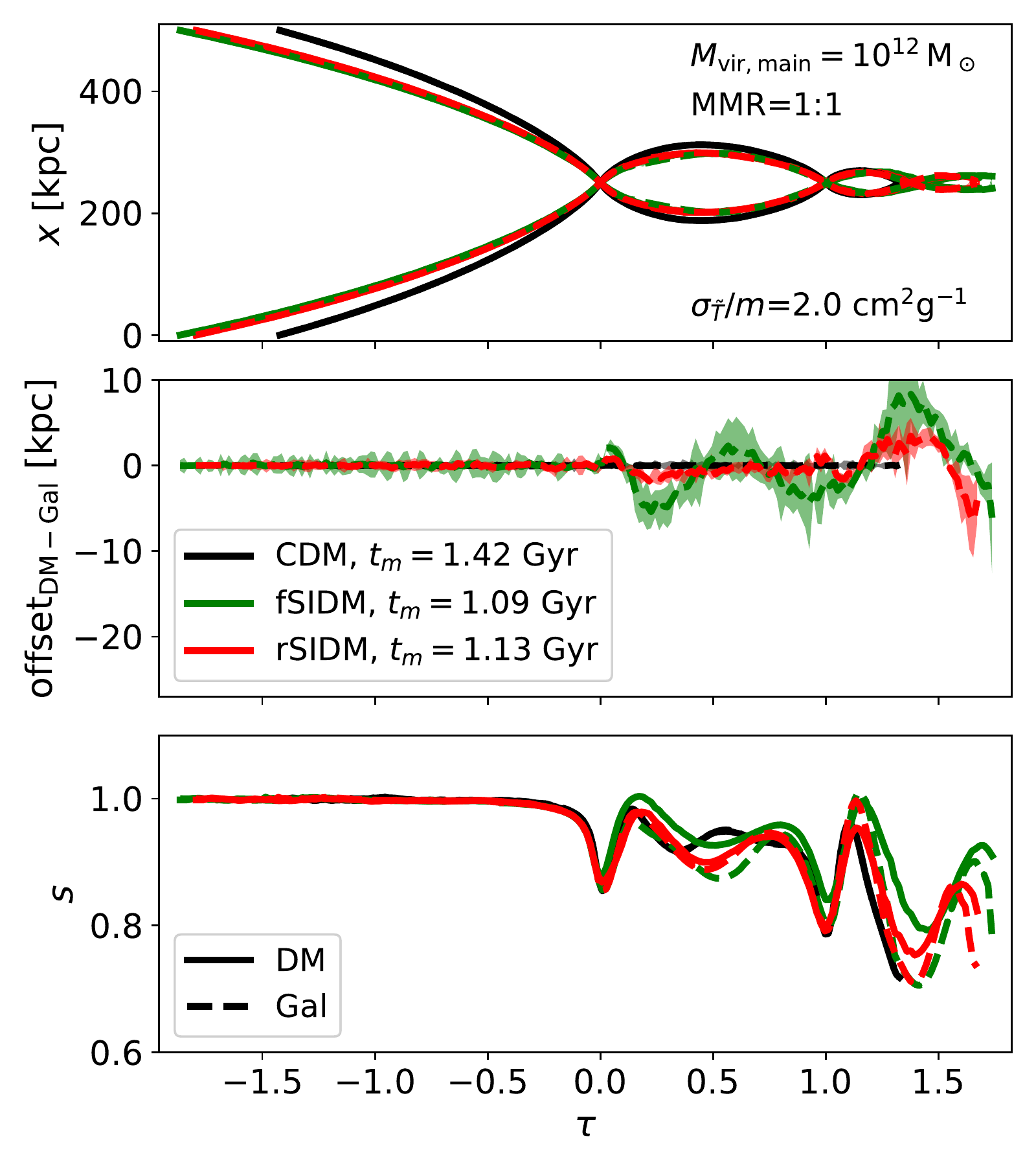}
    \caption{The same as in Fig.~\ref{fig:main_R1_S5}, but for the galaxy-scale runs and with a momentum transfer cross-section of $\sigma_\mathrm{\Tilde{T}}/m = 2.0 \, \mathrm{cm}^2\,\mathrm{g}^{-1}$.}
    \label{fig:main_R1_S20}
\end{figure}

\begin{figure}
    \centering
    \includegraphics[width=\columnwidth]{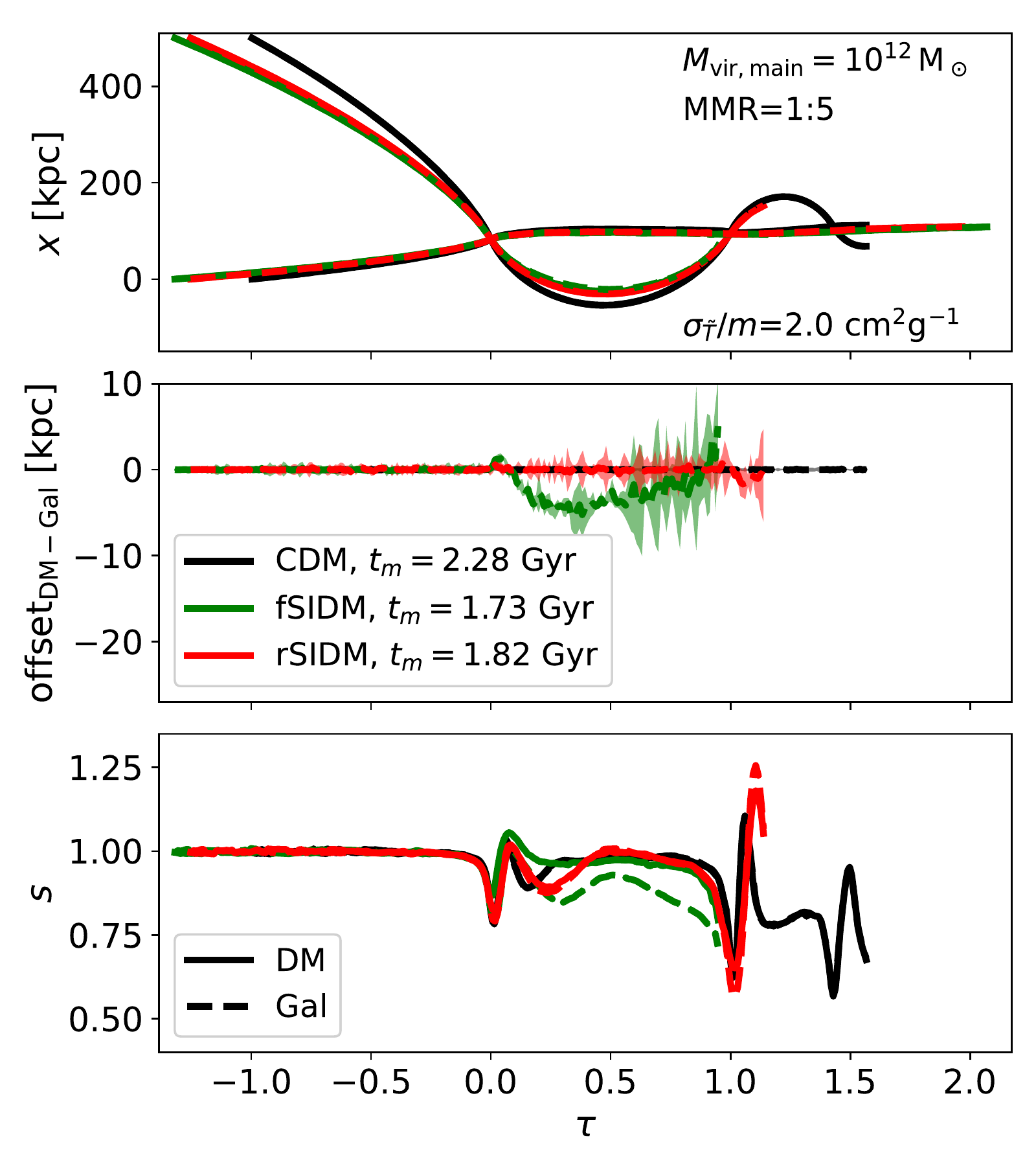}
    \caption{The same as in Fig.~\ref{fig:main_R1_S20}, but for an MMR of 1:5.}
    \label{fig:main_R5_S20}
\end{figure}

\begin{figure}
    \centering
    \includegraphics[width=\columnwidth]{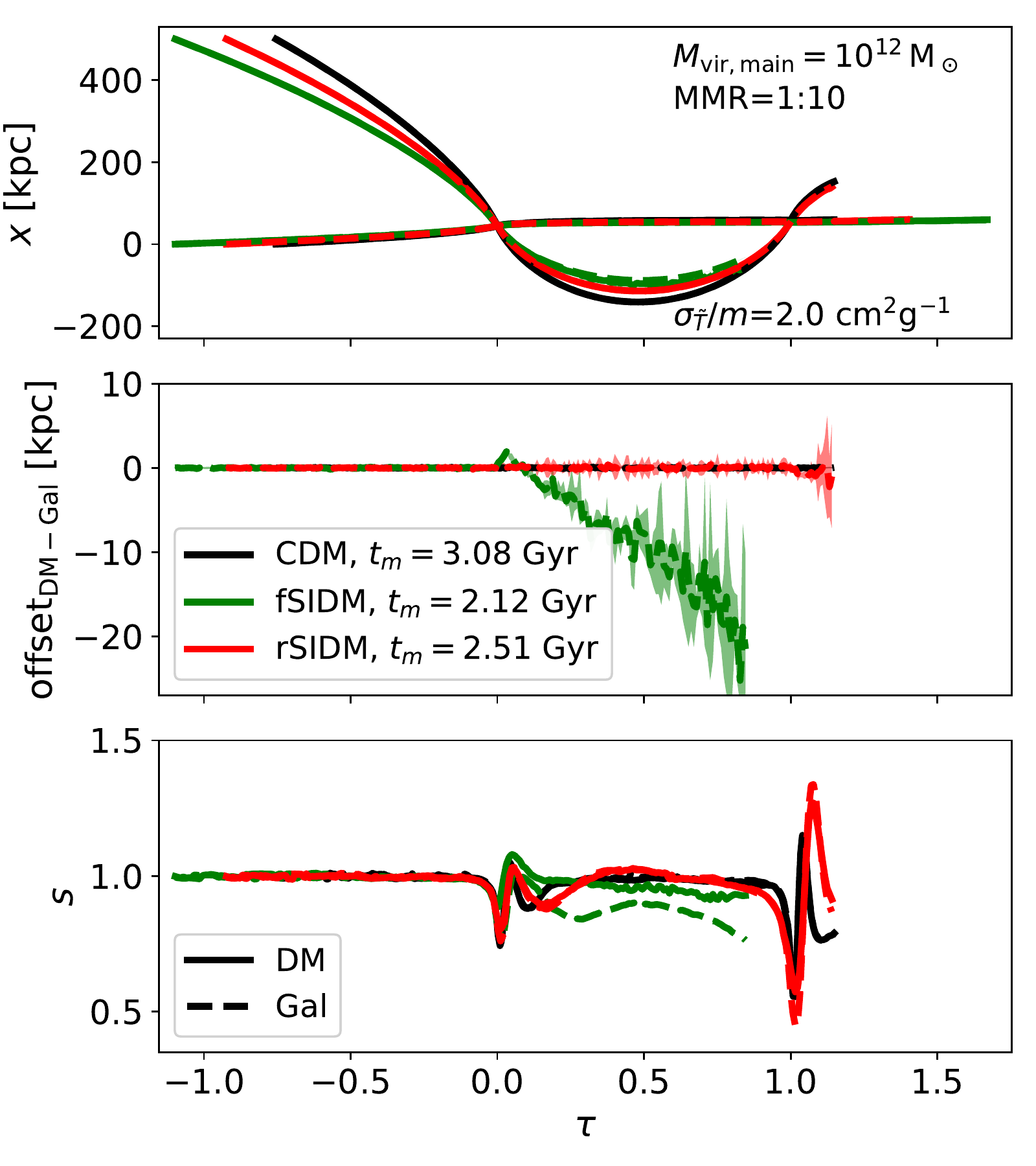}
    \caption{The same as in Fig.~\ref{fig:main_R1_S20}, but for an MMR of 1:10.}
    \label{fig:main_R10_S20}
\end{figure}

In this appendix, we provide further plots of the merger simulations with the highest cross-section we modelled.
In Fig.~\ref{fig:main_R1_S5} - \ref{fig:main_R10_S20}, we show the potential based peak position together with the offset (Eq.~\eqref{eq:offset}) and the halo shape (Eq.~\eqref{eq:shape}).

\bsp	
\label{lastpage}
\end{document}